**UNIVERSITY OF PATRAS**
DEPARTMENT OF ELECTRICAL AND
COMPUTER ENGINEERING
DIVISION: SYSTEMS AND CONTROL
LABORATORY OF AUTOMATION AND ROBOTICS

# Diploma Thesis
of the student
of the Department of Electrical and Computer Engineering
of the School of Engineering
of the University of Patras

Sapsanis A. Christos

Registration Number: 7119

Topic

**«Recognition of basic hand movements using Electromyography»**

Supervisor
Anthony Tzes

**Diploma Thesis Registration Number:**

Patras, June 2013

# CERTIFICATION

It is certified that the Diploma Thesis with topic

## «Recognition of basic hand movements using Electromyography»

of the student of the Department of Electrical and Computer Engineering

Sapsanis A. Christos

Registration Number: 7119

was publicly presented and examined in the Department of Electrical and Computer Engineering on the
21/06/2013

| **The Supervisor** | **Division Head** |
| --- | --- |
| Tzes Anthony<br>Professor | Kousoulas Nikolaos<br>Professor |

**Diploma Thesis Registration Number:**

**Topic: «Recognition of basic hand movements using Electromyography»**

Student: Christos Sapsanis

Supervisor: Anthony Tzes

Abstract

The aim of this work was to identify six basic movements of the hand using two systems. Being an interdisciplinary topic, there has been conducted studying in the anatomy of forearm muscles, biosignals, the method of electromyography (EMG) and methods of pattern recognition. Moreover, the signal contained enough noise and had to be analyzed, using EMD, to extract features and to reduce its dimensionality, using RELIEF and PCA, to improve the success rate of classification. The first part uses an EMG system of Delsys initially for an individual and then for six people with the average successful classification, for these six movements at rates of over 80%. The second part involves the construction of an autonomous system EMG using an Arduino microcontroller, EMG sensors and electrodes, which are arranged in an elastic glove. Classification results in this case reached 75% of success.

# Acknowledgments

At this point, I should thank those people who helped me with their experience, knowledge and support to complete this diploma thesis in the best possible way due to time constraints that always exist.

Initially, I would like to thank the supervisor Prof. Mr. *A. Tzes* for his support and guidance not only in the thesis but also in the final stage of my studies.

I would also like to thank the co-supervisor Prof. Mr. *P. Groumpos* for his support and for his valuable advice for my later life.

Special thanks to Dr. Mr. *G. Georgoulas*, whose contribution was crucial for achieving the publication of two research papers in this thesis.

I would also like to thank Prof. Mr. *E. Dermatas*, who helped greatly with his knowledge and experience of giving advice that was always accurate.

In addition, I owe a thank to PhD student *K. Andrianesis*, whose help in the initial part of the thesis was catalytic to understand how to address the issue of diploma thesis with the experience since a part of his doctoral work was to build a robotic hand. Special thanks to *D. Tsipianitis*, who was always available to help with the initial implementation. Also, I owe a thank you to Prof. of Medicine *E. Chroni* because as expert in electromyography helped to fill the gap that existed on the part of medicine.

At the same time, I could not forget to thank the faculty and staff of the Aalto University in Finland. Initially, I would like to thank Prof. *R. Sepponen* for his guidance and supervision. Moreover, I owe a big thank to Prof. *R. Vigario* not only for helping in the diploma thesis but also for the advice he gave me for my later life. It would be an omission not to thank Mr. *M. Linnavuo* and especially Mr. *H. Ruotoistenmäki* for assisting in the part of implementing the system.

Of course, it would be impossible not to thank all my friends who helped me with a practical way in this effort by dedicating personal time in order to get the necessary measurements - data.

Finally, I would have achieved nothing without my family and its continuous support all the years of my life.

# Contents

# List of figures

# List of Tables

# Abbreviations

| | |
|---|---|
| *EMG* | Electromyography |
| *EEG* | Electroencephalography |
| *ECG* | Electrocardiography |
| *SVM* | Support vector machines |
| *EMD* | Empirical Mode Decomposition |
| *IMF* | Intrinsic mode function |
| *PCA* | Principal Component Analysis |
| *k-NN* | K-nearest neighbor |
| *ANN* | Artificial Neural Networks |
| *PNN* | Probabilistic Neural Networks |
| *WAMP* | Willison Amplitude |
| *SMO* | Sequential Minimal Optimization |

# Chapter 1: Introduction

## 1.1 Motivation for the diploma thesis

Controlling an exoskeleton robotic hand is a challenging problem that should be addressed in order to implement a standalone system for cases of hand amputation. The use of electromyography signals (EMG) appears to be a feasible solution since each grasping has a unique "signature" of the generated signal. If the classification of the EMG signal is precise, it can lead to an effective control of a robotic hand [1] taking advantage of the developments in pattern recognition, biomedical signal processing and biosensors [2][3]. Furthermore, for an amputee, it is more convenient to use a glove in which the EMG electrodes are embedded than the use of, still promising, electroencephalography (EEG) with electrodes on the head region [4]. This problem is even more important for using a system on a daily basis.

There are several solutions to the motion recognition using EMG signal that have been proposed and achieved classification of high accuracy. They experiment with movements that are not necessarily typical of a human daily life and multiple electrodes (more than four in most of the cases) [5]. The use of multiple sensors can have a negative impact on the system's cost. Moreover, a high number of electrodes can lead to a complicated and hard-to-use system. Therefore, for a successful prosthesis, it would be preferred to assist the amputee in performing typical daily life movements by using a robotic hand. For instance, grasp a pencil, a glass of water, a credit card, a ball or hold a hammer [6].

## 1.2 Brief description of the structure of the experimental steps

The approach is oriented to healthy subjects for evaluating the success rates followed by the implementation of an autonomous EMG-system for the master-slave operation of a robotic hand [7]. Initially, Support vector Machines (SVM) are used [8] achieving high success for classification. Then, the interest is turned to the analysis of the signal, where the method of Empirical Mode Decomposition (EMD) [9] is selected. EMD brakes down the initial signal into multiple intrinsic mode functions (IMFs), followed by selecting specific characteristics of the raw signal and the IMFs. The features should assist in distinguishing the classes of hand movements in an optimal manner. Thus, two algorithms of dimensionality reduction of the vector of characteristics are used: the Principal Component Analysis (PCA) method [8] and the RELIEF algorithm [10], which exhibited similar results. Classification is achieved by using a simple linear classifier [11][12]. In the final stage, because there is a limitation in the available memory of the microcontroller, these methods cannot be programmed. Finally, the algorithm K-nearest neighbor (k-NN) [13] is programmed in Matlab for training and k-means in microcontroller for classification, which has very low system requirements and its classification results demand a low computational load and data length. The use of a short data window of 150 ms and the high rates of successful classification show promising potentials for a real application.

## 1.3 Structure of the diploma thesis

In **Chapter 2,** the anatomy of the human forearm is analyzed. Then, the structure of skeletal muscle and their location in the region of the forearm is described. Finally, the types of objects grasping is mentioned.

**Chapter 3** refers to basic principles and classes of signals and biosignals, the process of converting analog to digital signal through the techniques of sampling, quantization and coding.

**Chapter 4** focuses on the EMD method of signal analysis, on features of signal in the time and frequency domain and on techniques of dimensionality reduction (PCA method and the RELIEF algorithm) of the features vector of the signal.

**Chapter 5** refers to the origin of biopotentials, which are essential for the creation of biosignals, and to indicative recording techniques of biosignals: the Electromyography (EMG), the Electrocardiography (ECG) and the Electroencephalography (EEG).

In **Chapter 6**, basic Pattern Recognition methods are presented: the Artificial Neural Networks (ANNs), the k-Nearest Neighbor (k-NN), the Support Vector Machines (SVMs) and the Parzen Probabilistic Neural Networks (Parzen PNNs).

In **Chapter 7**, applications using EMG are introduced in three key areas: control of computer games, exoskeleton parts and rehabilitation, especially for the elderly.

In **Chapter 8**, the hardware used in both parts of the diploma thesis is described. The first part is the Delsys' EMG system, which consists of the main amplifier module, the NI USB-6009 DAQ device (A/D converter) and the DE-2.1 EMG sensors. The second part is an implemented EMG system consisting of an Arduino microcontroller and EMG sensors.

In **Chapter 9**, the experimental procedure is analyzed using both systems and acquiring initially measurements from one person and then from six people in the age range of 20-22 years.

**Chapter 10** comprises the conclusion and the future work with the view of improving the existing application.

**Appendix** includes all developed code in LabView, Matlab and Arduino Microcontroller in "Wiring" and "Processing" language.

# Chapter 2: Human Forearm

The forearm (πῆχυς in ancient Greek or antebrachium in Latin) in the terminology of anatomy is the part of the human upper limb from the elbow to the wrist. Previously, the forearm constituted a unit of length measurement as one of the first efforts of people to determine how to measure length. Its average length is 25 cm and it is linked with arm and hand with the joint of elbow and wrist, respectively.

## 2.1 Anatomy

The forearm has two bones, the radius and the ulna, as illustrated in Figure 1. The radius and the ulna are connected with the wrist from the side of thumb and small finger, respectively.

The ulna is longer, and it is connected more strongly with the brachial than the radius. The radius, however, creates a stronger contribution to the movements associated with the wrist joint.

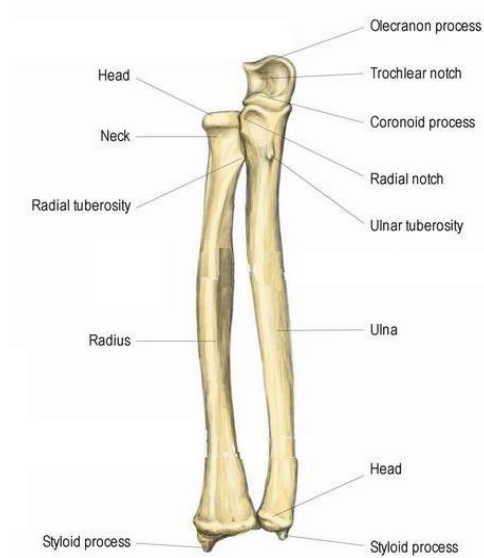

**Figure 1: Bones of forearm.**

*Ulna*: The proximal end of the ulna is joined with both the brachial and radius. The trochlear notch then creates a fusion with the trochlea of the brachial. The front edge of the trochlear notch is formed by the coronoid process and the trailing edge is formed by olecranon.

*Forearm diaphragm*: The radial notch, which is positioned on the side and lower part of the coronoid process, receives the head of the radius. The distal end of the ulna is conical, and, in that end, there is a portion that is gnarly, the head, and a knob-shaped bulge which is the styloid process. The ulna joins the radius in both ends.

*Radius*: The radius consists of a smaller proximal end and a larger distal end with a body between them. The shape of disk of near head meets the head the brachial, and the radial notch belonging to ulna. The radial camber is obvious and designed for attachment of the biceps muscle. It is positioned just below the head on the medial side. A double faceted surface at the distal end of the radius meets the proximal wrist bones. The styloid process is supported at the distal end of the radius. It is located in the lateral edge with an ulna notch based on the medial side, which is responsible for the receiving of the remote side. The styloid process of both the ulna and radius is designed for lateral and medial stability for the motion of the wrist.

## 2.2 Muscles and tendons

In the forearm there are several muscles that are directly related to the bending, the extent of fingers and moving the hand up and down. At the same time, of course, there are nerves and arteries, which support these movements. Before referring to the muscles and nerves of the forearm, it is suitable to refer to the structure of skeletal muscle.

## 2.2.1 Structure of skeletal muscle

The number of muscles in the human body is about 600 and is 2/5 of the weight of the human body. The muscles based on their structure, contractile properties and mechanisms of control, are divided into three categories: skeletal, smooth and cardiac muscles. This work deals only with the skeletal muscles, which will be analyzed below [14].

The skeletal muscles adhere to bone and they have significant contribution in the movement and support of the human skeleton. The expansion and contraction of skeletal muscle is made for the most part by voluntary control. In addition, through their movement, they act as a heat source because they produce 85% of the total body heat.

The connection of skeletal muscle is accomplished by tendons, collagen fiber bundles, which are located at the ends of each muscle. There are cases of tendons that have large enough length and the attached edge to the bone is remote from the end of the muscle. To understand the power transfer from muscle to bone, it can be compared to a person (muscle) pulling a rope (tendon).

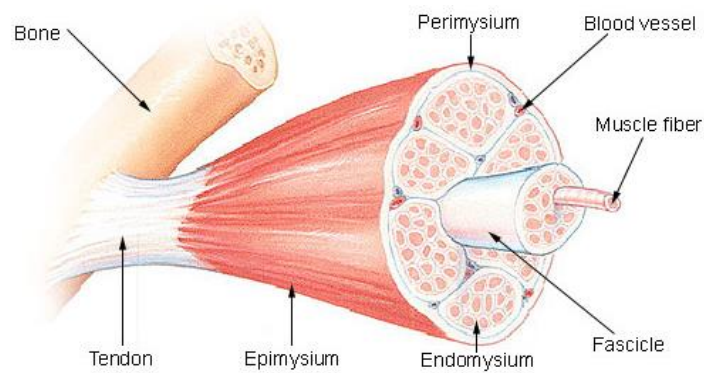

Figure 2: Basic components of a skeletal muscle.

The basic unit, by which the skeletal muscle is made of, is the muscle bundles and those of elongated cells, muscle fibers. Each of these is a separate cell containing several hundred nuclei. The space between the muscle fibers, including fascicle, is covered by connective tissue called endomysium. Furthermore, each muscle bundle is covered by a strong connective tissue, called perimysium. Moreover, epimysium is called an even stronger connective tissue, which covers the muscle. The epimysium is covered by a band of connective tissue (fascia), which surrounds and distinguishes the muscles. The connective tissue supports and protects the muscle fibers. It enables them to withstand the contraction forces and provide corridors for blood vessels and nerve cells. Figure 2 shows structure of the skeletal muscle.

The main function of the muscle fibers is the conversion of chemical energy to create motion and force. Their formation is completed at the time of birth and grows as the children develops, acquiring a diameter of 10-60 um and a length up to 30 cm long for muscle.

As they can be distinguished in Figure 3, the main parts of the muscle fiber are the sarcolemmal, the sarcoplasm, the sarcoplasmic reticulum and myofibrils.

*Sarcolemmal*: The cell membrane of the muscle fiber.
*Sarcoplasm*: The cytoplasm of muscle fiber which surrounds the myofibrils.
*Sarcoplasmic reticulum*: the network that encapsulates the myofibrils. One part surrounds the zone A and the other part of it networks the zone I. The zones are explained below.
*Myofibrils*: Cylindrical beams with diameter 1-2 um extending from one end of the fiber to another and constitute the bulk of the sarcoplasm. The main part of the myofibril is sarcomere. Each muscle fibril is composed by thick, consisting of myosin (expansion protein), and thin filaments consisting of actin protein (expansion) and other two other proteins, troponin and tropomyosin. All these yarns are placed in a repetitive way along the muscle and compose the myofibril [15].

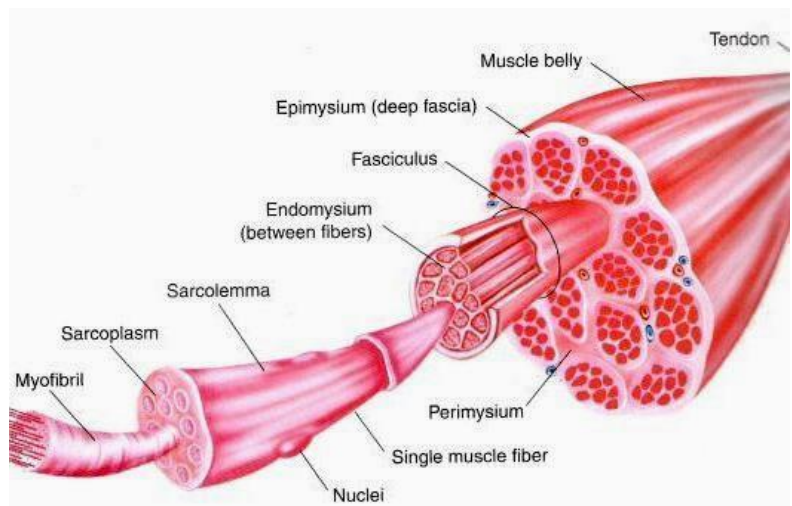

Figure 3: Individual parts of a skeletal muscle.

Under a light microscope and with the existence of a quite large number of thick and thin filaments, the characteristic of the sequence of light and dark stripes is observed, which is vertical to the longitudinal axis of the fiber. This leads to name the skeletal muscle as striated muscles.

The zones and the lines observed in sarcomere of the muscle fiber, as shown in Figure 4, are:

*Zone A*: A thick, dark stripe is produced by the parallel arrangement of the thick filaments in the middle of each sarcomere.

*Line or disc Z*: Two successive rows Z define the ends of a sarcomere, which contains two rows of thin filaments, one at each end. The one end of each thin filament is attached to a network of interconnected proteins, the line Z, while the other end connects partly with the thick yarn.

*Zone I*: The almost yellow band is located in between two zones A of adjacent sarcomeres and contains thin filaments, which are autonomous and are not connected with the thick yarn. In the middle of this zone is the line Z.

*Zone H*: The narrow almost yellow stripe in the center of the zone A.

*Line M*: The very narrow dark stripe in the center of the zone H corresponds to proteins, which perform the task of keeping the thick stripes together.

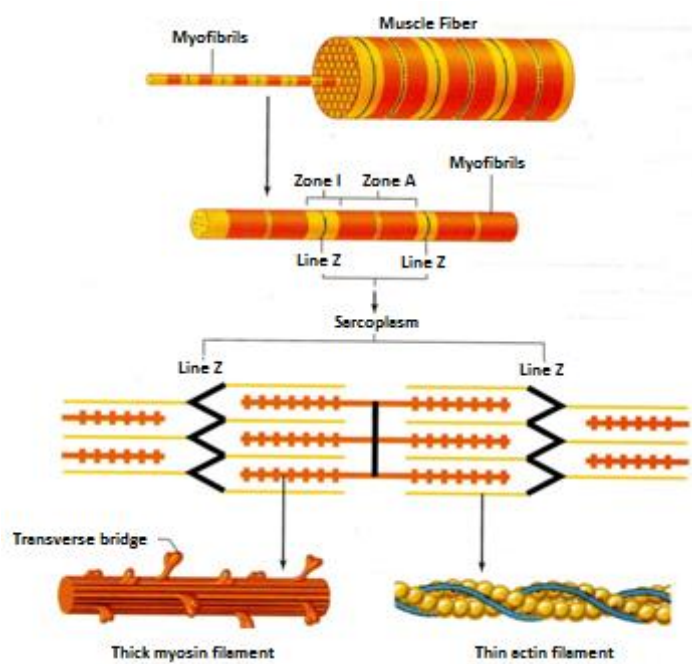

Figure 4: Depiction of filaments (lines and zones) in a skeletal fiber.

### 2.2.2 Categories of muscles and nerves

*Antibrachial Fascia* (known also as fascia antibrachium; deep fascia of the forearm): is extended with the brachial fascia and is a thick and membranous investment forming a sheath of all this region muscles. The dorsal border of the ulna and the olecranon are attached behind it. Its surface provides multiple intermuscular septa encompassing each muscle individually. It becomes denser above the flexor tendons while it is in close proximity to the wrist creating the volar carpal ligament. Its extension to the transverse carpal ligament develops a cover for the tendon attached to palmaris longus passing over the transverse carpal ligament and concluding into the palmar aponeurosis. It becomes even more dense in the area behind the wrist-joint by adding multiple fibers forming the dorsal carpal ligament. Antibrachial fascia is denser on the dorsal compared to the volar surface and at the lower compared to the upper forearm. Tendinous fibers originated from the biceps brachii in front and from the triceps brachii behind enhance its strength. Antibrachial fascia provides the origin to the muscular fibers at the upper part of the medial and lateral sides of the forearm and sets the cone-shaped cavities margins, where the muscles are enclosed. The vertical septa divides the individual muscles while the transverse septa separates the interior from the exterior muscles layers. There are some open spaces as a passageway for the vessels and nerves. One of these gaps is located in the front side of the elbow and it is used a passageway for exchange information among the superficial and deep veins [16][17].

The antibrachial or forearm muscles are separated into two groups: a volar and a dorsal.
1. *The Volar Forearm Muscles* can be split into two groups: superficial and deep.
   i. *Superficial group*: This category includes the following muscles: Pronator Teres, Palmaris Longus, Flexor Carpi Radialis, Flexor Carpi Ulnaris, Flexor Digitorum Sublimis. They are originated from the medial epicondyle of the humerus by a common tendon. Other fibers are also connected from the antebrachia's investing fascia close to elbow and from the septa passing via this fascia between the distinct muscles.
      a. The *Pronator Teres* is originated from both humeral and ulnar heads. The humeral head is the superior and closer to the surface and emerges above the medial epicondyle and from the common tendon of the surrounding muscles. Moreover, it also arises from the intermuscular septum between it and the flexor carpi radialis and from the antibrachial fascia. The ulnar head is a thinner fiber bundle compared to humeral emerging from the middle side of the coronoid process of the ulna and linking to the preceding at an acute angle. The two heads allow the median nerve to pass in between them and divide the ulnar artery from the ulnar head. Pronator teres crosses indirectly the antebrachium and concludes in a flat tendon. The side part of the muscle creates the central border of a triangular hollow, which is located in front of the elbow-joint, encompassing the brachial artery, median nerve, and tendon of the biceps brachii.
      b. The *Flexor Carpi Radialis* is located in the middle side of the pronator teres. It originates from the medial epicondyle, where the common tendon is, via the common tendon and from the intermuscular septa between it and crosswise the pronator teres, in the middle of the Palmaris longus and below the flexor digitorum. It is thin and without nerves in its origin but in thickens in size and concludes to a tendon, which constitutes almost its half-length. The tendon runs via a canal in the sideways part of the transverse carpal ligament and passes via a channel on the larger multangular bone. This channel is transformed into a canal using fibrous tissue and covered by mucous sheath. The tendon is attached to base of the 2$^{nd}$ and 3$^{rd}$ metacarpal bone. The radial artery is located in the antibrachium's lower side in between this tendon and the brachioradialis.
      c. The *Palmaris Longus* (PL) muscle is a thin and spindle-shaped and is placed on the middle side of flexor carpi radialis. It emerges from the medial epicondyle of the humerus by the common tendon, from the intermuscular septa between it and the neighboring muscles, and from the antibrachial fascia. It ends in a thin and flattened tendon, which runs over the higher side of the transverse carpal ligament. It goes into the middle part of the transverse carpal ligament and lower section of the palmar aponeurosis, which occasionally transfers a tendinous slip to the thumb's short muscles.
      d. The Flexor Carpi Ulnaris is located around the antibrachium's ulnar side. It is originated by the humeral and ulnar heads connected by a tendinous arch. The humeral emerges from the humerus'

medial epicondyle and from the common tendon of the surrounding muscles. The ulnar head emerges from the middle boundary of the olecranon, from the upper two-thirds of the posterior border of the ulna by an aponeurosis, which is shared with the extensor carpi ulnaris and Flexor digitorum profundus, and from the intermuscular septum among it and the flexor digitorum sublimis. The fibers conclude in a tendon, which takes over the frontal side of the muscle's lower half part and enters the pisiform bone. It is also expanded from this bone to others, such as the hamate and 5th metacarpal, via the pisohamate and pisometacarpal ligaments. The ulnar vessels and nerve are located on the lateral side of the tendon.

e. The *Flexor Digitorum Sublimis* (FDS) is located below beneath flexor carpi ulnaris and is the bulkiest muscle of this category. It originates from humeral, ulnar, and radial heads. The humeral head emerges from the humerus' medial epicondyle by the common tendon, from the ulnar collateral ligament of the elbow-joint, and from the intermuscular septa between it and the preceding muscles. The ulnar head emerges from the medial side of the coronoid process, on top of the ulnar origin of the pronator teres. The radial head emerges from the oblique line of the radius, spreading from the radial tuberosity to the pronator teres insertion. Flexor digitorum sublimis muscle splits into two planes of muscular fibers: superficial and deep. The superficial plane is separated into two segments concluding in tendons for the middle and ring fingers. The deep plane provides a muscular slip to connect the part of the superficial plane related to the ring finger's tendon. It splits into two smaller parts concluding in the index and little fingers tendons. The four tendons run below the transverse carpal ligament into the palm of the hand. They are organized in two pairs: the superficial, which passes to the middle and ring fingers, and deep, which leads to the index and little fingers. Each tendon is divided into two parts and recombines again creating a channel for receiving the tendon of the flexor digitorum profundus to pass. The tendon splits and enters the sides in the middle of the second phalanx.

ii. *Deep group*: This category includes the following muscles: Flexor Digitorum Profundus, Flexor Pollicis Longus, Pronator Quadratus.
   a. The *Flexor Digitorum Profundus* is located in the ulnar side of the forearm, exactly below the superficial Flexors. It emerges from the uppers side of the volar and medial surfaces of the ulna holding the insertion of the Brachialis on top and expanding beneath to the Pronator quadratus. Moreover, it emerges from a suppression on the medial side of the coronoid process, from the upper ulna's dorsal border, which is shared with the flexor and extensor carpi ulnaris, and from the ulnar half of the interosseous membrane. The muscle concludes in four tendons which pass below the transverse carpal ligament posterior to the tendons of the flexor digitorum sublimis. The tendons run via the openings in the tendons of the flexor digitorum sublimis and conclude to enter the bases of the last phalanges. The tendons related with the middle, ring, and little fingers are joined by areolar tissue and tendinous slips till the hand palm.
   b. The *Flexor Pollicis Longus* is located in antibrachium's radial side. It emerges from the grooved volar surface of the radius' body and expands from exactly beneath the tuberosity and oblique line close to the pronator quadratus. Furthermore, it emerges from the neighboring part of the interosseous membrane by a thick slip from the medial margin of the coronoid process or from the medial epicondyle of the humerus. The fibers conclude in a flattened tendon, which runs below the transverse carpal ligament. It is fixed among the flexor pollicis brevis' lateral head and the adductor pollicis' oblique part and, passing through a canal (osseoaponeurotic), it enters the base of the thumb's distal phalanx. The volar interosseous nerve and vessels run below the frontal part of the interosseous membrane among the flexor pollicis longus and flexor digitorum profundus.
   c. The *Pronator quadratus* (PQ) is a small, plane and four-sided muscle, expanding to the front of the lower parts of the radius and ulna. It emerges from the pronator ridge on the volar's surface lower part of the ulna's body, from the volar's surface medial part of the ulna's lower fourth and from a strong aponeurosis covering the medial third of the muscle. The fibers run laterally and in a downward direction, so they can enter the lower fourth of the lateral border and the volar surface of the radius' body. The deeper fibers enter the triangular area on top of the radius' ulnar notch, which is attached in a similar origin of the supinator from the triangular area below the ulna's radial notch.

*Nerves*: The median nerve controls all the muscles of the superficial layer with the exception of flexor carpi ulnaris, which is controlled by ulnar. The 6$^{th}$ cervical nerve mainly controls the palmaris longus derive, flexor carpi radialis and the pronator teres. The 7$^{th}$ and 8$^{th}$ cervical and the 1$^{st}$ thoracic nerves control the flexor digitorum sublimis but the flexor carpi ulnaris only from the two latter nerves. As for the deep layer, the 8$^{th}$ cervical and the 1$^{st}$ thoracic nerves via ulnar and the volar interosseous branch of the median control the flexor digitorum profundus. The 8$^{th}$ cervical and the 1$^{st}$ thoracic nerves through the median volar interosseous branch control the flexor pollicis longus and pronator quadratus.

2. *The Dorsal Antibrachial Muscles* can be split into two groups: superficial and deep.
   i. *Superficial group*: This category includes the following muscles: Brachioradialis, Extensor Digitorum Communis, Extensor Carpi Radialis Longus, Extensor Digiti Quinti Proprius, Extensor Carpi Radialis Brevis, Extensor Carpi Ulnaris, Anconeus.
      a. The *Brachioradialis* muscle is the most superficial on the antibrachium's radial side. It emerges from the higher part of the humerus' lateral supracondylar, and from the lateral intermuscular septum, which is constrained on the top by the channel for the radial nerve. The radial nerve and the anastomosis between the anterior branch of the profunda artery and the radial recurrent interfere in between brachioradialis muscle and the brachialis. The fibers conclude on top of the middle part of the antibrachium in a flat tendon, which is attached in the side part of the radius' styloid process base. This tendon intersects with the abductor pollicis longus and the extensor pollicis brevis tendons.
      b. Part of the *Extensor Carpi Radialis Longus* is located below the brachioradialis. It emerges from the lower third of the humerus' lateral supracondylar ridge, from the lateral intermuscular septum, and by a small number of fibers coming from the common tendon, the beginning of the antibrachium's extensor muscles. The fibers conclude at the higher part of the antibrachium in a flat tendon, which passes via the radius' lateral border, below the extensor pollicis brevis and abductor pollicis longus. Afterwards, it runs below the dorsal carpal ligament and enters the dorsal surface of the 2$^{nd}$ metacarpal bone base via its radial side.
      c. The *Extensor Carpi Radialis Brevis* is located below the aforementioned muscles. It has less length and more width compared to them. It emerges from the humerus' lateral epicondyle by a shared tendon and by three muscles from the radial collateral ligament of the elbow-joint, from a strong aponeurosis and from the intermuscular septa. The fibers conclude in the antibrachium's middle part in a flat tendon, which is attached to the extensor carpi radialis longus muscle and goes with it till the wrist. It runs below the extensor pollicis brevis and the abductor pollicis longus, then below dorsal carpal ligament and it finally enters the dorsal surface of the third metacarpal bone base on its radial side. The tendon is located in the back of the radius below the dorsal carpal ligament. The tendons for the two aforementioned muscles run via the same segment of the dorsal carpal ligament in a single mucous sheath.
      d. The *Extensor Digitorum Communis* emerges from the humerus' lateral epicondyle by the common tendon, from the intermuscular septa between it and the neighboring muscles, and from the antibrachial fascia. It splits underneath into four tendons, which run along with the extensor indicis proprius via a separate part of the dorsal carpal ligament covered with mucous sheath. Afterwards, the tendons deviate on the back side of the hand and enter the 2$^{nd}$ and 3$^{rd}$ fingers' phalanges. Each tendon across the metacarpophalangeal articulation is connected by fasciculi to the supplementary ligaments serving as the joint's dorsal ligament of this. It extends to a broad aponeurosis after having crossed the joint covering the dorsal surface of the 1$^{st}$ phalanx and it is strengthened by the Interossei and the Lumbricalis tendons. This aponeurosis splits across the 1$^{st}$ interphalangeal joint into three parts: an intermediate and two collateral. The intermediate enters the 2$^{nd}$ phalanx base

and the two collateral (linked along the sides of the 2$^{nd}$ phalanx) connect by their connecting boundaries and enter the dorsal surface of the last phalanx. The tendons supply the interphalangeal joints, while crossing them, with dorsal ligaments. The index finger's tendon goes with the extensor indicis proprius, which is located in its ulnar side. The tendons connected to the middle, ring, and little fingers are linked by two obliquely placed bands on the back side of the hand. The one moves lower and laterally to the second tendon and the other moves from the same tendon lower and medially to the fourth.

e. The *Extensor Digiti Quinti Proprius* muscle is thin and located and connected with the medial side of the extensor digitorum communis. It emerges from the common extensor tendon by a thin tendinous slip, from the intermuscular septa between it and the adjacent muscles. Its tendon passes via a part of the dorsal carpal ligament behind the distal radio-ulnar joint, then splits into two as crossing the hand. In the end, it connects with the expansion of the extensor digitorum communis tendon on the dorsum of the little finger's 1$^{st}$ phalanx.

f. The *Extensor Carpi Ulnaris* is located in the antibrachium's ulnar side. It emerges from the humerus' lateral epicondyle by the common tendon, from the ulna's dorsal border in common with the flexor carpi ulnaris and the flexor digitorum profundus by an aponeurosis and from the deep fascia of the forearm. It concludes in a tendon, which passes through a groove between the head and the ulna's styloid process, running via a distinct part of the dorsal carpal ligament. It enters the prominent tubercle on the ulnar side of the 5$^{th}$ metacarpal bone base.

g. The *Anconeus* is a minor triangular muscle located in the elbow-joint back seeming to be a continuance of the triceps brachii. It emerges by a single tendon from the back part of the humerus' lateral epicondyle. Its fibers enter the olecranon side and upper part of the dorsal surface of the ulna's body.

ii. *Deep group*: This category includes the following muscles: Supinator, Extensor Pollicis Brevis, Abductor Pollicis Longus, Extensor Pollicis Longus, Extensor Indicis Proprius.
  a. The *Supinator* is a wide muscle, which surrounds the upper radius part. It contains two fibers planes, where the deep branch of the radial nerve is located in between them. The two planes emerge similarly: the superficial by tendinous and the deeper by muscular fibers, from the humerus' lateral epicondyle, from the radial collateral ligament of the elbow-joint and the annular ligament, from the ulna's ridge, which pass indirectly downwards from the dorsal end of the radial notch, from the triangular depression beneath the notch, and from a tendinous extension surrounding the muscle surface. The superficial fibers cover the radius upper part and enter the lateral edge of the radial tuberosity and the radius tilted line, as low down as the insertion of the pronator teres. The upper fibers of the deeper plane constitute a sling-like fibre bundle, which encloses the radius neck on top of the tuberosity and is connected with the back part of its medial surface. The larger part of this portion of the muscle enters the dorsal and lateral surfaces of the radius body, midway among the tilted line and the bone head.
  b. The *Abductor Pollicis Longus* is located exactly below the Supinator with which is sometimes connected. It emerges from the lateral part of the dorsal surface of the ulna's body below the insertion of the Anconæus, from the interosseous membrane and from the middle part of the dorsal surface of the radius' body. Running indirectly lower and laterally, it concludes in a tendon, which passes via a groove on the lateral side of the radius' lower end, along with the extensor pollicis brevis tendon. Then, it is inserted into the radial side of the base of the 1$^{st}$ metacarpal bone.
  c. The *Extensor Pollicis Brevis* is located in and connected with the medial side of the abductor pollicis longus. It emerges from the dorsal surface of the radius's face below that muscle, and from the interosseous membrane. Its direction is comparable to that of the abductor pollicis longus. Its tendon runs the same groove on the lateral side of the lower radius end and enters the thumb's 1$^{st}$ phalanx base.
  d. The *Extensor Pollicis Longus* is superior compare to extensor pollicis brevis covering partially its origin. It emerges from the sideways part of the middle third of the dorsal surface of the ulna's body beneath the abductor pollicis longus origin and from the interosseous membrane. It concludes in a tendon running via a distinct segment in the dorsal carpal ligament, which is located in a thin and tilted groove on the back of the radius' lower end. Afterwards, it passes indirectly from the extensores carpi radialis longus and brevis tendons and gets detached from the extensor brevis pollicis by a triangular interval, where the radial artery is found. Eventually, it enters the base of the last thumb phalanx. The radial artery is intersected by the abductor pollicis longus and the extensores pollicis longus and brevis tendons.

e. The *Extensor Indicis Proprius* is a thin and extended muscle. It is located medial to and parallel with the extensor pollicis longus. It emerges from the dorsal surface of the ulna beneath the beginning of the extensor pollicis longus and from the interosseous membrane. Its tendon runs below the dorsal carpal ligament in the same segment that spreads the tendons of the extensor digitorum communis, and across the head of the 2$^{nd}$ metacarpal bone linking to the ulnar side of the extensor digitorum communis tendon, which connects to the index finger.

*Nerves*: The Brachioradialis is provided by the 5$^{th}$ and 6$^{th}$, the Extensores carpi radialis longus and brevis by the 6$^{th}$ and 7$^{th}$, and the Anconeus by the 7$^{th}$ and 8$^{th}$ cervical nerves, through the radial nerve; the other muscles are innervated through the deep radial nerve, the Supinator being supplied by the 6$^{th}$, and all the other muscles by the 7$^{th}$ cervical.

The cross-section and the muscles of the forearm are depicted in Figure 5 and Figure 6, respectively.

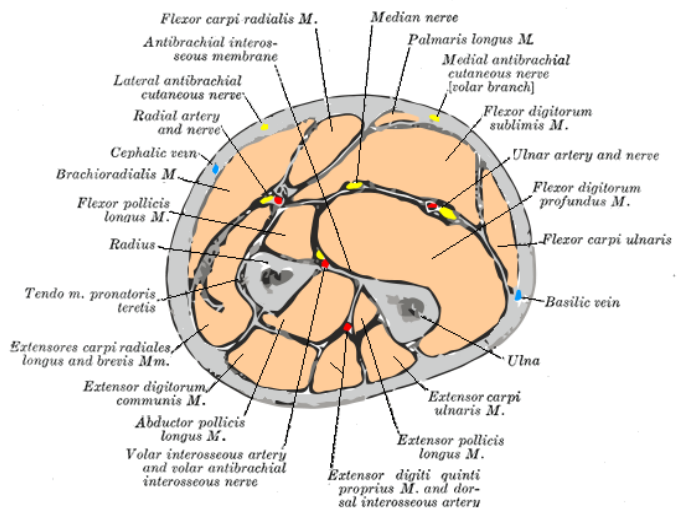

Figure 5: Vertical section of forearm.

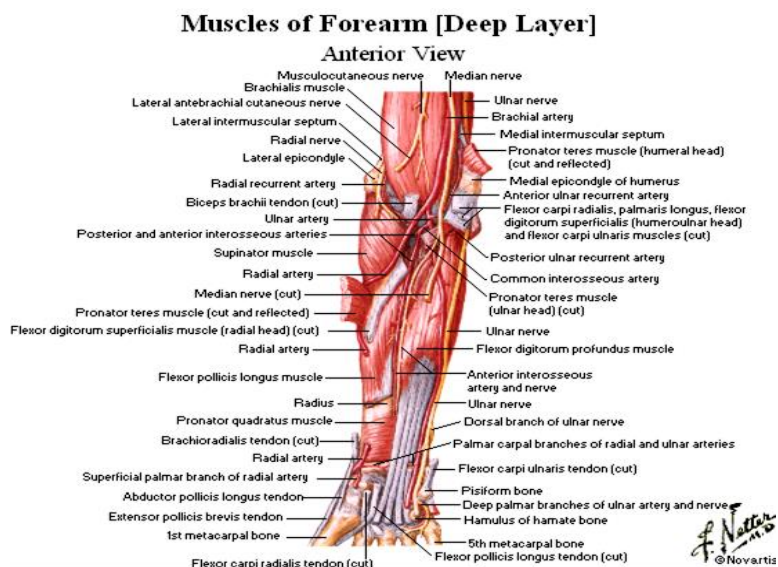

Figure 6: Muscles of forearm.

## 2.3 Grasping objects

The grasping types that a hand performs are classified into two categories: the precision and the dynamic. *Precision grasping* is defined as the combination of processes and functions required to maintain an object in a particular position relative to the hand. *Dynamic grasping* is defined as the coordinated movement of the fingers with the view of handling objects on the inner side of the hand.

These two grasping categories are separated mainly by the position and mobility of the CMC joint of the thumb and MCP joints, which can be seen in Figure 7. The basic grasping types of everyday life are depicted in Figure 8, distinguishing where the hand holds or lifts objects.

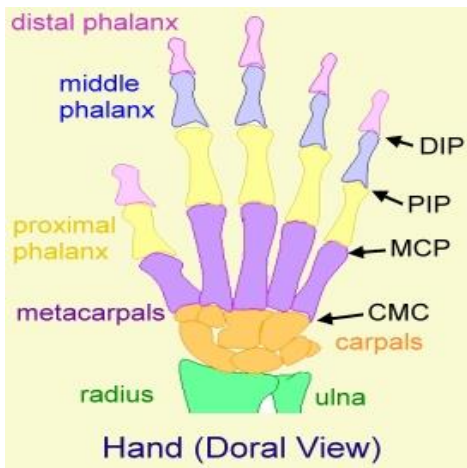
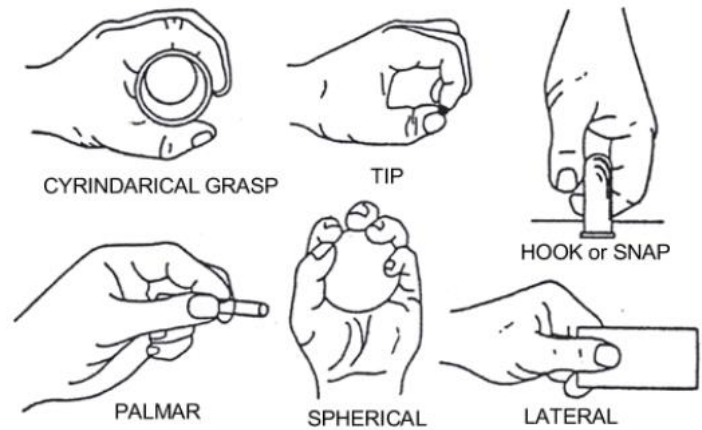

Figure 7: Joints.

Figure 8: Basic hand grasping.

In the category of *dynamic grasping*, the adductor of the thumb stabilizes an object against the palm with the hand still. This class includes the following grasping types:

i. Cylindrical (e.g. grasping a glass)
ii. Spherical (e.g. grasping a spherical object that fits in the palm while the fingers are closed)
iii. Hook (e.g. lifting a dumbbell)

In the category of *precision grasping*, the muscles are activated by abduction of thumb and its opposite position of the others fingers[1] as the hand is in a dynamic position. This class includes the following grasping types:

i. Palmar (e.g. grasping the pencil)
ii. Tip (e.g. the touch of the thumb to the index)

There is also the lateral grasping which can belong to both categories. It is accounted as dynamic grasping when the thumb is in adduction[1] and as precision grasping if the thumb is in opposite position of the others[1].

---

[1] Flexion and extension of the thumb are performed in the CMC joint. These are the movements with which the thumb approaches or removes the index and movements with which the thumb is brought to the palm and placed opposite the other fingers.

# Chapter 3: Signal and Biosignal

## 3.1 Signal

### 3.1.1 Definition

Signal is defined as a physical quantity, which contains information and changes in relation to one or several independent variables [18].

Mathematically is expressed as the function of one or several independent variables, of which the most common are the time and space:

$$t \rightarrow x(t)$$

The signals appear through natural processes (e.g. speech signal, seismic signals, biosignals) or artificially as a result of technological advances (e.g radio, econometric, satellite signals).

The problem that arises is that we cannot extract from the signal the information we need as it contains the element of noise [19]. This is because processes the signal to obtain the "pure" information.

### 3.1.2 Types of signals

The signals can be classified in multiple ways. Based on the purpose of use, the appropriate classification is chosen.

Based on the **range of values**, the signals are divided, as illustrated Figure 9, into:

1. *Continuous-time (analog)*: the set of values of the independent variables is the continuous time domain and which is infinite in size. Typical examples of continuous signal are:

$$x(t) = A \cdot sin(\omega t), y(t) = A \cdot cos(\omega t + \varphi),$$

   where $A$, $\omega$ and $\varphi$ is defined as the width, frequency and phase respectively.

2. *Discrete time*: the set of values of the independent variables is the discrete set of time, which is finite in size. Typical examples of discrete signals are:

$$x[n] = A \cdot sin[\omega n], y[n] = A \cdot cos[\omega n + \varphi],$$

   where $A$, $\omega$ and $\varphi$ is defined as the width, frequency and phase respectively. Subcategory of discrete signal is a digital signal, which has specific values, 0 or 1.

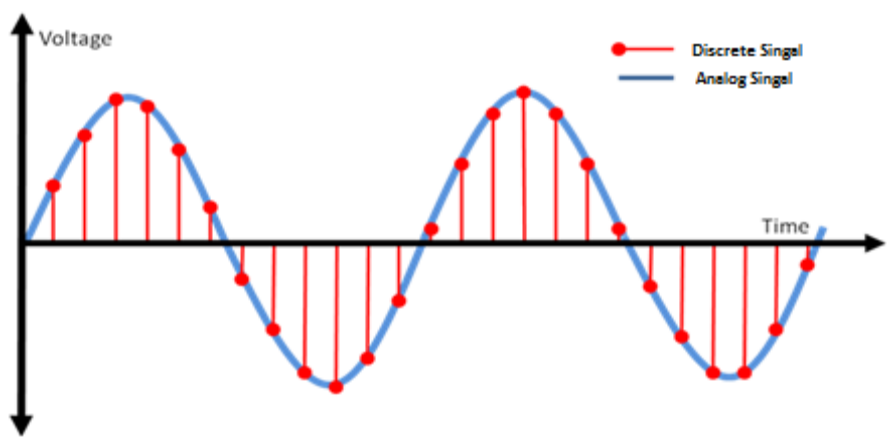

Figure 9: Analog and Discrete Signal.

Here should be mentioned the advantages and disadvantages of discrete signals over analog [20]:

1. *Advantages*:
   - Uniformity (all kinds of information can be converted into digital format and edited in the same way and material).
   - Less sensitive to noise.
   - Easier to encrypt information.
   - Multimedia sources (voice, video, data) can coexist and be transmitted through a common digital system.

2. *Disadvantages*:
   - Distortion of the signal due to the sampling and quantization.
   - Need for greater bandwidth.

Based on the **dimensions**, the signals are classified into:

1. *one-dimensional*: they have only one independent variable, which is usually the time. A typical example is the speech where time is the independent variable and acoustic pressure the dependent [21].

2. *multidimensional:* An example of two-dimensional signal is an image where the spatial coordinates $(x, y)$ are the independent variables and the dependent brightness. Example of three-dimensional signal is the video, which is a sequence of images in time. Therefore, it differs to the image only in the independent variables where now the time is included, i.e. $(x, y, t)$ [22].

Based on **the type of description** [23], the signals are classified into:

1. *deterministic*: are called the signals that can be described by a mathematical equation. In the real world no signal belongs to this category as their form is affected by noise and unexpected changes in parameters. Nevertheless, it is quite convenient to try to model or approximate a signal with the help of deterministic functions. The periodic signals belong in this category.

2. *stochastic*: are called the signals that cannot be expressed mathematically, but with probabilities. In stochastic processes that are stationary, the dispersion of the random variables is the same for each value of the variable parameter. Therefore, they are processes whose statistics remain unchanged over time. In the real world, most signals are not static.

Based on the **property**, the signals are divided into:

1. *periodic or not*: periodic is called the signal whose waveform is repeated after a certain amount of time named period. The formula expressing this property is:

$$x(t) = x(t + nT)$$

   where $n$ is an integer which expresses the number of repetitions and $T$ is the period. Otherwise, it is called non-periodic.
2. *even or odd*: a signal is called even if $x(-t) = x(t)$ and odd if $x(-t) = -x(t)$.
3. *causal or not*: a signal is called causal if $x(t) = 0$ for t<0 and non-causal if $x(t) = 0$ for t>0.

Several examples of biological signals depending on the dimension of the signal are depicted in Figure 10, Figure 11, Figure 12 and Figure 13.

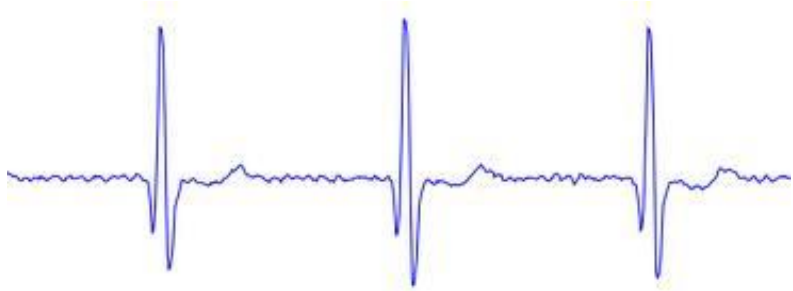

**Figure 10: One-dimensional signal (1D): Electrocardiogram of healthy person without noise.**

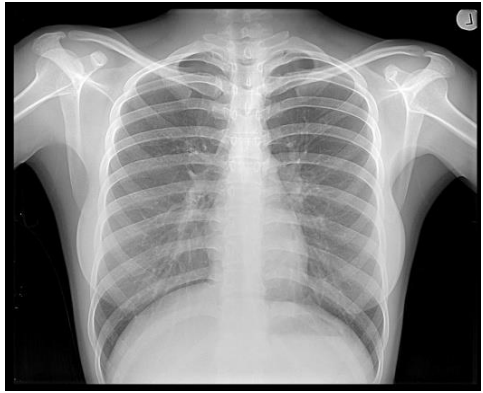

**Figure 11: Two-dimensional signal (2D): X-ray picture (radiograph).**

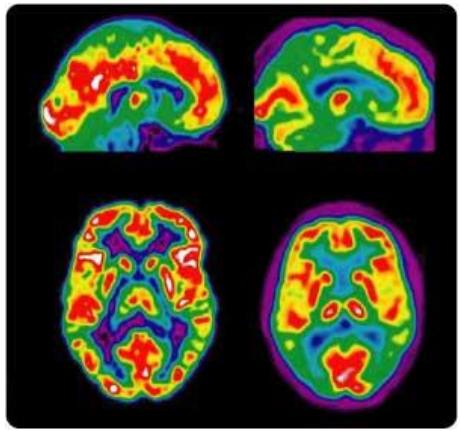

**Figure 12: Three-dimensional signal (3D): Picture PET (positron).**

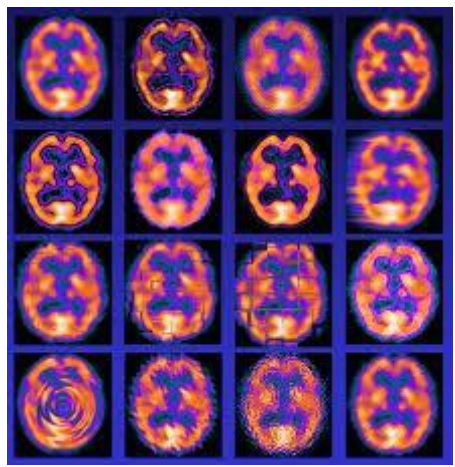

**Figure 13: Four-dimensional signal (4D): Picture SPECT (single photon emission tomography).**

## 3.2 Biosignal

### 3.2.1 Definition

As *biological signal* or *biosignal* can be defined the description of a physical phenomenon where chemical or physical quantities characterize the biological condition of a person [3][24]. The largest number of them is continuous-time signals. The most common process is their conversion into discrete through the technique of sampling. Those signals may be used to explain the physiological mechanisms that underlie a particular biological event or a system. The biosignals can be acquired in multiple ways, based on the type of biosignals and will be reported in the following section. The use of the stethoscope by the doctor for the heartbeat is a fairly simple example making biosignals. Of course, with the advances in technology, quite complex medical equipment has been built with multiple features for acquiring and using biosignals.

### 3.2.2 Types of biosignals

The biosignals, as the simple signals, can be classified in multiple categories. Based on the purpose of use, the appropriate classification can be chosen.

Based on the **way of activation** [25], signals are classified into:

1. *Active*: The source for the measurement is derived from the patient himself ("internal source"). This category can be divided into two subcategories, the:

    a) *electrical*: known as biopotential is the most widespread category of biosignals. Examples the belong to this type of signal are the signals generated by the techniques of Electrocardiography, Electroencephalography, Electromyography, Electrogastrography etc.

    b) *non-electrical*: Although it is common when we refer to biosignals to mean the bioelectrical signals, in fact there are also non-electric. Examples of this category are the body temperature and blood pressure.

2. *Passive*: The energy source is outside of the patient ("external source"), for example in X-ray CT scanner.

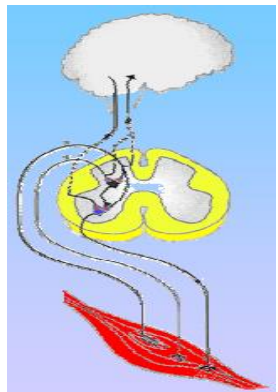

Figure 14: Acquisition mode of bioelectric signal.

Based on the **source or the physiological origins** [23], the signals are classified into:

1. *Bioelectrical*: Nerve and muscle cells are responsible for their generation. The membrane potential, based on specific conditions, is the source of an action potential generation. The action potential is a biomedical signal and can be measured in a single cell by using specific microelectrodes as sensors. If the measurement is not targeting only one cell and, thus, simple surface electrodes can be employed as depicted in Figure 14. In that case, the acquired signal is obtained by multiple cells and its due to the electric field in the area that the electrode covers. The cells act as propagation medium for this field and, hence, the action potential can be captured at parts that are convenient to use surface electrodes avoiding the need for invasive sensors. The sensor for the action potential is relatively simple.

2. *Bioimpedance*: Measuring the tissue impedance can provide significant information regarding the tissue composition, the automatic nervous system activity, the endocrine activity, blood volume and distribution. The generation of the bioimpedance signal is conducted by injecting sinusoidal currents into the tissue under evaluation. The current frequency and amplitude can vary from 50kHz to 1MHz and from 2mA to 20A respectively. Those ranges are determined based on the minimization of the polarization issues of the electrodes for the frequency and the tissue damage because of heating effects.

Bioimpedance measurements normally use 4 electrodes, as shown in Figure 15: i) The two source electrodes are used to inject the current into the tissue via a current source and ii) the other two electrodes are used for measurements and positioned on the tissue under test measuring the voltage drop that is created due to the induced current and the tissue impedance.

Typical examples of this signal measurement are the impedance plethysmography and the thoracic bioimpedance [23].

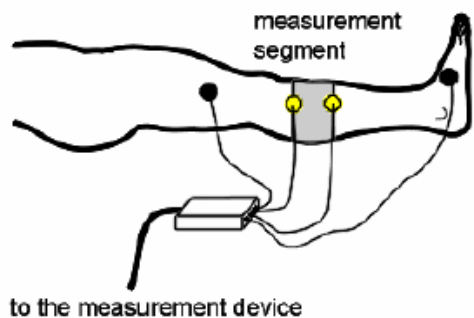

Figure 15: Acquisition mode of bioelectrical signal.

3. *Biomagnetic*: Numerous organs, such as the heart, brain, and lungs, generate extremely weak magnetic fields (from nano (nT) to micro (uT) Tesla). Measuring these fields can provide information that cannot be obtained from other biosignals. One important issue with this type of measurement is the low signal-to-noise ratio because of their weak level [26]. Typical examples are the magnetocardiogram and the magnetoencephalogram.

The design of these acquisition system should be considered carefully. For example, a sensor that is used for this type of record is the superconducting quantum interference devices or SQUID, as depicted in Figure 16. They are very sensitive magnetometers employed to acquire extremely weak magnetic fields. The working principle depends on superconducting loops containing Josephson junctions. Any other effect that could be performed by the environment as a necessary condition, which is full of magnetic signals, should be excluded in order to acquire accurately the signal.

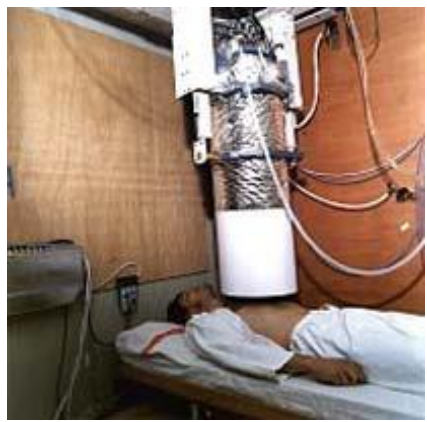

Figure 16: Acquiring mode of biomagnetic signal.

4. *Biomechanical*: This term contains all the biomedical signals that are associated with a mechanical function of the biologic system. Signals of this type can be those that are related with tension, pressure, displacement, motion, and flow. Multiple sensor types, not always simple and low cost, can be used for measuring them. This is because the mechanical phenomenon does not follow the propagation property that occurs in acoustic, electric and magnetic fields. Therefore, the measurement should be precise in the location, which complicates the acquisition and leads to invasive systems, for instance, blood pressure, Carotidography. The bioacoustic signal, that follows below, also belongs in this category since it is in the form of vibration (movement).

5. *Bioacoustic*: Acoustic noise can be generated by multiple biomedical phenomena, for example, the heart's blood flow via the valves, the air flow via the airways and the lungs. Sounds known as snores, coughs, lung sounds are attracting the interest. Moreover, the digestive tract, the joints, the contracting muscle can create an acoustic sound. The sound propagates via the biologic medium and, hence, it is acquired on the surface by microphones as shown in Figure 17. An illustrative example of that category is the phonocardiogram.

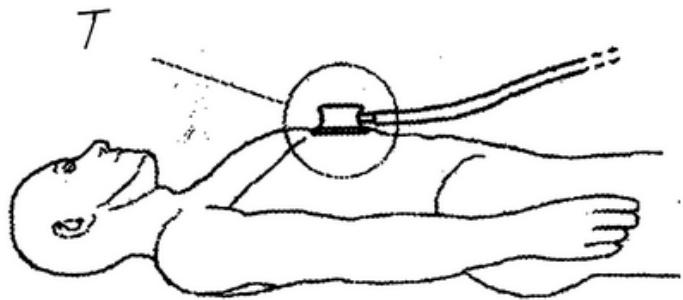

Figure 17: Acquiring bioacoustic signal.

6. *Biochemical*: They are obtained from the biochemical measurements extracted from a living tissue or from a sample evaluated in a lab environment. For instance, a chemical signal can be extracted using ion electrodes for measuring the ions concentration inside and near a cell. Other examples are the carbon dioxide ($pCO_2$) in the blood or respiratory system and the partial pressures of oxygen ($pO_2$). This class of signals can be used for different purposes, such as the determination of glucose, lactase and metabolites and providing information on the functioning of various physiological systems. The frequency domain of these signals is in the lower range with most of them to be purely DC.

7. *Bio-optical*: This signal type occurs due to the optical functions of the biologic system and arises in a natural way of by a movement. A typical example is oximetry, where the measurement in the tissue of the transmitted and backscattered light in multiple wavelengths can provide an estimation of the blood oxygenation. These signals can occur naturally or, in some cases, can be caused by using a biomedical technique. The fluorescence features extraction of the amniotic fluid can provide information about the fetus condition. The dye dilution technique can be employed to estimate the heart output by observing the recirculated dye's presence in the bloodstream. Multiple applications of bio-optical signals take advantage of the fiberoptic technology.

8. *Thermal*: This signal type conveys information related to the temperature distribution throughout the body surface. The biochemical and physical processes progressing in the organism are reflected by the temperature. A sensor using a contact measuring technique is called thermometer and a non-contact can be a 2D thermographic camera, as illustrated in Figure 18.

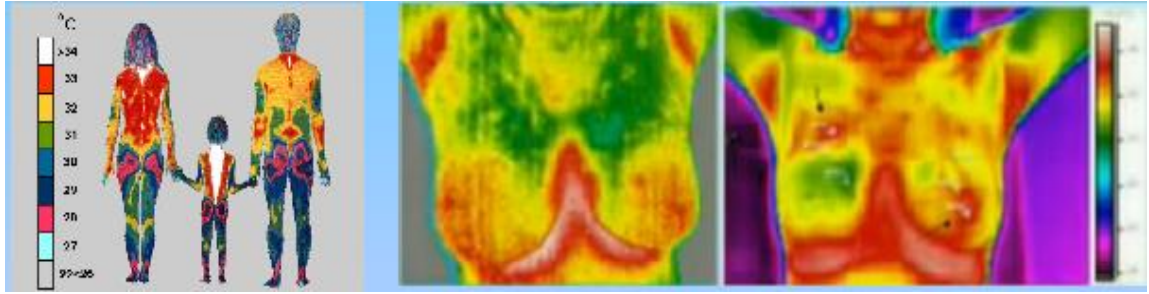

Figure 18: Thermal signals in the human body.

9. *Radiological*: The ionizing interaction with biological structures generates this signal type. They convey information about the structure of the inner anatomy and play a significant role in diagnostics and therapy, as depicted in Figure 19.

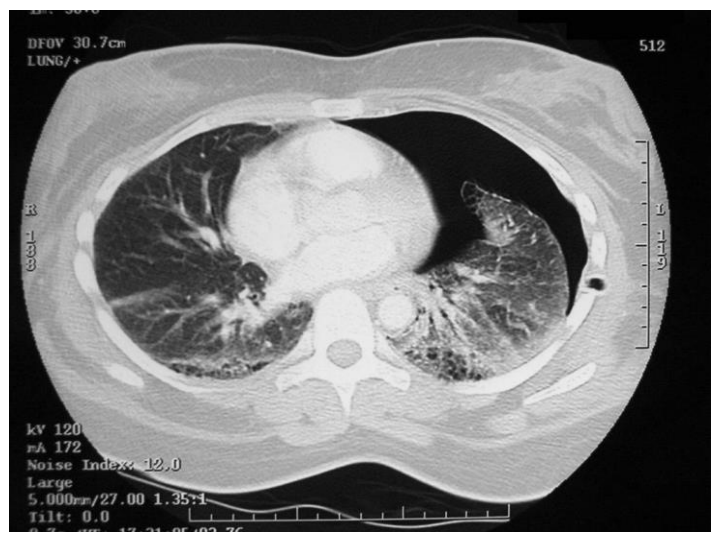

Figure 19: Computed tomography chest.

10. *Ultrasonic*: The interaction of ultrasonic waves with organism tissues generate this type of signal. The acoustic tissue impedance and the changes in anatomy are the main information that they provide. A piezoelectric transducer is used in the probes used for acquisition. A typical example is to detect the condition of the fetus, as shown in Figure 20.

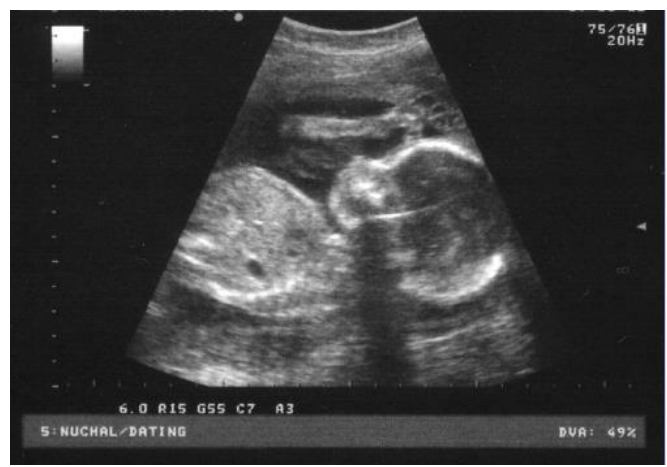

Figure 20: Application of ultrasound in fetal examination.

## 3.3 Conversion of analog signal to digital

The analog signal contains infinite information unlike the discrete which contains a certain amount of information and is part of a corresponding analog. Of course, because many signals in nature and more specifically biological signals are generally analog, there is a need to convert them into digital signals, in a sequence of numbers of finite precision to achieve easy processing. This process is called analog to digital conversion. To create a discrete signal from an analog, a system that converters the analog into digital signal (A/D) is needed with sampling period $T_s$. It should be noted here that a discrete signal can come from many different analog signals and be identical in all cases. The only change is the frequency or sampling rate, which is denoted as $f_s = 1/T_s$ [19][27].

As can be observed in Figure 21, the conversion of analog to digital signal passes through three stages, sampling, quantization and encoding, which will be explained below in detail [21].

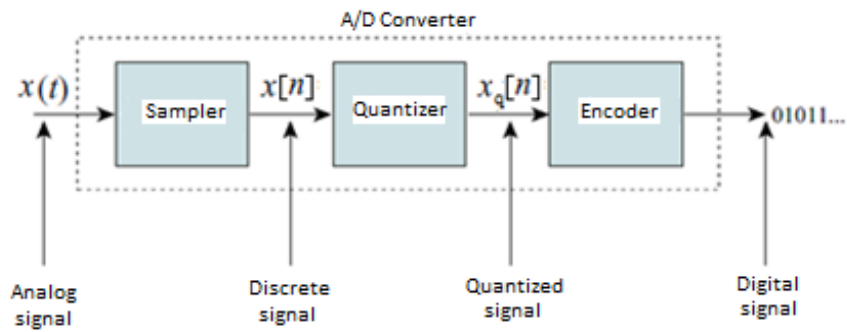

Figure 21: Conversion of analog signal to digital.

Before using the A/D converter, an *anti-aliasing filter* and a *sample and hold module* (S/H) can be employed [19]. This type of filter is necessary to be employed to the incoming analog signal to prevent the undesired phenomenon of frequency aliasing. Furthermore, because it is impossible to convert the analog signal into a digital instantaneously, it is important to preserve a constant value to the sampler for the time the conversion lasts. This is performed by the S/H module.

The creation of analog signal from discrete needs a digital to analog converter and a reconstruction filter to unite the discrete points. The later conversion cannot have the original information as it has been lost and so it takes an approximation of the original.

The digital to analog conversion will not be further analyzed as it is not part of this thesis. However, the conversion of analog signal to digital as the input to the digital processor can be distinguished through Figure 22. Then, the processor output is a digital signal, which inserts in the digital-to-analog converter to be converted back to analog.

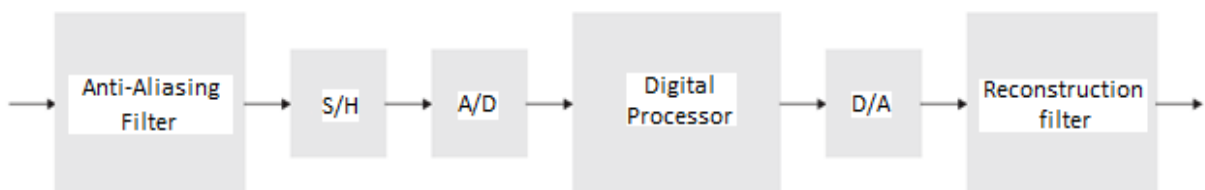

Figure 22: Main parts of analog to digital and digital to analog conversion.

### 3.3.1 Sampling

Sampling is the first step of the conversion of a signal from analog to digital. The sampler, as depicted in Figure 23, is the system which performs this process and extracts samples (at equal time defined by $T_s$) from a continuous signal. A theoretically ideal sampler produces samples, which correspond to instantaneous value of a continuous signal at specific time points. Let analog signal $x(t)$ and a discrete $x[n]$. The relationship that connects them is: $x[n] = x(nT_s)$, $-\infty < n < +\infty$.

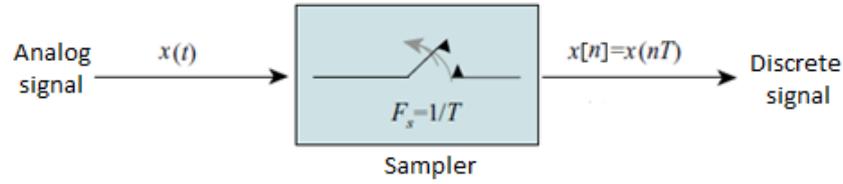

Figure 23: Sampler.

The variable of time for the analog signal is t while for the discrete is n and the relationship that connects them is $t = nT_s$. This results in that logically there will be a link that connects the frequency $f$ of analog signals with frequency $\lambda$ of discrete-time signals. If the signal $x(t)$ is sinusoidal, then:

$$x[n] = x(n \cdot T_s) = A \cdot \cos(2\pi f nT_s + \theta) = A \cdot \cos\left(2\pi n \frac{f}{f_s} + \theta\right)$$

Comparing the above equation and the type of discrete sinusoidal signal, it is concluded that $\lambda = f/f_s$ is a *normalized* or *relative frequency* and, thus, to know the frequency of a discrete signal we need to know the sampling frequency. Furthermore, the frequency of the analog sinusoidal signals can take values in the range $0 < f < +\infty$. In discrete sinusoidal signals, the shortest possible period that allows rotation is $N = 2$. Therefore, we have $2 < N < +\infty$, which means because of the relationship $\lambda = 1/N$ that $0 \leq \lambda \leq 1/2$. If negative frequencies are allowed, then the relationship is $-0.5 < \lambda \leq 0.5$ and for $\Omega = 2 \cdot \pi \cdot \lambda$ is $-\pi < \Omega \leq \pi$. It is easily understood that the basic frequency interval in discrete-time signals is limited, $(-1/2, 1/2]$, but for corresponding analog signals is the entire real line (infinite) [19].

According to the above, since we have the maximum value for the discrete frequency signals, which is 0.5, then we conclude that the maximum value for $f$ for a given sampling rate will be found by the formula: $f_{max} = f_s/2$. To basic question that arises and has to do with the sampling is: *Under which conditions can we reconstruct (fully) the sampled signal? So how often we should take samples in order to have a replica of the analog signal?* The answer can be given by the sampling theorem or Nyquist - Shannon. The sampling theorem guarantees that narrow spectrum signals (e.g. signals which have the frequency domain values to a maximum frequency) can be fully reconstructed from sampled form in the case the sampling frequency is equal or higher than two times the maximum frequency, $f_s \geq 2f_{max}$. Otherwise, the phenomenon of aliasing appears and leads to distortion of the content of the signal. Of course, the negative side of the sampling, which is inevitable, is some loss of signal information.

### 3.3.2 Quantization

*Quantization* is the process of converting the samples obtained from the output of the sampler in random sequence of discrete values, which belong to a finite set of amplitude levels. The values that a digital signal can receive are named *levels of quantization*. The difference among two consecutive levels is called *resolution or quantization step*. If the levels equidistant, then the quantization is called *uniform* otherwise *non-uniform*. So, through quantization, the nearest quantization level of each price resulting from sampling is found. After this point the signal is digital. The difference between the original unquantized sample $x[n]$ and the quantized output $x_q[n]$ is the sequence $e_q[n]$, which is called the *quantization error* or *quantization noise* and it often is denoted by $\Delta$. With the quantization, the range is limited to a finite number of values M, which determines the sharpness of the signal (The greater it is, the less the distortion-quantization error). These values are represented by a series of binary numbers of 0 and 1 [21].

Having as main purpose the reduction to a desired number of digits, the rounding method is used. During the rounding, the quantizer gives in $x[n]$ the value of the nearest quantization level. The error resulting $e_q[n] = x[n] - x_q[n]$ ranges in the values $-\Delta/2$ and $\Delta/2$, $-\Delta/2 \leq e_q[n] \leq \Delta/2$. The quantization step is defined as $\Delta = (x_{max} - x_{min})/(L-1)$, where $x_{max}$, $x_{min}$ is the largest and the smallest possible value of $x[n]$ respectively, the difference is called dynamic range of the signal and $L$ is quantization levels number. Thus, the quantization leads to loss of information. It is a necessary and an irreversible process since all samples within $Δ/2$ of a quantization level are represented by the same value. Therefore, it is impossible to remove the quantization noise. What can be achieved is limiting it with an increase in the quantization level $L$. The quality of the quantization is evaluated by signal to noise ratio (SNR): $SNR = 1.76 + 6.02b$.

### 3.3.3 Encoding

*Encoding* is called the conversion of the quantized samples in a suitable form, which means that each value of quantized samples is represented by a number consisting of b bits. Therefore, the levels to which we want to represent the signal are selected and then assigned each quantized sample with a codeword. Usually it is implemented in practice with binary codes, where the codewords comprise two distinct values of 0 or 1.

More specifically, if we have L quantization levels, then L different binary numbers are required. A codeword length of b bits can describe $2^b$ levels and so the required number of quantization levels should be less than or equal to this number or equivalent to $b \geq log_2 L$. The selection of the number of channels is based on the desired accuracy (compromise between precision of representation of signal, repository and processing time). In market, we can find A / D converters with accuracy up to b = 24 bits.

The quantization and encoding process of a signal is illustrated in Figure 24.

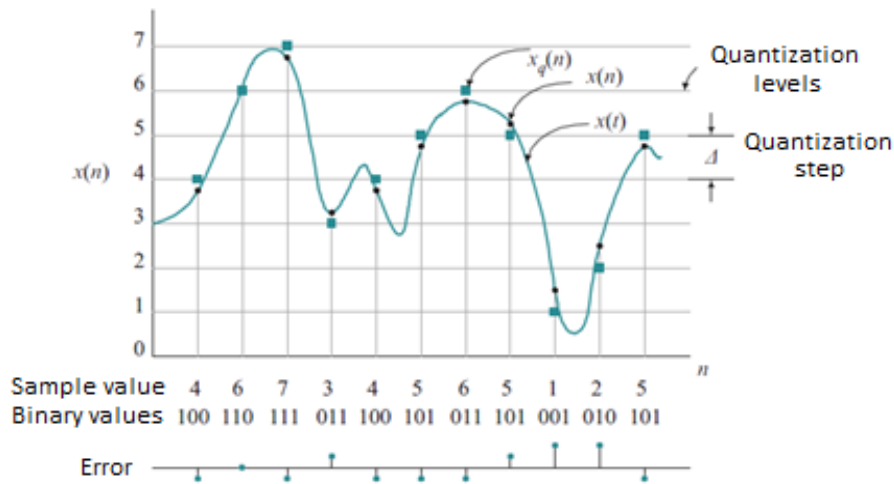

Figure 24: Quantization and signal encoding.

# Chapter 4: Signal processing, feature extraction and methods of dimensionality reduction

## 4.1 Empirical Mode Decomposition (EMD)

The EMG signal related with hand movements can be assumed to be a non-stationary and non-linear process. Thus, the use of decomposition techniques for stationary and linear processes will provide misleading results. For this reason, the use of EMD [9] seems ideal for this case, since it is an adaptive method targeting ideal for non-stationary and non-linear signals, which happens with most of the real-life applications.

EMD relies on the following assumptions: (a) there are at least two peaks in the signal (one maximum and one minimum value), (b) the time interval between the peaks defines the representative time measure. In the case in which there are no peaks but includes only inflection points, the extrema will be revealed after differentiating once or multiple times the initial signal. The result will be gained by integrating the components.

EMD is a vital part of the Hilbert-Huang transform. This method can break down any complex dataset into multiple (finite) components named intrinsic mode functions (IMF). Each IMF portrays a simple oscillatory mode which complements the simple harmonic function. The constrains set in the IMF definition ensures a well-behaved Hilbert transform of it. EMD has proved to be highly efficient and adaptive and its operation on the local characteristics of the signal in the time domain, which makes it ideal for nonstationary and nonlinear processes [28].

There are two conditions to be satisfied for a function to be considered an intrinsic mode: (a) the number of peaks should be equal or vary mostly by one to the number of zero crossings in the full waveform under test, and (b) the average value of the two envelopes based on the local maxima and minima is zero [9].

The initial fact of EMD is to realize oscillatory signals at the level of their local oscillations and to form the notion that the signal is the superposition of fast oscillation on the slow ones, as observed in Figure 25 [29].

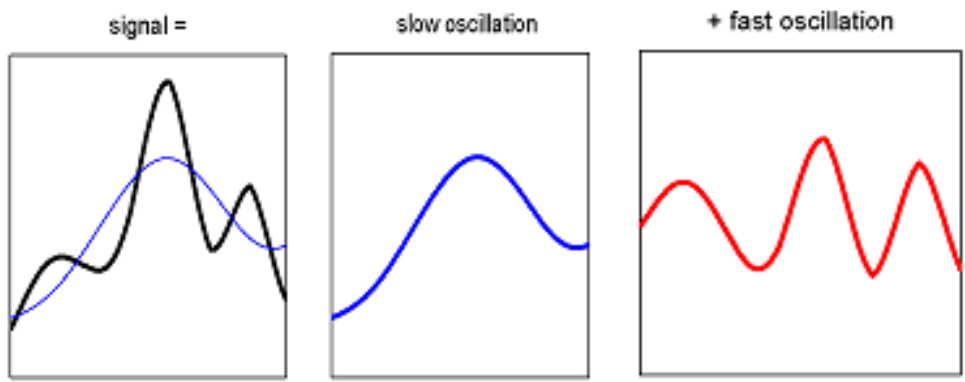

Figure 25: Components of fast and slow oscillations of the signal.

**EMD Algorithm**

The signal $x(t)$ is decomposed by using EMD into multiple IMFs using the following technique:

Step 1  All the local extrema of $x(t)$ are found to form two envelopes: one for the local maximum ($e_{max}(t)$) and one the local minimum ($e_{min}(t)$) values through interpolation (typically a cubic or spline).

Step 2  The running mean is estimated via the envelopes created in the previous step: $m(t) = (e_{min}(t) + e_{max}(t))/2$.

Step 3  The running mean is subtracted from the signal creates the detail signal, $d(t) = x(t) - m(t)$.

Step 4    The overall procedure is repeated by replacing $x(t)$ with $m(t)$ until reaching a monotonic function, which will be the final residual (or a specific number of IMFs set by the user is reached).

Step 3 will not produce necessarily an IMF based on the constraints set before. Thus, the detail signal, $d(t)$, will meet these criteria after sifting [9][30]. The summation of IMFs and the residual term recover the initial signal $x(t)$:

$$x(t) = \sum_i IMF_i(t) + r(t)$$

where the $IMF_i$ can be called as the "high frequency" part and the residue $r$ as the "low frequency" part.

After the EMD algorithm usage, the Hilbert transform can be used to each IMF separately to calculate the instantaneous frequency, which is the phase function's derivative. Thus, Hilbert transform estimates the amplitude and frequency of each IMF as a function of time. The initial signal can also be expressed as the real part (RP) of the Hilbert transform [31], using the following equation:

$$x(t) = RP\left(\sum_j a_j(t) \cdot e^{i\theta_j(t)}\right) = RP\left(\sum_j a_j(t) \cdot e^{i\int \omega_j(t)dt}\right)$$

The Hilbert-Huang spectrum $(H(\omega, t))$ is the amplitude's time-frequency distribution. The function of EMD can be easily understood by the Figure 26 showing step by step the process for creating an IMF.

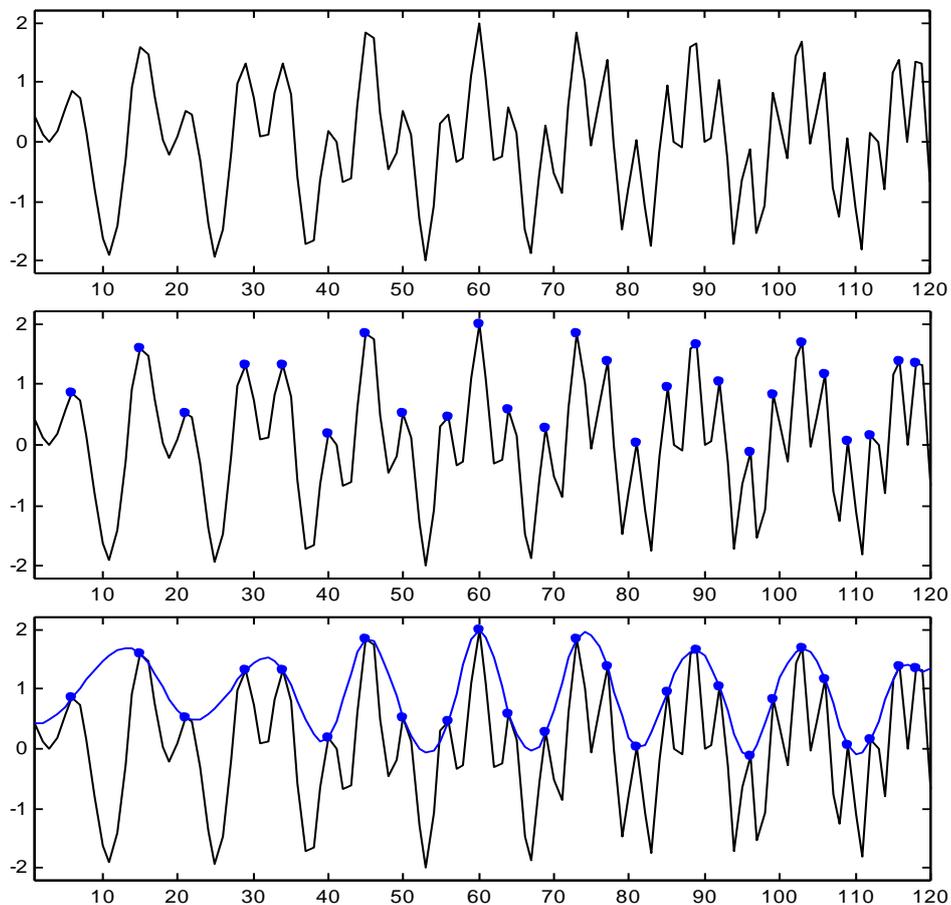

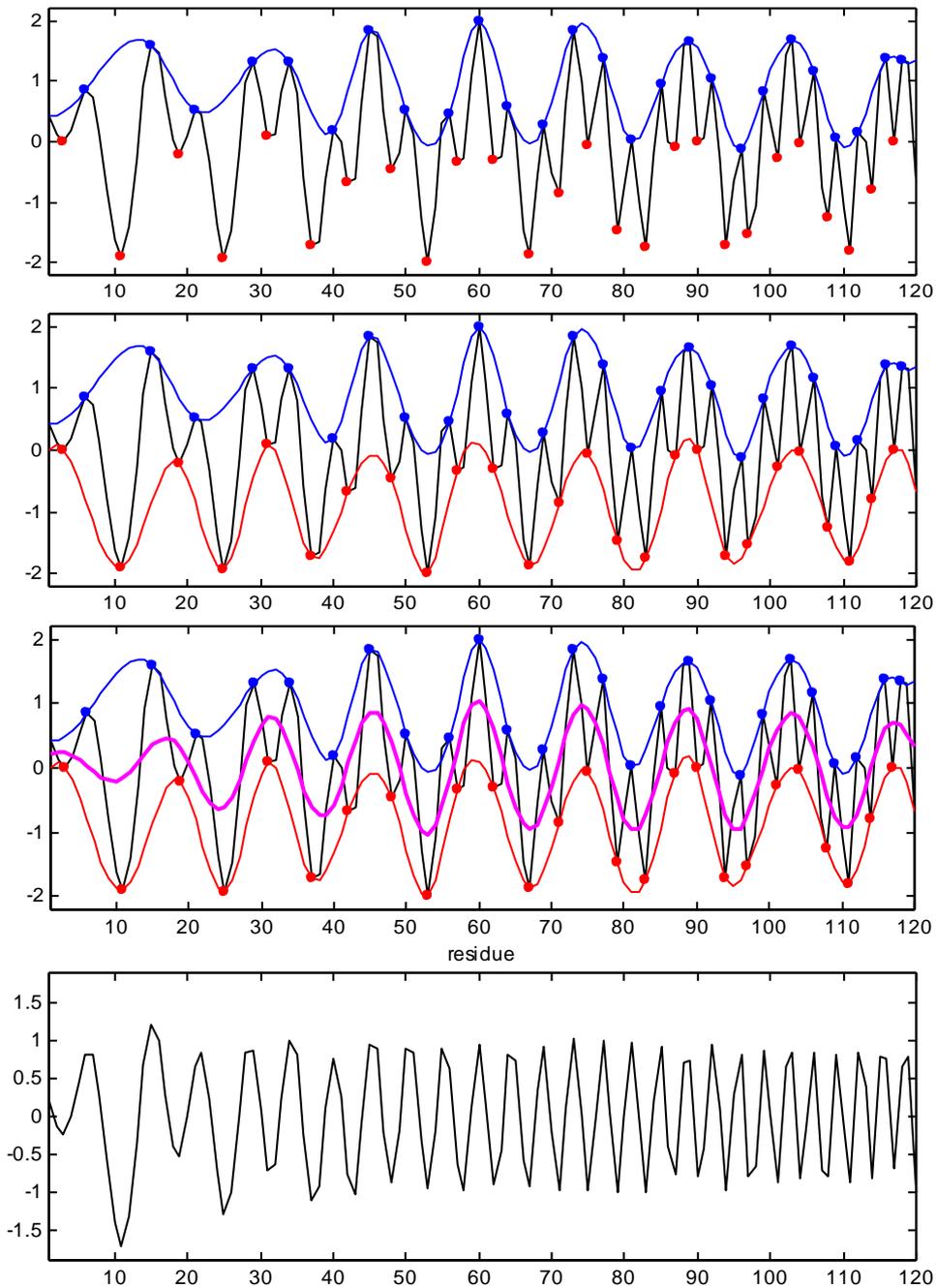

Figure 26: One iteration of the EMD algorithm.

## 4.2 Signal Features

Acquiring a raw signal in most cases has no direct use. For this reason, even if there is no processing in our raw signal, there are some features that are extracted. The feature extraction conducts the condensation of the information and deal with the dimensionality problem, which is solved by the methods listed in the next section.

The data that will be produced by taking the features of each signal can be used quite easily in matrixes and inserted as inputs for algorithms. For instance, such algorithms are pattern recognition techniques, which will be discussed in Chapter 6 and will be used in this thesis, to separate the classes with the maximum accuracy. In these following sections, the most widely used features in time and frequency domain are to be explained.

### 4.2.1 Features of the time domain

The **time domain** features [32] are:

1. *Average value*: is the summation of the signal samples values divided by the number of samples. If the signal takes negative values, it is quite often to take the absolute signal value. The mathematical expression is:

$$\mu = \frac{1}{N} \sum_{k=1}^{N} |x_k|$$

   when $x_k$ is the $k^{th}$ value from $N$ samples of the signal.

2. *Variance*: is a power density measure of the signal and its square root is called *standard deviation* ($\sigma$). Variance is given by:

$$VAR = \frac{1}{N-1} \sum_{k=1}^{N} x_k^2$$

3. *Zero Crossing (ZC)*: counts the number of times that the signal passes through zero or changes sign and, thus, expresses the frequency. A threshold needs to be set to reduce the noise induced at zero crossing. Using two contiguous samples of EMG signal $x_k$ and $x_{k+1}$, the ZC can he calculated as:

$$ZC = \sum f(x), \text{ where } f(x) = \begin{cases} 1, & if, (x_k > 0 \text{ και } x_{k+1} < 0) \\ & or \ (x_k < 0 \text{ και } x_{k+1} > 0) \\ 0, & otherwise \end{cases}$$

   for $k = 1, 2, 3, \ldots, (N-1)$

4. *Slope Sign Changes (SSC)*: is the measure of the times a slope of the signal changes sign. For three consecutive samples of the signal $x_{k-1}$, $x_k$ and $x_{k+1}$, the SSC can be calculated as:

$$SSC = \sum f(x), where \ f(x) = \begin{cases} 1, & if, (x_k < x_{k+1} \text{ και } x_k < x_{k-1}) \\ & or \ (x_k > x_{k+1} \text{ και } x_k > x_{k-1}) \\ 0, & otherwise \end{cases}$$

   for $k = 1, 2, 3, \ldots, (N-1)$

5. *Waveform Length (WL)*: is a measure of the cumulative variation of the signal. For two adjacent samples of EMG signal $x_k$ and $x_{k+1}$, it is given by:

$$WL = \sum_{k=1}^{N-1} (|x_{k+1} - x_k|)$$

6. *Willison Amplitude (WAMP)*: is the number of times that two consecutive samples exceed a specific threshold. Given two contiguous samples of EMG signal $x_k$ and $x_{k+1}$, it is given by:

$$WAMP = \sum_{k=1}^{N-1} f(|x_{k+1} - x_k|), where \ f(x) = \begin{cases} 1, & if \ x > threshold \\ 0, & otherwise \end{cases}$$

7. *Kurtosis*: is a measure of the "peakedness" of a real-valued random variable probability distribution [33]. A high value in the chart is illustrated by thick lines and low dispersion and a low value is by thin lines and the dispersion concentrated near the middle. The values obtained can be classified into three categories: leptokurtic, mesokurtic or platykurtic, see Figure 27. The kurtosis can be defined as:

$$k = \frac{E(x - \mu)^4}{\sigma^4}$$

where $E$ is the expected value and $\sigma$ is the standard deviation.

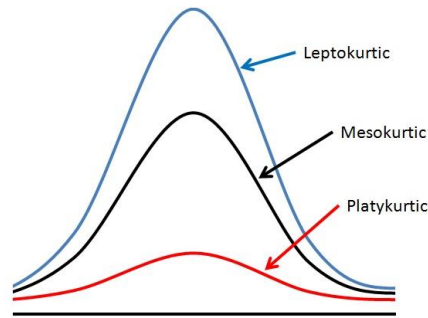

Figure 27: Diagrams of kurtosis with different values.

8. *Skewness*: measures the asymmetry of probability distribution of a real-valued random variable meaning towards which side of the mean it "leans". The skewness can have either a positive or negative value or be undefined. As can be easily observed in Figure 28, the asymmetry in the negative left tail is longer and the most volume of the distribution is to the right side of the diagram [34]. The exact opposite happens in the positive asymmetry. The mathematical expression for the asymmetry is:

$$s = \frac{E(x - \mu)^3}{\sigma^3}$$

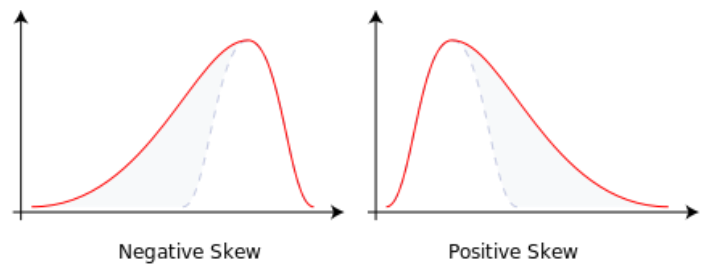

Figure 28: Negative and positive skew.

9. *RMS value*: known also as the root mean square is a statistical measure of the magnitude of a variable quantity, which is expressed by the following formula:

$$RMS = \sqrt{\frac{1}{N} \sum_{n=1}^{N} x_n^2}$$

10. *Auto Regressive model (AR)*: is the representation of a type of random process and describes each signal's sample as a sum of linear combination of former samples and the error of the white noise. It is expressed as:

$$x_n = -\sum_{i=1}^{p} a_i \cdot x_{n-i} + w_n$$

where $x_n$ are the samples of the signal, $a_i$ is AR factor, $w_n$ is the white noise and $p$ is the degree of AR model.

### 4.2.2 Features of the frequency domain

The **frequency domain** features [35] are:

1. *Median Frequency (FMD)*: is the frequency for which the spectrum is split into two areas with equal power. It is expressed mathematically as:

$$\sum_{j=1}^{FMD} P_j = \sum_{j=FMD}^{M} P_j = \frac{1}{2}\sum_{j=1}^{M} P_j$$

   where $P_j$ is the signal's power in the frequency $j$.

2. *Mean Frequency (FMN)*: is the weighted sum calculated by the product of the frequency and the power spectrum and divided by the spectrogram intensity summation, as in:

$$FMN = \sum_{j=1}^{M} f_j \cdot P_j \bigg/ \sum_{j=1}^{M} P_j$$

   where $f_j$ is the frequency spectrum in frequency $j$.

3. *Total Power*: is the sum of the power spectrum subtracted by a constant amount SMO and is defined as:

$$TP = \sum_{j=1}^{K} P_j - SMO$$

   where $SMO$ is zero spectral moment, $P_j$ is the power for frequency $j = 1, \dots, K$

4. *Mean Power (MNP)*: is the mean power in the power spectrum of the signal given by the following equation:

$$MNP = \sum_{j=1}^{K} P_j \bigg/ K$$

5. *Power spectrum Ratio (PSR)*: is defined as the ratio of power $P_0$, which is the maximum value of the power spectrum and $P$, which is the total energy of the power spectrum:

$$PSR = \frac{P_0}{P} = \sum_{f_0-n}^{f_0+n} P_j \bigg/ \sum_{-\infty}^{+\infty} P_j$$

## 4.3 Methods for dimensionality reduction of extracted features

Dimensionality reduction of features is useful for two main reasons. Firstly, it is common for some features to be correlated meaning they provide redundant information. Secondly, if some specific features are irrelevant, they can affect negatively the classifier's discriminative ability. Thus, it is common to use dimensionality reduction techniques for conveying the information as pure as possible to the classifier and eventually improving its performance.

The dimensionality reduction techniques can be divided in two main families: the one focuses on transforming the data mapping them to a lower dimensional space and the other on selecting a features subset. Algorithms from both families are selected in this work; principal component analysis (PCA) [8] and RELIEF algorithm [10].

### 4.3.1 Principal Component Analysis (PCA)

PCA is commonly used in pattern recognition for dimensionality reduction and feature generation. The original space is linearly transformed into a lower dimensional space by projecting the *N*-dimensional data onto the M eigenvectors (*M ≤ N*) extracted from their covariance matrix that corresponds to the larger eigenvalues. Even in the case where all the eigenvectors are selected and, thus, there is no dimensionality reduction, it is possible to achieve an improvement in the classification performance in the case where PCA generates uncorrelated features [8].

PCA is a technique used for data patterns identification since it transforms the data in order to emphasize their similarities and differences. The PCA usage becomes even more important in high dimensional data, where the graphical representation is not possible. Moreover, PCA assists in data compression since it achieves dimensionality reduction without losing information [36].

The original data space is transformed into a new one, which is rotated so that its axes will point towards the direction with the highest variance of the data. These axes are called as principal components (PCs) and are sorted by variance, which means that the first principal component (PC 1) represents the direction of the highest variance of the data. Moreover, PC 2 represents the direction of highest value of the remaining variance, which is orthogonal to PC 1. This is expanded to all the other PCs, which can create the new space covering the desired variance level. It should be noticed that each of the principal components is a linear combination of all the original variables. The transfer from the three-dimensional space (x, y, z) in the two-dimensional (PC 1, PC 2) can be easily understood via Figure 29 [37].

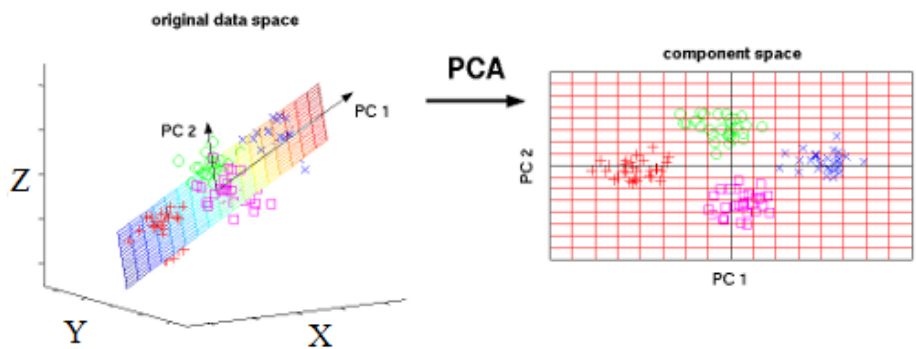

Figure 29: Transfer from three-dimensional world to two-dimensional using EMD.

**PCA transform** [36]

Step 1    Get characteristics in $N$ dimensions and perform a pre-processing of characteristics subtracting the mean for each dimension (attribute). So, if each feature has vector $x_i$ for $i \in [1, N]$, where all the values has mean $\tilde{x}_i$. This creates a data set with zero mean. The table is called $DataAdjust$.

Step 2    Calculation of matrix, which is $N \times N$.

Step 3    Calculation of eigenvalues and eigenvectors (the covariance matrix is square). Sort eigenvectors from the highest to lowest price giving an order of importance to the eigenvalues.

Step 4    Extraction of the maximum eigenvector, which is the principal component (PC). Ignore those of the lowest significance, especially if they have low price without losing essential information. Therefore, if there are $N$ dimensions of eigen-values and eigenvectors and choose the first $P$ eigenvalues, it will conclude at a data set with only $P$ dimensions.

Step 5    Construction of a feature vector: $FeatureVector = (eig_1 \; eig_2 \; \ldots \; eig_P)$

Step 6    Get the new data set. Multiply the transpose matrix FeatureVector with the inverse table of the initial data set:

$$FinalData = FeatureVector^T \times DataAdjust^{\,T}$$

## 4.3.2 RELIEF Algorithm

RELIEF is considered as one of the most successful algorithms due the simple and effective assessment of the quality of the features. The algorithm is a feature weight based inspired by instance-based learning [10].

The RELIEF algorithm explores features that have a statistical relevance to the target concept using training data $D$, a sample size $m$ and a $\tau$-threshold ($\tau \in [0,1]$). If the two instances are different, it can be expressed with the function $diff$, which uses the two instances $x, y$ as input.

The function $diff$ in the numerical case (integer or real) is:

$$diff(x_k, y_k) = (x_k - y_k)/no_k$$

where $no_k$ is a number employed for normalization of the values that are differed and is bounded to [0, 1].

RELIEF selects a sample in $m$ triplets ($X$, Near-Lose, Near-hit) format. In order to select the last two instances, the weight update employs the Euclidean distance. The selected features are those which have scored higher above a user predefined threshold.

**RELIEF Algorithm** [38]

Step 1    Separate $D$ into positive ($D^+$) and negative ($D^-$) instances,
$$W = (0, 0, \ldots, 0)$$

Step 2    For $i = 1$ to $m$
    Select randomly an instance $X \in D$
    Select randomly one $Z^+ \in D^+$ closest to $X$
    Select randomly one $Z^- \in D^-$ closest to $X$
    if ($X$ is a positive instance)
      then $Near - hit = Z^+ \; ; Near - miss = Z^-$
      else $Near - hit = Z^-; Near - miss = Z^+$
    For $i = 1$ to $p$ ;update weights
    $w_i = w_i - diff(x_i, Near - hit_i)^2 + diff(x_i, Near - miss_i)^2$

Step 3    Relevance $=. (1/m)W$
    For $i = 1$ to $p$
      if ($relevance_i \geq \tau$)
        then $f_i$ is a relevant feature
        else $f_i$ is an irrelevant feature

RELIEF, as defined above, chooses a sample of m triplets of an instance $X$, its Near-hit instance and a Near-miss instance. RELIEF uses a p-dimensional Euclidean distance for the selection of Near-hit and Near-miss. It updates the weight vector of characteristics $W$ [10] for each triplet and computes the average feature weight vector Relevance (of all the features of the target concept). Eventually, the algorithm chooses the features with an average weight («relevance level») higher that the set threshold $\tau$ [10].

The aforementioned formulation targets binary classification problems. In this work, we are using a modification, RELIEF-E, for multiclass problems. In that case, the near miss of an instance X is calculated by employing the nearest neighbor belonging to any other class [38].

## Chapter 5: The origin of biopotentials and biosignal acquisition techniques

This chapter will analyze the generation of various biological signals, which are usually acquired in clinical practice. Many different types of bioelectric phenomena can be acquired with relative ease from various body parts with specific techniques. Such techniques, which are described below, are for example the electromyogram (EMG), the electrocardiogram (ECG) and the electroencephalogram (EEG) [39][40].

### 5.1 Electrical activity of excitable cells

The electrical activity is present in our body and essential for sustaining life. This electrical activity has source at the cellular level, for example, the neurons in the brain, cardiac cells in the heart, and cells of the pancreas. This electrical activity is a result of the distribution and movement of electrically charged ions [2].

Bioelectric potentials are generated due to the electrochemical activity of excitable cells, as observed in Figure 30. They are components of muscular, glandular, or nervous tissue. When they are stimulated properly, they exhibit an *action potential*, otherwise they remain or fluctuate around their *resting potential* [39].

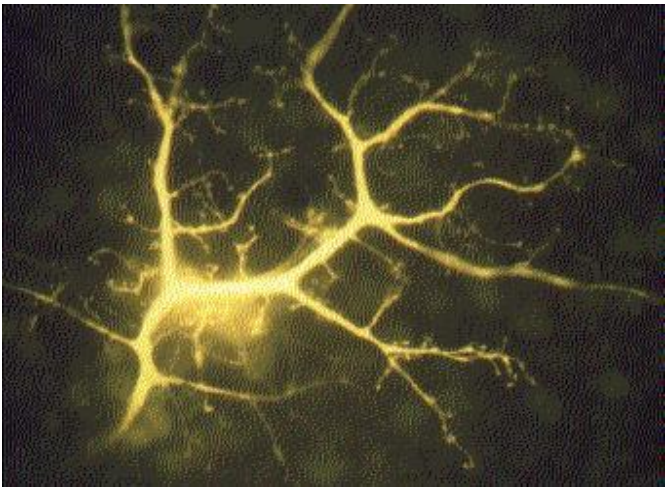
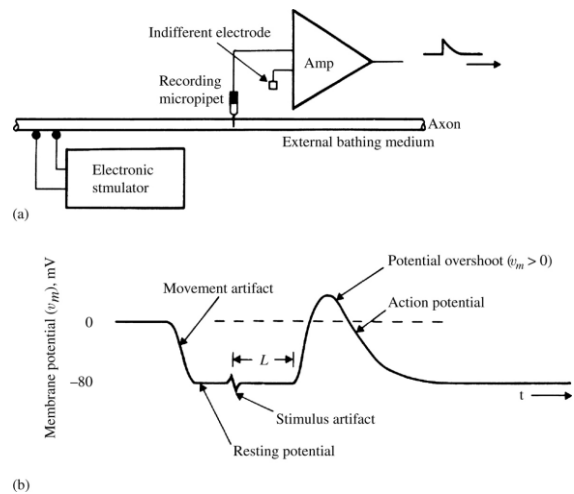

Figure 30: Single interneuron in the rabbit's retina.

Figure 31: Action potential recording of an invertebrate nerve axon.

Figure 31 (a) depicts the method of measuring the resting potential is usually measured. A microelectrode should be placed near the surface of an excitable cell. A short current pulse can be produced in the axon by an electronic stimulator. The recording of this activity is performed through an electronic stimulator and enhanced by the amplifier. Figure 31 (b) illustrates a recording of the electrical activity of a single nerve fiber along with its resting potential (DC offset). Moreover, it depicts the transient behavior action potential after applying an adequate stimulus. L is the latent period from stimulus to recording site.

### 5.1.1 Resting Potential

The difference between intracellular and extracellular potential remains almost constant in a single excitable cell. This resting potential of the internal medium is in the range -40 to -90 mV in comparison to the external medium. This potential difference between the interior and the exterior of the cell membrane is expressed by a negative value. This is because the potential outside the cell is defined zero and, hence, there is a relative excess of negative charges through the membrane.

The cell has a very thin membrane (7 to 15 nm) consisting of lipoprotein complex, which assist in making it not accessible to intracellular protein and other organic anions ($A^-$). The resting state membrane slightly permeable to $Na^+$ and rather freely permeable to $Cl^-$ and $K^+$ [39][42]. Both in the extracellular and the intracellular space are negative and positive charges (ions) equal to each other. In the extracellular space, the primary cation is $Na^+$, which has a ten times higher concentration there than in intracellular space, and the main anion Cl⁻. The calcium ions ($Ca^{2+}$) and chloride ions ($Cl^-$) concentrations are also preserved at higher levels outside the cell apart from some intracellular membrane-enclosed parts that it is possible to have high $Ca^{2+}$ concentrations (green oval), as observed in Figure 32. The main concentration for the intracellular space is $K^+$, which has more than twenty times greater concentration there than in extracellular space, and large organic anions. The typical $K^+$ concentration of the cell internally (cytosol) is 140 mmol/liter, whereas externally (bathing) is 2.5 mmol/liter.

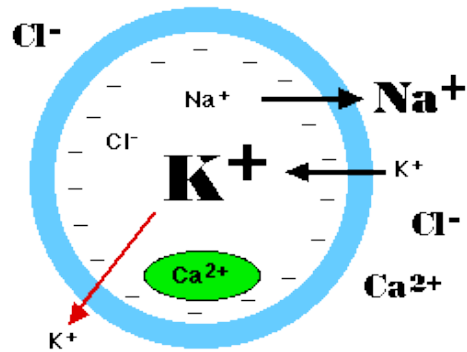

Figure 32: Ionic Relations in the Cell.

There is an excess of negative charged ions through the cell membrane while outside there a concentrate with an equal quantity of positively charged ions. The difference of concentration generates a diffusion gradient with an outward direction from the membrane. There is a $Na^+$ - $K^+$ pump that constantly pulls molecules $Na^+$ outwardly and leaves equal number of anions in the cell, since the anions are large molecules and cannot cross the cell membrane. These anions are pulled for electrostatic reasons within the cell $K^+$ ions that can penetrate the cell membrane. $Na^+$ ions are concentrated outside the cell and $K^+$ ions therein during the electrochemical equilibrium of the cell, in which there is no net movement of ions anymore. The $K^+$ movement towards this diffusion step (while the anion component of the non-diffusion remains within the cell) makes the cell interior more negative compared to the outside. The result is a *polarization* that is a potential generated by the difference in concentration between the interior and the exterior of the cell membrane. This potential is called the *resting potential* which is characteristic of each cell type and has a size of a few mV, because the genesis involved only a very small number of ions.

The membrane can be modelled electrically as a leaky capacitor due to its structure, where the lipoprotein complex acts as a thin dielectric material and charge separator. The pores (transmembrane ion channels) are those who provide the leakage property since ions can move inwards and outwards the cell membrane. The corresponding electric field to the membrane capacitor at rest has an inward direction from positive to negative across the membrane. It usually constrains the flow of cations ($K^+$) and anions ($Cl^-$) outside and inside the membrane, respectively. Hence, a balance is eventually reached between the opposed forces, electrical and diffusional, across the membrane. The voltage for which this balance occurs is known as equilibrium potential for the potassium ions ($E_k$), in the case where the potassium ions are the main species responsible for the resting condition. $E_k$ is measured in Volts and expressed via the Nernst equation:

$$E_k = \frac{RT}{nF} \cdot ln\frac{[K]_o}{[K]_i} = 0.0615 \cdot log\frac{[K]_o}{[K]_i} \text{ (V)}$$

at 37 °C (body temperature). The n is the valence of the potassium ions, $[K]_o$ and $[K]_i$ are the extracellular and intracellular $K^+$ concentrations in moles per liter, respectively, R is the universal gas constant, T is absolute temperature measured in Kelvin, and F is the Faraday constant [39].

In order to maintain a steady-state with the intracellular and extracellular ionic concentration imbalance, a repetitive active transport of ions is required to cancel the effect of their electrochemical gradients. The sodium–potassium pump is the mechanism which allows the active transport of 3 $Na^+$ outside the cell in exchange of 2 $K^+$ inside. The $i_{NaK}$ is the current associated with this mechanism with an outward direction and the tendency to increase the negativity of the membrane potential. The pump uses adenosine triphosphate (ATP), which is produced by mitochondria in the cell, for covering its energy needs.

Therefore, the ions flow across the membrane are affected by (a) the available pores of the membrane, (b) diffusion gradients, (c) ions active transport opposing to the established electrochemical gradient and (d) the electric field with inward direction. The resting potential depends on the ions conductivities ($g_K$, $g_{Na}$, $g_{Cl}$) and the charge separated by the cell membrane. The K$^+$ ions have an outward diffusion direction based on their concentration gradient. On the other hand, the organic anions that are non-diffusible remains inside the cell, which creates a membrane potential difference. Although the electroneutrality is preserved among the internal and external bulk media, there is a monolayer of concentrations scattered on the outer surface of the membrane and anions spread in the inner surface because of the membrane capacitance. The number of ions causing the membrane potential is a small portion of the total existing in the bulk media. Due to the $g_{Na} \ll g_K$ in the resting state, K$^+$ efflux cannot be compensated by the Na$^+$ influx. The diffusion of chloride ions is inwards against its concentration gradient, which is balanced by the electrical gradient [39].

### 5.1.2 Action Potential

The excitable cells can generate an action potential if they are properly triggered (Figure 31b) and can conduct it leading to its propagation. The stimulation needed for its generation should be adequate to force the membrane potential to overcome the threshold potential. In that case, the action potential is generated and travels thought-out the medium without attenuation and with a constant conduction velocity. The cell even in the resting state has a potential, which means its polarized. The action potential is characterized by the all-or-none property meaning that the membrane potential undergoes a typical cycle: a potential change from the resting level of a certain amount a specific time period. For the case of a nerve fiber, the potential change is around 120mV and the duration is 1 msec approximately. Even if a stimulus provides more than the required for an action potential generation, the result will be the same.

The action potential originates from the dependencies in voltage and time of the membrane conductivities/permeabilities of specific ions, especially for N$^+$ and K$^-$. Due to depolarization, the sodium conductance $g_{Na}$ is substantially increased. Thus, N$^+$ moves rapidly inside the cell causing even further depolarization and leading to a further increase of the conductance. In the case where $υ_m$ exceeds the threshold, depolarization occurs due to the positive feedback between the N+ influx and the $g_{Na}$ increase. In this case, $υ_m$ reaches the sodium equilibrium Nernst potential, $E_{Na}$, with value around +60 mV.

Nevertheless, there are two reasons that $υ_m$ cannot achieve this level: (1) the sodium conductance ($g_{Na}$) is voltage and time dependent with shorter period compared to the duration of action potential, and (2) the potassium conductance ($g_K$) increases with a delay causing hyperpolarizing effect, which drives the $υ_m$ towards resting potential. When the membrane is in the resting state, the $g_K$ remains higher than its normal value for that state. It returns to it after some slow exponential decay of time. This means that there is an outflow of K$^+$ leading to hyperpolarization and a $υ_m$ undershoot.

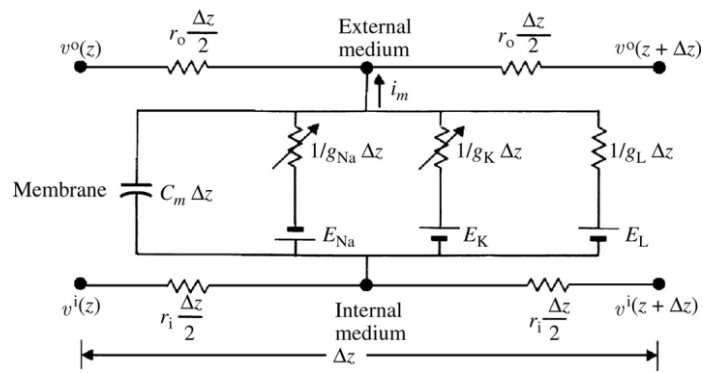

Figure 33: Equivalent circuit that simulates a cylindrical cell.

Figure 33 shows an equivalent circuit of the network that describes the electrical behavior of a membrane thin segment. The whole membrane nervous axon can be characterized with a distributed way using a repetitive structure of this same basic form. The membrane properties are specific membrane conductances $G_{Cl}$, $G_K$ and $G_{Na}$ (mS/cm$^2$) and the specific membrane capacitance $C_m$ (uF/cm$^2$). It is common to eliminate $G_{Cl}$ conductance. If it is assumed that the cell cytoplasm and the external bathing medium are purely resistive, they can be modelled as $r_i$ and $r_o$ (V/cm), respectively. Moreover, the transmembrane current per unit length (A/cm) is annotated as $i_m$ and the internal and external potentials at point z as $υ^i$ and $υ^o$, respectively. The transmembrane potential at each point in z is given by $υ_m = υ^i - υ^o$ [44].

When an initial stimulus cause change in the membrane potential and the potential level reaches the firing, all channels Na$^+$ open and Na$^+$ enters the cell by simple diffusion mechanism. Due to the high accumulation of positive charges inside the cell, the polarity of the membrane reverses and reaches a maximum voltage value for the type of cell. This phenomenon is called *depolarization*. The Na$^+$ channels remain open for few milliseconds. Because of the membrane potential displacement to positive values, opening of K$^+$ channels and moving of the accumulated K$^+$ outside the membrane is induced. This leads to the return of the membrane potential in the initial state and the phenomenon is called *repolarization* [39][42]. The energy potential is illustrated in Figure 34.

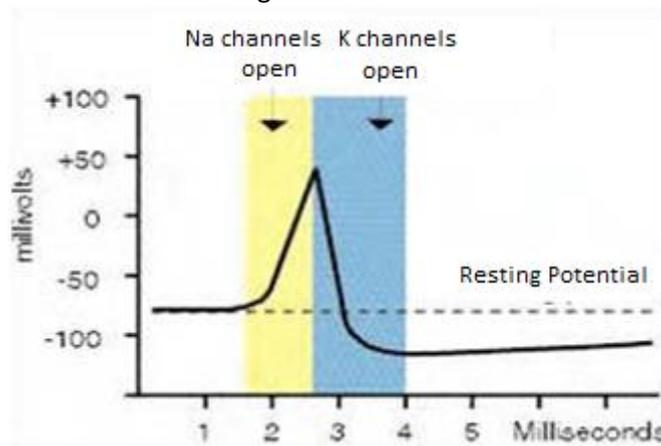

Figure 34: The energy potential.

The generation of an action potential after receiving an adequate stimulus alternates the ability of the excitable membrane to give any response to another stimulus right after the first. In fact, the membrane will be impervious to any other, even very strong, stimulation in the initial phase of the action potential. This time segment is known as absolute refractory period. The period after this is called relative refractory, in which an action potential is possible is there is a strong stimulation. These periods set an upper limit to the firing rate of an excitable cell. For instance, in the case where the absolute refractory period of a nerve axon is 1 msec, then the upper limit of firing is less than 1000 impulses/sec.

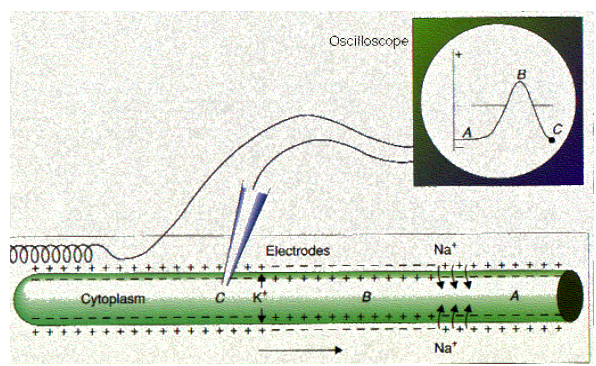

Figure 35: The impulse of the nerve.

As it can be observed from Figure 35, the inner part of the axon membrane is charged negatively compared to the outer part (A) in the case of the resting neuron. In the case where the action potential propagates (B), the polarity is inverted, which changes again to normal polarization (C) due to K+ ions rapid outflow [40].

The movement of the electrical signal after the stimulus can be observed along the axon of the nerve cell. By carefully observing the Figure 36, we see that in the stimulus area is initially created depolarization and is negatively charged in the extracellular region. Then the positive charges are attracted to the right, and so we have a shift of the stimulus to the right. The displacement speed of the stimulus is a few meters per second. Besides, in the case of faster speeds, confusion could have been caused in the brain from their processing [43].

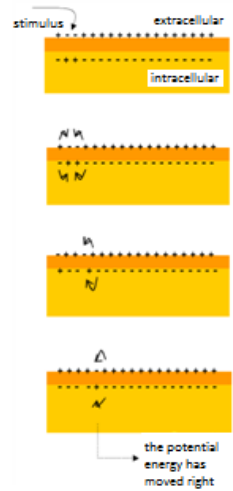

Figure 36: Electrical signal moving along the nerve cell after stimulus.

For an action potential that propagates via a single unmyelinated nerve fiber, the fiber segment, that undergoes a change to the active state (active region), is small relatively to the fiber length and happens momentarily. Figure 35 depicts schematically the load distribution along the fiber near the active area. Note that the action potential propagates towards right, and the membrane which is located in front of the active region is polarized, as in the rest condition. The active region is reversed polarized due to the membrane depolarization reaching positive voltage values. On the back side of the active region, the membrane is in the re-polarized phase.

Based on the specified charge distribution, solenoidal (closed-path) current streams in the pattern depicted in Figure 35. There is a close path (solenoidal) of current flowing in the right direction due to the charge distribution. This current causes a drop in the membrane potential $υ_m$ in the region after the active area causing a depolarization to this segment. This area becomes activated when the $υ_m$ reaches the threshold, which is around 20 mV higher than the resting potential. This current has no effect in the regions behind the active area since they are in refractory state. Thus, the full process is considered self-excitatory. The action potential can travel throughout the fiber without any attenuation.

## 5.2 Biosignal Acquisition Techniques

The biosignals have different recording techniques, which are dependent on the body location and form found. They could have been also classified based on **the way of recording** the signal. Some techniques are listed below [27].

### 5.2.1 Electromyography (EMG)

The *Electromyography (EMG)*, also mentioned as myoelectric activity, is a recording method of the muscle's electrical activity. The distribution of electricity is spread into the muscle when an action potential is at a muscle fiber. The signal generated by the superposition of the induced action potentials by all the muscle's active motor.

There are two **recording methods** the EMG signal:

I. *intramuscular or depth* (needle ή fine-wire) type of electrodes has as basic part a needle which is made of stainless steel and is wholly insulated except the edge, as seen on Figure 37. It enters the muscle interior to measure the potential of using extracellular fluid. It provides fine movements and records the activity within the muscle. The value of the signal resulting from moving beyond the muscle, relies on the electrodes size and the distance between them. The electrode's size is determined by the diameter of the circular disk (1-3 cm). It is important because it is proportional to the monitored muscular volume and inversely proportional to the resistance. Depending on the muscle targeted for recording, the proportional size electrode is used. Before placement of the electrodes of this type, a specific skin preparation should be performed. This includes the removal of dead cells with light rubbing using rough material and cleaning with alcohol solution to reduce the resistance connection with skin electrodes.

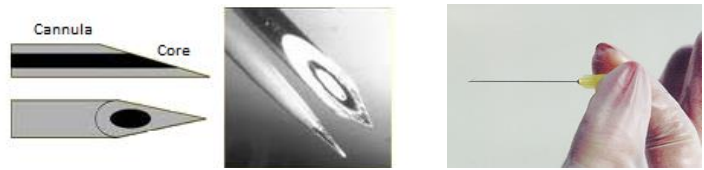

Figure 37: Concentric needle electrodes and fine wire respectively.

II. *surface* type of electrodes is placed on the skin surface of the muscle under test and can be split into two categories [14]:

   i. *passive surface* type of electrodes is connected to an amplifier circuit with the assistance of cables for suitable signal acquisition. They consist of a metal disc, usually silver or chlorargyrite (Ag or AgCl), an adhesive disk, contain insulation everywhere except the point of contact with the skin and detect the average muscle activity on the surface, as observed in Figure 38. They can be disposable or reusable.

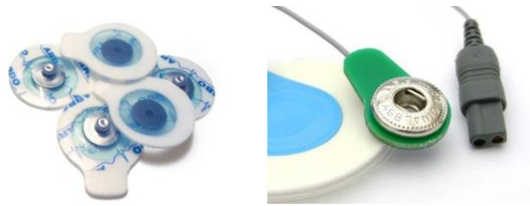

Figure 38: Passive electrodes.

   ii. *active surface* type of electrodes [45] has attachment containing a pre-amplification for surface electrodes and are commonly referred as dry electrodes because there is not demanding preparation of the skin surface needed in the area where we want to measure, as shown in Figure 39 and Figure 40. In fact, the high input impedance amplifier is placed close enough so there is no need for an electrolyte cream.

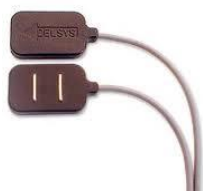 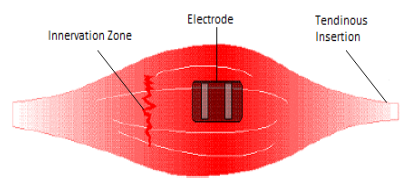

Figure 39: Active electrodes.

Figure 40: The ideal position of electrode is between innervations zone and ventricle muscle.

The basic difference of these two types of electrodes can be found in the frequency range. Assuming that the signal recorded by needle electrodes do not pass through the skin and fat tissue that act as a low pass filter, they have a much higher frequency range, i.e., up to 5000 Hz, in contrast to that of 500 Hz of surface electrodes. Of course, the placement of the intramuscular electrodes to the patient is a much more painful process compared to the surface. Furthermore, intramuscular electrodes may cause bleeding or infection.

In each case, the electrode is either *unipolar* or *bipolar*.
- An electrode of *monopolar* configuration is placed over the belly of the muscle, an electrode further away is used as reference and the signal generated between both electrodes is amplified and recorded as shown in Figure 41.

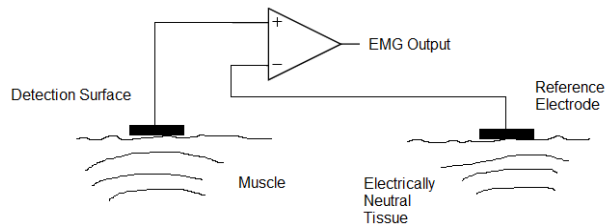

Figure 41: Recording signal monopolar electrode.

- In the case of *bipolar* electrode configuration, as depicted in Figure 42, two recording electrodes are positioned on top of the belly muscle within 1 to 2 cm putting a reference electrode further away but equidistant from the two recording electrodes. The signal is generated by the subtraction of the two recording electrodes and the output signal, which is relative to the reference, is amplified and recorded. The bipolar electrode configuration has the advantage that removes the common noise between the two electrodes and, thus, the signal obtained has less noise.

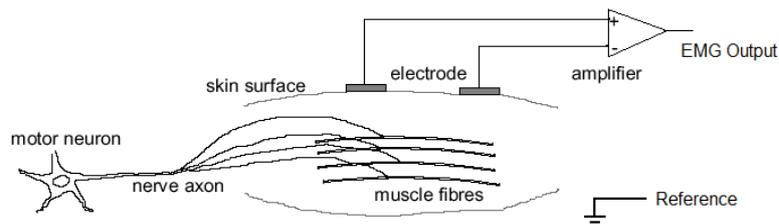

Figure 42: Recording signal of bipolar electrode.

The recorded signal has low intensity and contains enough noise. So, it needs some amplification in order to be processed and used. For this reason, it should be strengthened linearly over the range of the amplifier and recording system.

Furthermore, the noise that the primary signal contains may be due to:
- the existence of dead cells or hairs on the skin and hence preventing the stable recording.
- a movement of the wire which connects the electrode and the amplifier, if they are not connected to a fixed connector contact.
- recording and other signals that cause the adjacent muscles to that we want to record.
- the equipment.
- ambient noise caused by electrical power lines, electromagnetic interference power of PPC in the frequency of 50 Hz, mobile phones and other devices.
- radio waves and magnetic fields that interact with the human body.

If we try to eliminate the noise, we will only achieve to limit it. Therefore, filters should be used except from the amplifier, as illustrated in Figure 43, to limit the noise signals.

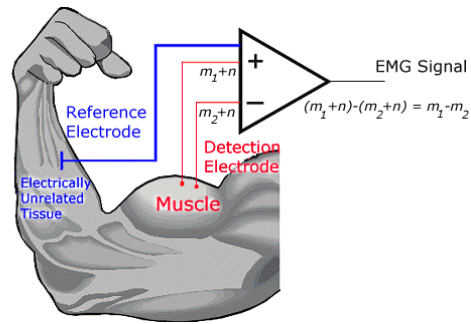

Figure 43: Schematic of the differential amplifier, where m is the signal and n the noise.

To record as much clean signal as possible, bio-amplifiers should have certain characteristics such as [14]:
- Amplifier gain / Dynamic Range; the gain is on the range of 100-1000 and there is a choice to the user depending on the desired gain.
- High Input Impedance to be much larger than the skin's.
- Frequency Response - Bandwidth, to enhance appropriately the signal to the frequencies that the EMG signal occurs (in surface electrodes 5-1000 Hz and in depth 20-2000 Hz).
- Common Mode Rejection (CMMR), since the differential amplifier takes the difference of the two signals and, thus, assuming that they have the same width, the one neutralizes one another.

According to the above, in order to have the optimal possible recording, we should make the correct choice in the size and type of the electrode, which will be determined by the muscle and the purpose of measurement. Furthermore, the skin surface should be cleaned in the case of passive electrode. Also, the appropriate locations for the electrodes' placement should be selected for recording and reference. Moreover, special attention should be paid to the distance between them to conform with the specifications.

The EMG signal is useful because we can study [46]:
- the functional role of muscles, i.e., we can learn what muscles are of primary motion for a particular movement.
- the simultaneous activation of agonist and antagonist muscles.
- muscle fatigue and its role in the functional role of the forearm.
- difference between a healthy muscle and a muscle that suffers from myalgia.

EMG signals have been used in numerous applications of human-machine interface, such as control of exoskeletons and developed in many clinical and industrial applications.

## 5.2.2 Electrocardiography (ECG)

The *Electrocardiography* is associated with the recording of changes in the electrical potentials produced by the electrical stimulation/activity of the heart, as depicted in Figure 44. The heart has a certain beating pattern depending on the current state of the body and pumps blood throughout the body. The signals that cause heart's fibers contraction are originated from the sinus node, which is considered the natural heart's pacemaker. Specifically, they are generated from the production system and treatment of stimulations that the heart has and transported to the final recipient, the myocardial cell which contracts. Then, the restoration of electric potentials follows in the initial state. Therefore, if electrodes are positioned on the skin two sides of the heart, it becomes possible to record the electric potential that is produced. The curve obtained in this way is called electrocardiogram and it is displayed on special paper, in which any problems in heart rate and the conductivity of the heartbeat are reported showing if there is an underlying cardiovascular disease.

Normally, the first electric potential of the heart is produced on sinus, which in turn diffuses within gulfs and performs the contraction of these. Then passes the atrioventricular node, diffuses to the ventricles through the left and right part of His, and becomes the contraction of the ventricles. The electrocardiogram records the above electrical potentials as arriving at the surface of the body traveling from the sinus node to the ventricles.

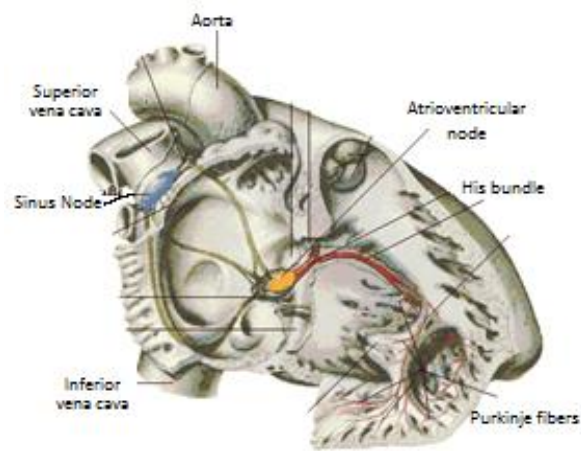

Figure 44: Veins and arteries in the heart.

A normal ECG pulse contains a P wave, a QRS complex and T wave [47], as illustrated in Figure 45:
- The P wave represents the successive depolarization[2] of the left and right atrium and has typically positive polarity. P wave has a maximum duration of 120 msec and is of low frequency, under 10-15 Hz.
- The QRS complex reflects the depolarization of the left and right ventricle. Its duration can vary from 70 to 110 msec at a normal heart rate, having the greatest amplitude of ECG waveforms. Because of the sharp rise and fall, the frequency range of the QRS complex is significantly higher than the other waves, ranging from 10 to 40 Hz approximately.
- The T wave corresponds to the ventricular repolarization[3] having duration around 300 msec under physiological conditions. The heart rate is greatly dependent on the T wave location since the closer it is to the QRS complex, the faster is the pace.

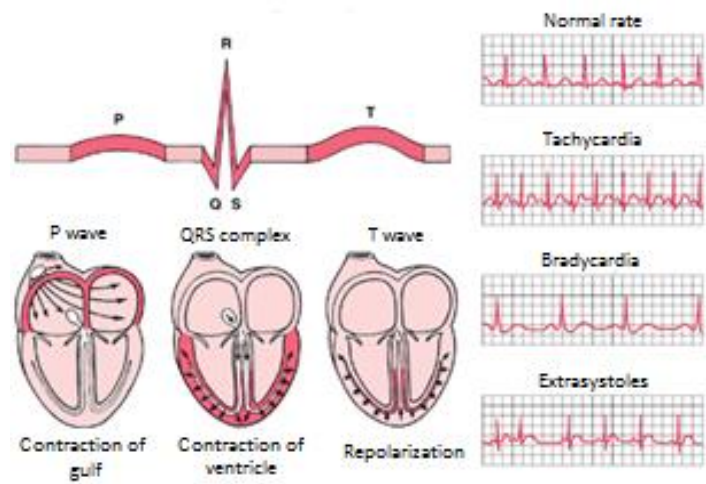

Figure 45: ECG signal of healthy person and different signals in specific situations.

There are certain positions on the chest for the placement of each electrode (ten in total). Possible incorrect placement may lead to inaccuracies that will lead to wrong medical diagnosis. $V_1$ and $V_2$ are positioned in the 4th intercostal space on both sides of the sternum and $V_3$ close to APEX of the heart. The $V_4$,

---

[2] Depolarization is the increase of the membrane potential of the cell reaching positive values. An action potential is possible to be generated in neurons by a large enough depolarization. The opposite phase is hyperpolarization, which inhibits the generation of an action potential.

[3] Repolarization is the continuous decrease of the membrane potential returning it in negative values. It occurs right after the depolarization phase and returns the membrane potential to its resting value.

$V_5$ and $V_6$ are placed at the same level and more specifically in the 5$^{th}$ intercostal space in mid-clavicular line, anterior and middle axillary line, respectively. Moreover, electrodes are placed at the ends, and more specifically at points RL, and LL (anywhere above the ankle and below the trunk), RA, and LA (anywhere between the shoulder, and elbow), wherein R is the right and L the left side. All of the above are shown in Figure 46.

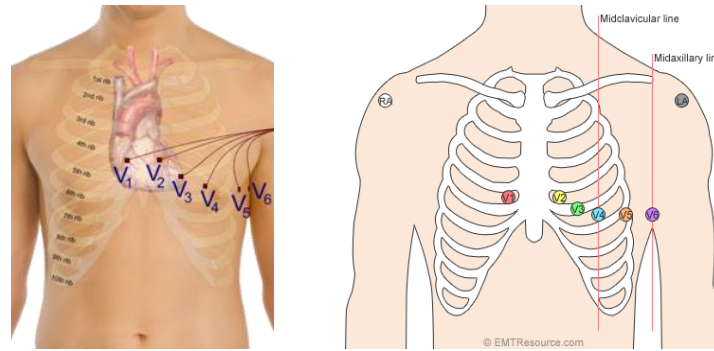

Figure 46: Positions of electrode placement in Electrocardiography.

Electrocardiography assists in diagnosing acid or old infarct, atrial fibrillation, atrial or ventricular bradycardia, atrial or ventricular tachycardia, atrial flutter, atrioventricular block of various types, hypertrophy of cardiac sinus and ventricular, dilatation of sinuses, anatomical lesions of the conduction system (exclusion of right or left part) of advanced electrolyte disturbances, toxic effect of cardiac drugs, the actual cessation of the heart.

## 5.2.3 Electroencephalography (EEG)

The electroencephalogram (EEG) is the recording of the signal produced by the brain's electrical activity. The brain, as depicted in Figure 47, is composed of billions of cells and each of them generates and transmits infinitesimal electric currents. In order to record, there must be larger signals which result from the sum of many signals of brain nerve cells. The recording of the electrical brain signals is achieved with electrodes placed on the scalp. Therefore, electrical signals are recorded from the cortex, which is much closer to the skull of the other structures of the cortex of the brain that are internal [48].

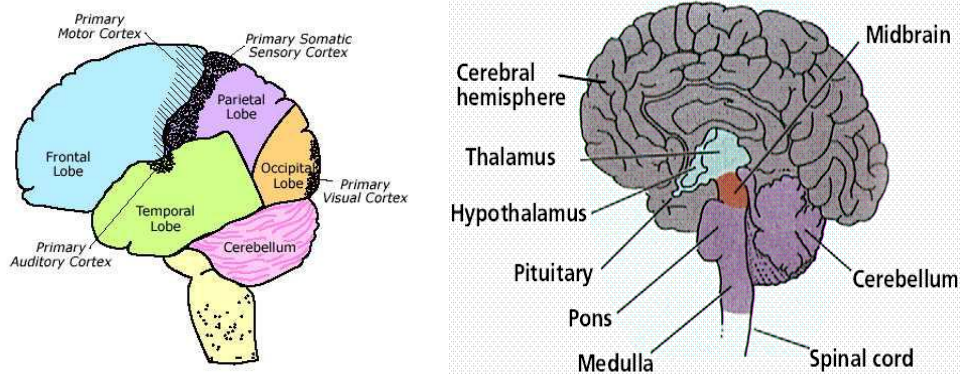

Figure 47: Parts of Brain.

The EEG consists of a straight horizontal line called isoelectric line. Its name stems from the fact of not changing the potential of the extracellular matrix recorded at the electrode, which implies non-crossing current of the membrane, so the nerve cells are in a resting state. The lines appear to diverge from the isoelectric line are called waves and reflect the change in the potential of the extracellular space of the brain. Therefore, the activation of nerve cells is captured by moving current through the membrane of nerve cells. The waves, then, are indicative of electrical activity of nerve cells in the cortex. They may be either positive or negative, i.e. above or below the isoelectric line.

There are five categories depending on the wave frequency band: delta, theta, alpha, beta, gamma.
- *(delta waves – δ)*: They range from 1 to 4 Hz and have usually a maximum amplitude at about 100 mV. These waves are displayed in normal state of the deep sleep of adults and in infants and in pathological state in alert mode.
- *(theta waves – ϑ)*: They range from 4 to 8 Hz with an amplitude greater than 50 mV. They are produced under normal conditions to children and pathological conditions in adults.
- *alpha waves – α*: They range from 8 to 13 Hz with a maximum amplitude of 50 mV. They are waves that occur naturally when the body and mind is at rest and pathologically when the person is in coma.
- *beta waves – β*: They range from 13 to 30 Hz and have amplitude of 5 to 10 mV. They can be observed in full alert. In this state beta waves replace alpha [49].
- *gamma waves - γ*: They are the frequencies of 30 Hz and above and occur during sleep REM. Although there is the theory that these waves related to consciousness, there is no absolute agreement, but it is argued that "Whether or not related waves – gamma with subjective knowledge is a question difficult to answer with certainty at the present time" [50].

Figure 48 shows the waves that are produced in various situations of human life and examples of the various wave categories which have been described above.

The placement of electrodes on the scalp has been standardized and there are various standards with the most popular to be the International System 10-20, which uses a total of 21 electrodes. The name is due to the choice of 20% of the distance between the two ears as the distance between any two electrodes and in the selection of 10% of the distance between the two ears as the distance from the ear to the closest to this electrode. The system can be observed in Figure 49 and Figure 50 [49][51].

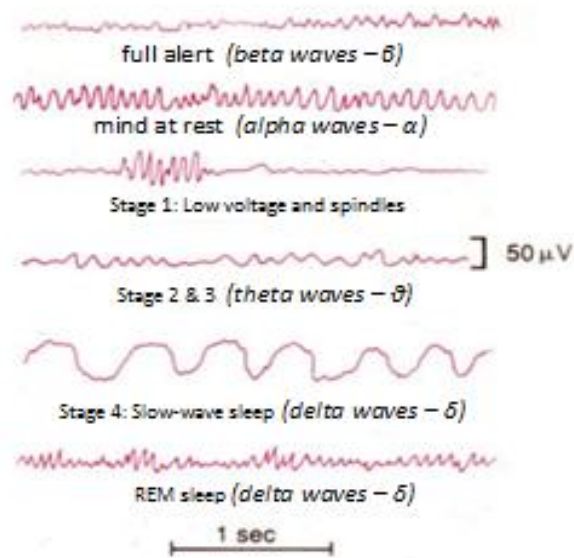

Figure 48: Category of waves according to the wave frequency band.

Originally, the tests were conducted based on the simple rhythms quantification (delta, theta, alpha and beta), but the EEG signal characteristics (varying from 5 to 300 mV in the range of 0.01 - 150 Hz) and the lack of the necessary equipment has been an obstacle for any development in this area. The entire expansion of BCI is related with the fact that the different brain activities are related to different patterns of neural activity shown in the distributions of EEG potential, which can be identified and quantified. Nowadays, there are systems that allow controlling a cursor of the screen and the software options just by thought.

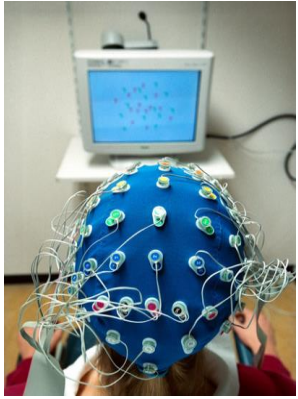
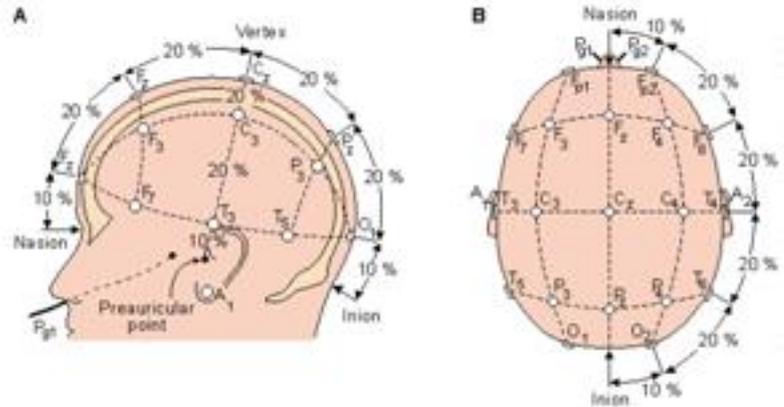

Figure 49: Electrodes placed on the patient.   Figure 50: Electrode placement on the 10-20 International system.

Besides the categories mentioned above, there are several more. The above methods were mentioned since one method (Electromyography) will be used in this thesis and the other two (Electrocardiography and Electroencephalography) approach it enough. In Figure 51 the voltage and frequency range are observed for the aforementioned signals.

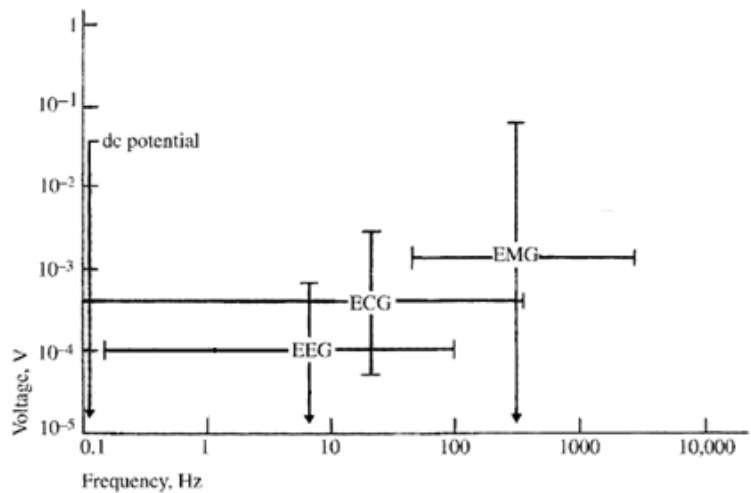

Figure 51: Voltage and frequency range from a few biopotential signals.

# Chapter 6: Pattern Recognition Methods

Pattern recognition is a scientific area with a main objective to classify objects into categories. These objects, also known as patterns, can vary from signal waveforms to images based on the application. Its history can be tracked even before 1960's, at which period it was part of statistics and in theoretical level. The demand for practical applications increased with the technological development in computation, which set new directions in the theoretical studies. There is an increasing demand for data management and automation in industrial production in the postindustrial phase of our society. This fact has set pattern recognition in the forefront of engineering research and development as one of the most vital parts for machine intelligence systems used for decision making applications [8]. In few words, the goal of pattern recognition is to provide a reasonable answer for all possible inputs and perform their most likely matching [52].

## 6.1 Artificial Neural Network (ANN)

ANN is influenced by the biological NN such as those in brain. It is a mathematical model of multiple interconnected neurons that constitute a neural network, which receives information, processes it and gives the processed data outputs. Therefore, they work together to solve specific problems. As the brain learns mostly from experience so the ANN do. They should be trained in order to operate and provide correct output. A neural network is an adaptive system based on a learning phase, which causes changes in the synapses and, thus, changing its structure. Each application may need a different arrangement, such as classification of information via a learning procedure [53][54][55].

### 6.1.1 Neuron

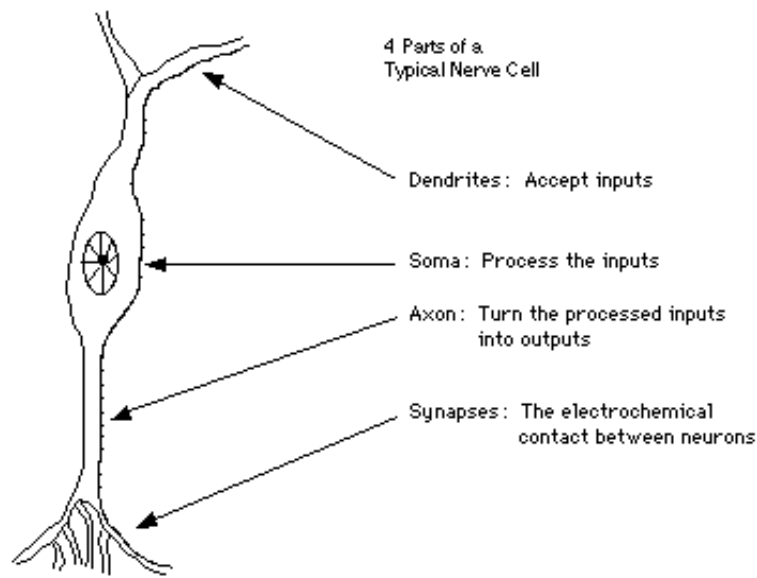

Figure 52: The four parts of a Nerve Cell.

A typical nerve cell [57], as observed in Figure 52, consists of:
- dendrites: receive inputs.
- soma: transforms the inputs.
- axon: provides outputs through the processed inputs.
- synapses: the electrochemical connection among neurons.

The neuron is the fundamental building block of any ANN, as can be seen in Figure 53. Studies on the physiology of nerve cells demonstrated that the output is a non-linear transformation of the inputs. After several approaches were conducted to determine the mathematical model that describes the operation of the nerve cell, the most important and yet the simplest model will contain a linear and a non-linear operator, which are connected in series as in Figure 54 [55].

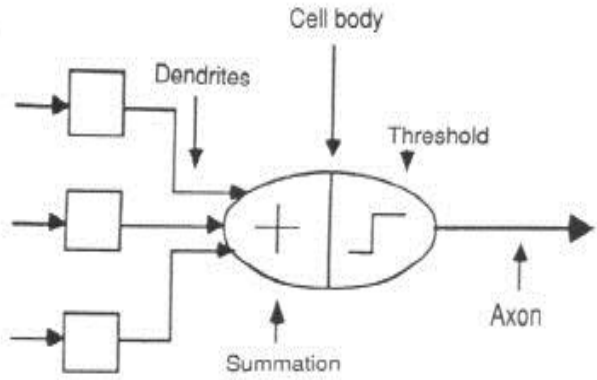
Figure 53: Neuron Model.

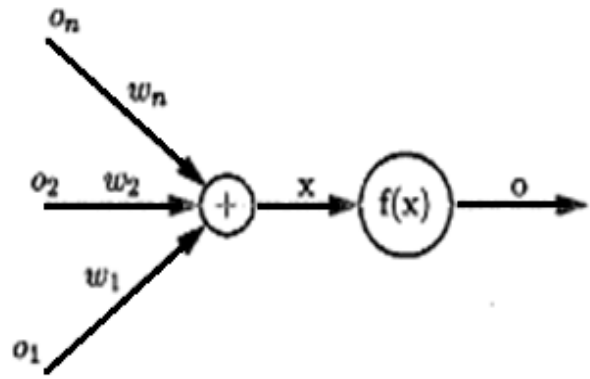
Figure 54: The cell body.

The linear operator is the inner product of the input vector *o* with weight *w* that represents the memory of the neuron represented by the formula:

$$x = o \cdot w = \sum_{i=1}^{N} o_i \cdot w_i$$

The operator *o* is a nonlinear function that receives as input the output of predecessor neurons:

$$o = f(x)$$

As observed, the input and output of the neuron has the same symbol as the output of one can be the input of another. The function of the nonlinear operator has the characteristics of the sigmoid functions, where the graph is shown in Figure 55, which:

1. is an *increasing function*, i.e. $\forall x_1, x_2 \in R, x_1 > x_2 \implies f(x_1) \geq f(x_2)$
2. has *infinitesimally finite limits*, i.e.

$$\lim_{x \to +\infty} f(x) = a, \quad a \in R - \{-\infty, +\infty\}$$
$$\lim_{x \to -\infty} f(x) = b, \quad b \in R - \{-\infty, +\infty\}$$

3. has as *domain all real numbers* and a *bounded range*, i.e.: $f: R \to [a, b]$. This property is a consequence of the previous two properties.

Usually, this function can take many forms such as the *exponential sigmoid* and *exponential tangent* respectively:

$$f(x) = \frac{1}{e^{-az+b} + c} \text{ and } f(x) = \frac{e^{ax} - e^{-ax}}{e^{ax} + e^{-ax}}$$

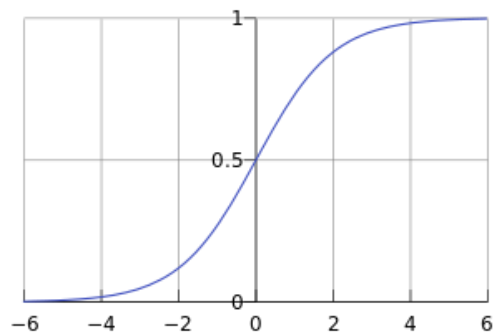
Figure 55: Sigmoid curve.

### 6.1.2 Structure

Generally, each neural network is divided into three basic levels, as distinguished in Figure 56:

- *input layer*: raw data fed to the network.
- *hidden layer*: dependent on the activity of the input units and the weights of the synapses between the input and the hidden units.
- *output layer*: dependent on the activity of hidden units and weights of the synapses between the hidden and output units.

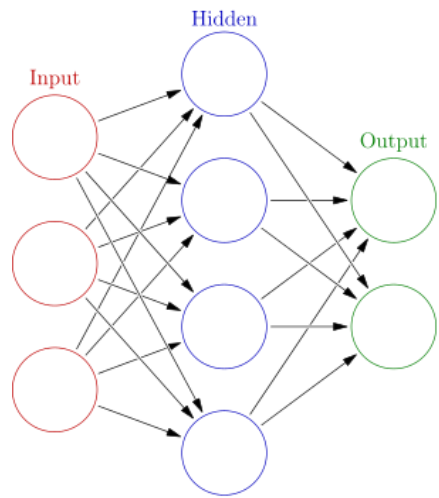

Figure 56: Structure of one hidden layer ANN.

Furthermore, neural networks can be divided, depending on how to propagate the information, into two classes: feed-forward and feedback networks [54][55].

#### 6.1.2.1 Feed-forward Networks

These networks allow direct signal propagation in one direction: from input to output. In this category, there is no feedback loop meaning the output of each layer is not affected by a previous value of it. They can be divided into a network (as shown in Figure 57) or multiple levels (which connects in series many networks of Figure 57 to create it) [54][56].

#### 6.1.2.2 Feedback Networks

The feedback network, as shown in Figure 58, can spread signals in both directions. It introduces feedback loops in the system. They are very powerful and can become very complicated. The special feature is their dynamic behavior. Their situation is alternating repeatedly until they converge to an equilibrium point. They remain in balance until changes are made at the entrance and a new balance to be found. Therefore, the opportunity that is given through these networks is the simulation of time-varying patterns.

### 6.1.3 Training

All the training methods that can be applied to ANN can be split in two classes [53][56]:

1. *supervised learning*: each output unit is trained with the desired response based on a specific input acting as an external teacher and providing continuous evaluation of the solutions' quality. Examples of this category are stochastic learning, reinforcement learning and error-correction learning [58].

2. *unsupervised learning*: there is no external teacher since it self-arranges the local information of the network and detects emergent collective properties. For this reason, it is also known as self-organization. Examples of this category are competitive learning and Hebbian learning [54][55].

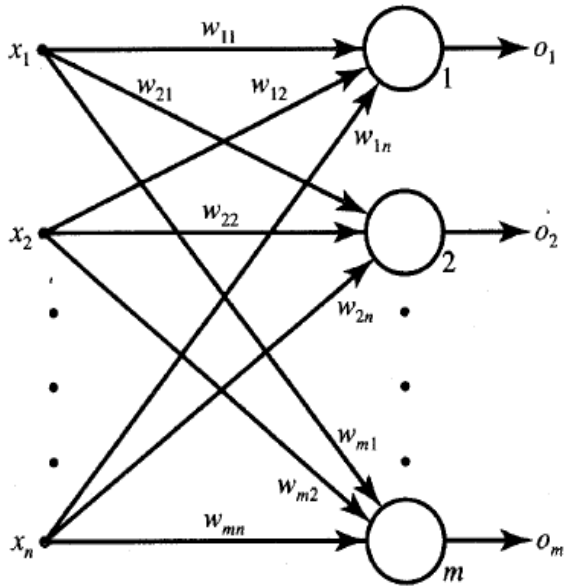
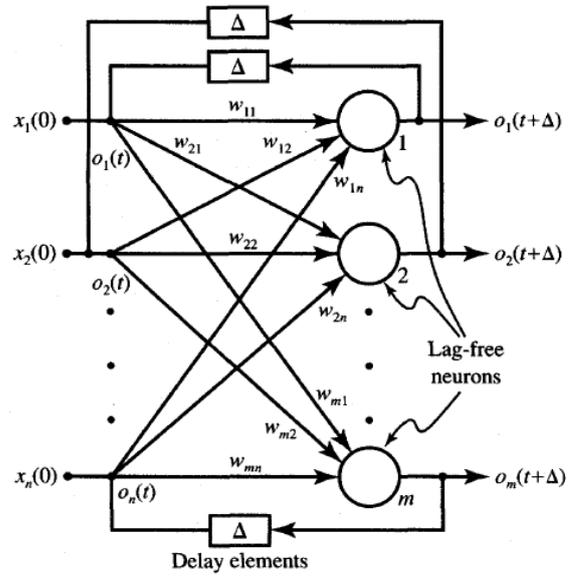

Figure 57: Feed-forward Network.   Figure 58: Feedback Network.

## 6.2 k - Nearest Neighbor (k-NN)

k-NN needs the data to be separated into training and testing set. For each row of the test set, the k nearest, based on Euclidean distance, provides the classification of it based on the training data that are superior in number in that area.

It is a nonparametric method which means that no assumptions are made about the distribution of data. This is quite important, since the majority of the real-life data does not follow completely theoretical assumptions.

At the same time, it could be described as 'lazy' learning since it does not use the train set to make any generalization. Thus, the training stage is quick enough, and the lack of generalization means that the k-NN reserves all the train set. More specifically, all the train set is necessary throughout the testing stage. This comes in contrast to other algorithms, such as the SVM, which can easily reject all data that are not support vectors.

Hence, there is a minimal training stage, whereas the testing part is computationally intensive and memory-hungry, since all data participate in the classification decision [59].

### 6.2.1 k-NN Algorithm

The training set incorporates feature vectors of a multidimensional space that are labeled under a specific class. During the training stage, the features vectors and class labels are stored.

During the classification stage, the constant k is defined and an unspecified sample or otherwise a test vector is classified by determining the class based on the most frequently occurring category of training samples among the k samples that are closest to the point.

A common method of distance measurement is Euclidean. Furthermore, the distance measurement can be improved significantly if trained with special algorithms, such as nearest neighbor of large margin [60].

**k-NN Algorithm** [61]

Step 1    Specify a positive integer k with defining a new sample.

Step 2    Choose k records in the dataset that are nearest the new sample.

Step 3    Using the most common class of these records to provide it to this sample.

## 6.3 Support Vector Machines (SVM)

The initial algorithm of SVM was developed by Vladimir Vapnik. He also suggested the soft margin modification in collaboration with Corinna Cortes. SVM builds a model during its training phase, using the marked labels of the training samples, in order to be able to predict in which class a testing sample belongs. It is a non-probabilistic binary linear classifier using an N-dimensional hyperplane to split in an optimal way the data into two classes [8][62]. There is strong connection between SVM and ANN since there is an equivalence between an SVM model with the use of a sigmoid kernel function and a perceptron neural network of two layers. In the SVM terminology, an attribute is a predictor variable and a transformed attribute is called a feature, while the set of features is a vector. Thus, the main objective of SVM is to generate the hyperplane which can optimally separate the two classes are completely separated and located on the two sides that are created. Support vectors are called the vectors that are in close proximity to the hyperplane [62].

### 6.3.1 Function

It is assumed that the data is linearly separable. Thus, there is a line on the $x_1, x_2$ graph that is able to isolate the two classes ($D = 2$) or a hyperplane on $x_1, x_2, \dots, x_D$ graphs ($D > 2$). The key element in designing an SVM is the concept of margin. Considering the linear classifier: $w^T x + w_o = 0$, the points closer to the separating hyperplane, known as support vectors, are in the space between the plane $H_1$ and $H_2$ which is $w^T x + w_o = d$ and $w^T x + w_o = -d$, respectively, as depicted in Figure 59. This interval is called margin. Defined as $d_1$ and $d_2$ the distance from the plane $H_1$ and $H_2$ respectively and must be equal if we want the same distance to the two levels of the level of the classifier, i.e. $d_1 = d_2 = d$. The Euclidean distance between any point on one of the two parallel hyperplanes and the hyperplane classifier which is equal to $d/\|w\|$, where $\|\cdot\|$ is the Euclidean norm. The margin should be maximized to have the hyperplane as possible far away the support vectors [62].

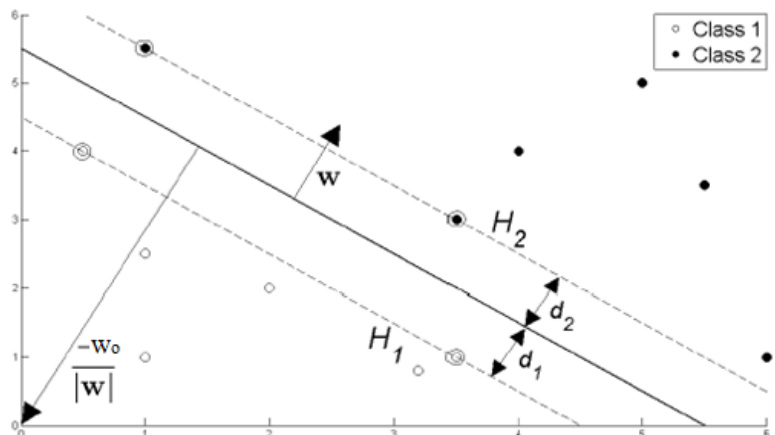

Figure 59: Hyper plane through two linearly separable classes.

Moreover, if there are $L$ number of training points, where are $D$ features for each $x_i$ input and belongs to one of the two classes $y_i$ = - $d$ ή $d$, then the form of training data is $\{x_i, y_i\}$, where $i = 1, 2 \dots, L$, $y_i \in \{-d, d\}$, $x \in \mathbb{R}^D$. Thus, if $w^T x + w_o \geq d$, $x$ belongs to $\omega_1$ category but if $w^T x + w_o \leq -d$ belongs to $\omega_2$ category. The above equations are summarized into the following equation:

$$y_i(w^T x + w_o) - d \geq 0 \; for \; y_i = \pm 1, for \; each \; i.$$

The issue of maximizing the margin can be solved by maximizing $d/\|w\|$ and if $d = 1$ then $1/\|w\|$. In a few words, the goal is to minimize the $\|w\|$ so $y_i(w^T x_i + w_0) - 1 \geq 0 \; for \; y_i = \pm 1, \forall i$.

The minimization of $\|w\|$ is equivalent to the minimization of $1/2 \|w\|^2$ (1/2 is used for mathematical convenience), which assists in performing Quadratic Programming (QP) optimization afterwards. The Lagrange multipliers constraints $\alpha$, where $a_i \geq 0 \; \forall i$, are assigned:

$$L_P = \frac{1}{2}\|w\|^2 - a[y_i(w^T x_i + w_o) - 1 \forall i] = \frac{1}{2}\|w\|^2 - \sum_{i=1}^{L} a_i[y_i(w^T x_i + w_o) - 1]$$

$$L_P = \frac{1}{2}\|w\|^2 - \sum_{i=1}^{L} a_i y_i(w^T x_i + w_o) + \sum_{i=1}^{L} a_i$$

$L_P$ is the *Primal* form. It should be differentiated with respect to $w$ and $w_o$ and set to zero. Then, the $w$ and $w_o$ should be found in order to minimize the derivative and $\alpha$ that maximizes it. This forms the following equation, which is dependent on $\alpha$, that should be maximized:

$$L_D = \sum_{i=1}^{L} a_i - \frac{1}{2} \sum_{i=1}^{L} a_i a_j y_i y_j x_i \cdot x_j \text{ and } a_i \geq 0 \; \forall i, \sum_{i=1}^{L} a_i y_i = 0$$

$$L_D = \sum_{i=1}^{L} a_i - \frac{1}{2} \sum_{i,j} a_i H_{ij} a_j \text{ where } H_{ij} = y_i y_j x_i \cdot x_j$$

$$L_D = \sum_{i=1}^{L} a_i - \frac{1}{2} \alpha^T H \alpha \text{ and } a_i \geq 0 \forall i, \sum_{i=1}^{L} a_i y_i = 0$$

$L_D$ is known as the *Dual* form of the *Primary* $L_P$. Therefore, this problem of quadratic optimization can be resolved by using a QP solver since the problem was moved from minimizing $L_P$ to maximizing $L_D$. When the variables $w$ and $w_o$ are extracted, the Support Vector Machine is derived [62].

### 6.3.2 Extensions

#### 6.3.2.1 Soft Margin

The soft margin is a modification of the maximum margin idea that allows some examples to be mislabeled. In the case where there is no hyperplane to split the examples in the two classes, the Soft Margin method will find a hyperplane that can divide them as optimally is possible allowing some misclassification. This method still searches for the maximum distance among the hyperplane and the nearest cleanly split examples. Soft Margin introduces a new type of variables, slack $\xi_i$, for evaluating the degree of misclassification of the $x_i$:

$$y_i(w^T x_i + w_o) \geq 1 - \xi_i \text{ για } 1 \leq i \leq n$$

There is a function for penalizing non-zero $\xi_i$ by increasing the objective function. The optimization transforms into a trade-off among a small penalty error and a large margin.

#### 6.3.2.2 Nonlinear classification and kernel function

The initial algorithm was developed as a linear classifier in 1963. In 1992, Vapnik in collaboration with Bernhard Boser and Isabelle Guyon proposed a modification of the algorithm by applying the kernel trick, which transforms it into non-linear classifier. The only difference is the substitution of every dot product with a nonlinear kernel function, which make the algorithm capable of fitting the hyperplane of maximum margin in the new feature space after the transformation.

When SVM is applied to linearly separable data, the dot product of the input variables can create the $H$ matrix:

$$H_{ij} = y_i y_j k(x_i, x_j) = x_i^T x_j$$

where $k(x_i, x_j)$ is a function from the family of *Kernel Functions*, where $k(x_i, x_j) = x_i^T x_j$ is the Linear Kernel.

The kernel functions basis is the computation of the inner products of two vectors. In the case where a function can be transformed into a space of higher dimensionality using mapping function of non-linear feature $x \to \varphi(x)$ determining the mapped inputs' inner products in the feature space, there is no need for calculating $\varphi$ explicitly. Thus, the *Kernel Trick* makes SVM useful even in multiple regression/classification problems, where inputs $x$ are not linearly regressable/separable in the initial space. Thus, it provides a suitable mapping $x \to \varphi(x)$ into a higher dimensionality feature space.

Some examples of kernel functions are the following:

- Polynomial (homogeneous): $k(x_i, x_j) = (x_i \cdot x_j)^d$.
- Polynomial (inhomogeneous): $k(x_i, x_j) = (x_i \cdot x_j + 1)^d$.
- Radial Basis Function (RBF) or Gaussian: $k(x_i, x_j) = \exp(-\gamma \|x_i - x_j\|^2)$, for $\gamma > 0$, which is ussually parametrized by $\gamma = 1/2\sigma^2$.
- Hyperbolic tagent: $k(x_i, x_j) = \tanh(k \cdot x_i \cdot x_j + c)$ for some (not every) $x > 0$ and $c < 0$.

### 6.3.2.3 The sequential minimal optimization (SMO) algorithm (Li He)

The SMO algorithm solves the dual problem emerging from the SVM derivation in an efficient way. The problem of dual optimization is:

$$max_a W(a) = \sum_{i=1}^{m} a_i - \frac{1}{2} \sum_{i,j=1}^{m} y^{(i)} y^{(j)} a_i a_i \langle x^{(i)}, x^{(j)} \rangle \text{ and } 0 \leq a_i \leq C, i = 1, \ldots, m$$

$$\sum_{i=1}^{m} a_i y^{(i)} = 0$$

If there is set of $a_i$'s to meet the constraints, the $a_2, \ldots, a_m$ will be held fixed while the objective will be reoptimized with respect to $a_1$. The existence of the second constraint means that $a_1$ cannot be modified since the remaining $a_i$'s are fixed. Hence, there should be an update of two $a_i$'s at the same time for meeting the constraints. So, the SMO algorithm follows the steps below:

- A pair of $\{a_i, a_j\}$ should be selected for updating (this pair should ideally provide the fastest step in the direction of global maximum).
- Optimize again the $W(a)$ with respect to $a_i$ and $a_j$ with all the remaining $a_k$ fixed.
- All the previous steps should be repeated until convergence.

The algorithm convergence can be tested if the Karush–Kuhn–Tucker (KKT) conditions are met to within a convergence tolerance parameter ($tol$), which ranges from 0.01 to 0.001. The efficiency of SMO algorithm lies in the efficient computation of the $a_i$, $a_j$ update [62].

### 6.3.2.4 Multiclass SVM

Multiclass SVM has an objective to allocate labels to examples with the SVM usage. The widely used method is by splitting the single multiclass problem into multiple binary classification problems [63]. Some of the methods are:
- binary classifiers, which can be split into two cases:
  1. one-versus-all: The new examples are classified by using the winner-takes-all strategy. In this case, the class is assigned to the classifier with the highest output function assuming that they are calibrated to provide comparable scores.
  2. one-versus-one: The new examples are classified via the max-wins voting strategy. They are assigned to the class that has accumulated the most votes after multiple binary classifications, where the classifier should select between two classes and give a vote to the selected one.

- error-correcting output codes.
- Directed Acyclic Graph SVM (DAGSVM).

## 6.4 Parzen Probabilistic Neural Network

The probabilistic neural net (PNN) has its base on Bayesian classification theory and the estimation of probability density function (PDF). PNN was introduced by Donald F. Specht in 1990. It is an effective algorithm for multiple classification problems due to the simplicity in its training and the complete statistical foundation in Bayesian estimation theory. If we replace sigmoid activation function with an exponential, the PNN will be able to compute nonlinear decision boundaries approaching the Bayes optimal [64].

PNN is associated with Parzen window pdf estimator. It contains multiple sub-networks, each of which is a pdf estimator of a Parzen window for each class. The Parzen windows classifier is the base for the probabilistic neural network development. It is a non-parametric method that generates a pdf estimate by multiple windows superposition. The Parzen windows classifier needs to calculate the pdf of each class based on the training examples and, then, it provides the classification decision. The classifier of multiple classes is expressed as follows:

$$p_k \cdot f_k > p_j \cdot f_j$$

when $p_k$ is the prior probability for an example from $k$ class to occur and $f_k$ is the estimated probability density function of class k. The term "neural network" originates from its structure (two-layer feedforward network).

### 6.4.1 Structure

There are three major layers in a PNN, the input, the pattern and the output layer, which is depicted in Figure 60. The first layer shows the input level of $d$ features. It does not perform calculations, but simply allocates the inputs to neurons in the fully connected pattern layer. The number of pattern layer nodes is equivalent to the number of training examples. At the pattern layer, there are aggregating results units from the use of the exponential function, which correspond to the category from which the model of education was chosen. Finally, the output level has $c$ nodes, one for each category.

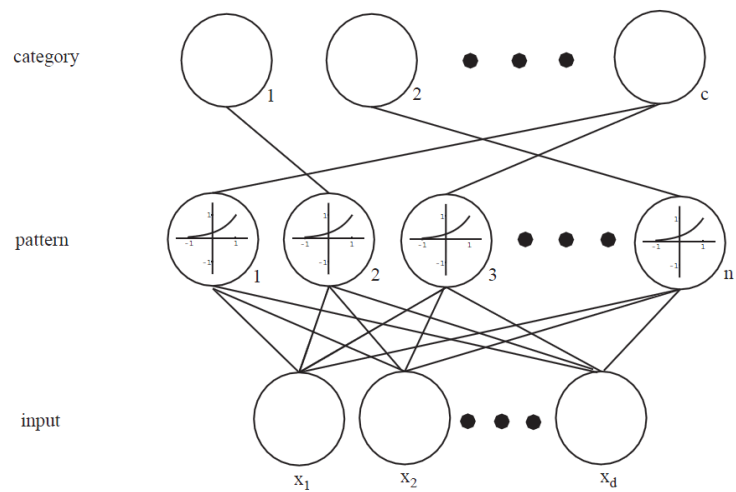

Figure 60: PNN structure.

### 6.4.2 Function

The network needs a parameter h that indicates the width of the window. Any given vector $x$ should be normalized using $\sum_{i=1}^{d} x_i^2$. This applies to data belonging both in training and testing samples. Below, the training and testing PNN algorithm are explained.

**Training PNN Algorithm** [65]

Step 1   Each object $x$ that belongs to the train set is normalized to 1.

Step 2   The first training pattern is set on the input units of PNN.

Step 3   The weights connected with the input and the first pattern units are set to: $w_1 = x_1$.

Step 4   A sole connection should be made from the first pattern unit to the category unit that corresponds to the class of that pattern.

Step 5   The process is repeated for all the other training objects with the weights set to: $w_k = x_k\ (k = 1, 2, \ldots, n)$.

Afterwards, the trained network will be used for classification as described below. A normalized sample $x$ is placed in the input plane. Each unit calculates the standard inner product $net_k = w_k^T x$ which is input to the non-linear (exponential) function $f(net_k) = \exp[(net_k - 1)/\sigma^2]$, where $\sigma$ is a parameter specified by the user and equal to $\sqrt{2}$ times the length of the applied Gaussian window. In order to understand the choice of non-linearity, it is sufficient to consider a non-normalized Gaussian window in the center position of a trainee patterns $w_k$. Working backwards, in order to achieve the desired function of the Gaussian window, a nonlinear transfer function should be inferred that can be used from the pattern units. This equation, if $h_n$ is constant, is:

$$\varphi\left(\frac{x_k - w_k}{h_n}\right) \propto e^{-(x-w_k)^T(x-w_k)/2\sigma^2} = e^{-(x^T x + w_k^T w_k - 2x^T w_k)/2\sigma^2} = e^{-(z_k - 1)/2\sigma^2}$$

where $x^T x = w_k^T w_k = 1$. The sum of these local estimations (calculated in the same category) gives the function $P_n(x|\omega_j)$, where the Parzen window calculates the dispersion. Afterwards, the maximum value is selected, which provides the desired category for this example.

**Testing PNN Algorithm** [65]

Step 1   The test pattern x should be normalized and placed at the input units.

Step 2   The inner product is computed for each pattern unit to produce the net activation: $net_k = w_k^T * x$, and emit a nonlinear function:

$$f(net_k) = \exp[(net_k - 1)/\sigma^2]$$

Step 3   Each output unit sums the outputs from all pattern units linked to it:

$$P_n(x|\omega_j) = \sum_{i=1}^{n} \varphi_i \propto P_n(\omega_j|x)$$

Step 4   The classification is conducted based on the maximum value of $P_n(x|\omega_j)\ (j = 1, \ldots, c)$

# Chapter 7: Applications using the EMG signal

The use of the EMG signal is quite prevalent in multiple applications. Some of the applications are controlling exoskeleton parts, playing computer games and aiding the elderly. Therefore, except from the research part, there are practical applications for the society in the above three categories, which occupy an important proportion of the world's economy. Apart from the issue of social responsibility of research, there is entrepreneurship that develops enough profit. This leads to continuous improvement of the systems that benefit from advancements in bio-sensors, pattern recognition in biomedical signal processing and embedded systems [2][3].

## 7.1 Applications in controlling exoskeleton parts

One of the practical applications of the EMG signal is the use for control exoskeleton limps of body, such as hand. The amputation of the patient is not always the same and over time the signal that is provided by the forearm decreases in intensity. This happens because the muscles and nerves disuse since there is no reason for a move and the brain is not activated to perform a hand grasping. Therefore, the use of exoskeleton parts should be used immediately as after a period of inaction, maybe less than a year, the resulting signal is very weak.

There have been several approaches to create an autonomous system that includes the robotic hand and the autonomous EMG system. The first successful application in the market is i-Limp of Touch Bionics [66]. This hand replacement looks and acts like a real human hand and represents a significant advance in bionics in patients' care.

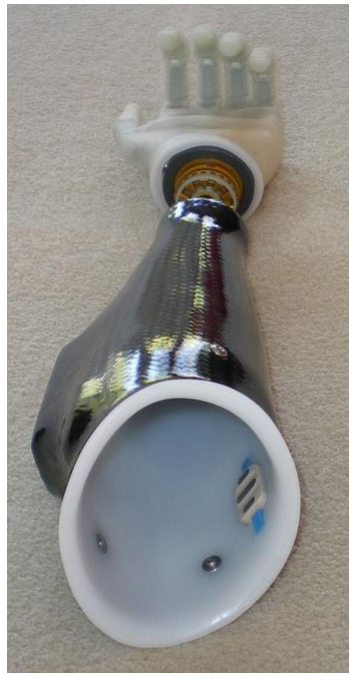

Figure 61: Panoramic view of i-Limp.

The hand controlled from a single, highly intelligent control system that uses two sensors for opening and closure of the hand with the battery included as a bulge in the left side, as can be observed in Figure 61. Figure 62 shows the i-Limp's parts.

The EMG signal of the patient is generated by the remaining portion of the arm muscles. The electrodes are placed on the skin surface having a specific location. Existing system users can adapt fast on system by using their data to control it within minutes [67].

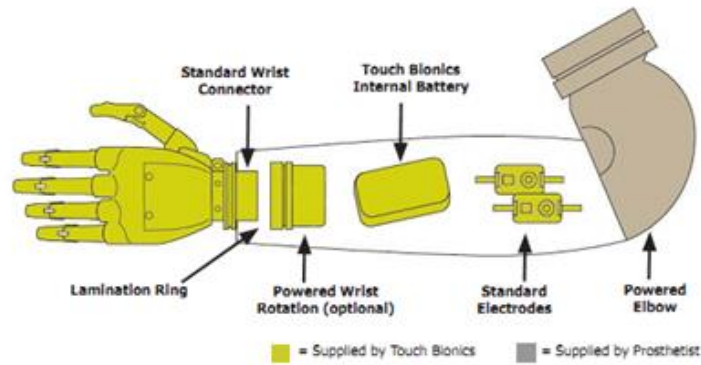

Figure 62: Parts of i-Limp.

It should be noticed that this thesis was triggered by the construction of the robotic hand from the PhD student Mr. K. Andrianesis. The main feature of the anthropomorphic robot hand is made of light weight biomimetic actuator and more specifically Shape Memory Alloys. Making use of the advantages of these non-conventional actuators, a prosthetic device was developed that meets all the demanding requirements as outlined by the users of such devices in similar studies [1]. The result of this construction is the development of a completely silent prosthesis, as shown in Figure 63, with a very small volume and mass and increased dexterity able to perform all normal daily work of a man with upper limb disabilities. Therefore, due to the promising results of the system developed in this thesis, including some system improvements and experiments in amputees, it may be combined with the robotic hand in a future work to create an autonomous exoskeleton system.

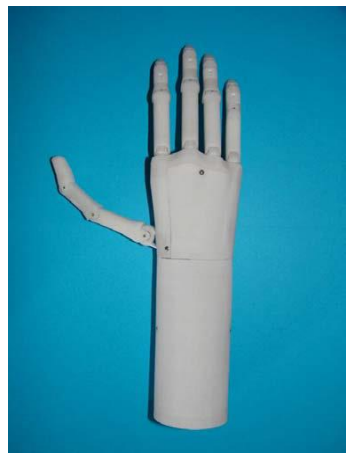

Figure 63: Construction of an innovative prosthetic hand of the University of Patras.

## 7.2 Applications in computer games

The field of entertainment and computer games is a developed industry especially in prosperous economies. The computer game companies, except from the improvement in the graphics and the game plot, try to improve the interaction between the user and the game. Therefore, they strive for different ways to change the control of the game, other than the conventional joystick, such are some simple body movement or even muscle construction. One option is the use of EMG signal because a simple movement of the hand with a muscle contraction can give a choice in a game. For this reason, the user can gain greater interaction in the environment of the computer game.

### 7.2.1 EMG ring of Microsoft and its use in Guitar Hero

One of the latest efforts is the creation of the EMG ring, as illustrated in Figure 64, which was patented in 2009 by Microsoft. It includes 10 electrodes and sends wirelessly signals to the computer via a novel low power wireless sensor module into muscle. They developed an application that allows the user to use the interface of muscle-computer to play the game of Guitar Hero. In Guitar Hero, the users have a controller like a guitar and push on buttons with both hands when the system presents stimuli associated with the music being played at that time [7].

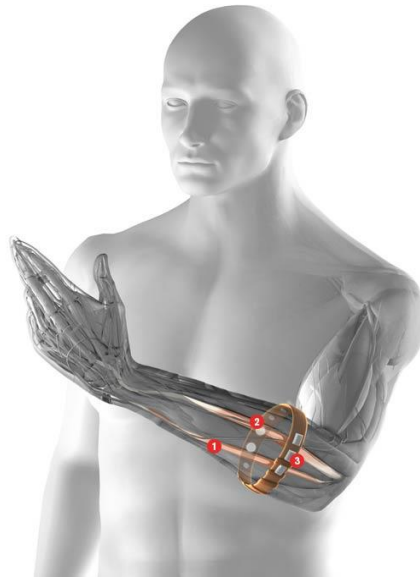

**Figure 64: (1) Electrical activity is generated by the gesture related muscles. (2) Activity is detected by electrodes placed on a ring. (3) Signals are transmitted wirelessly to a computer.**

### 7.2.2 Play Super Mario by using EMG signals

A simpler system was developed using the same sensors of those used in the present thesis. It aims to control the popular game Super Mario, as seen in Figure 65. The control was achieved by putting in place of buttons the power of signals. When, one value just exceeded a certain threshold, the movements in the game were triggered with the same energy [68]. In short, it was corresponding with the push of a button. Of course, it is much more comfortable and much more interactive for the user to participate with the muscles in the game creating greater pleasure than to handle a conventional joystick.

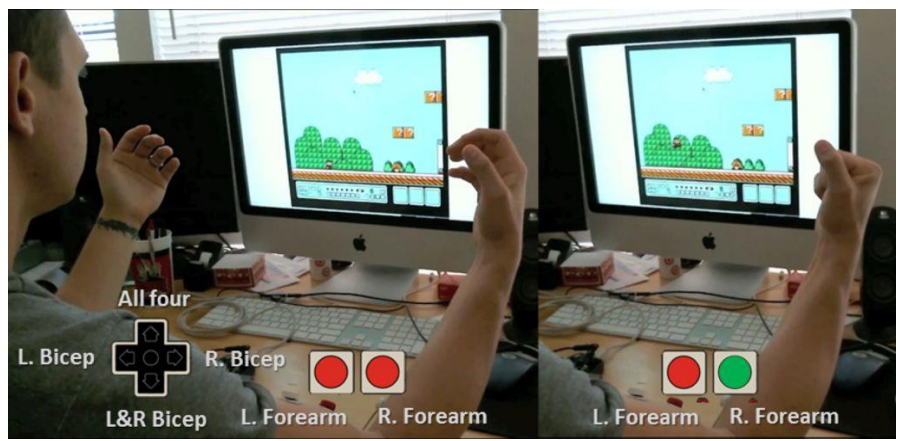

**Figure 65: Play Super Mario by using via EMG sensor.**

### 7.2.3 Checking grimace through EMG signal

Finally, apart from the use of the forearm for recording the EMG signal, the development of interesting EMG signal applications derived from the facial muscles has been observed. This technique can be used more by designers of computer games as a way of feedback from the user to various points of the game in order to receive feelings from grimacing facial. In this way, they are constantly improving their applications and improve a stage of the game that can cause repeated dissatisfaction by many users who perform similar grimace [69].

## 7.3 Applications in elderly

Applications for the elderly could not be excluded from the field of survey using the EMG signal. In most cases, it is used for rehabilitation reasons after an illness.

### 7.3.1 Restoration of muscle function after stroke

Recovery from dysfunctions of the human body, such as stroke, which usually affects the elderly is necessary and painful. Therefore, the treatment for muscle paralysis as part of stroke rehabilitation can be achieved with advanced devices, such as the Biomove's devices [70]. It is common after a stroke for the control signals to be unable to reach muscles located in the limbs, such as foot or hand. Thus, the electrical activity is reduced in these muscles leading them to contraction failure, which deteriorates through the time.

The system designed by Biomove uses three electrodes, which are placed either on the leg or forearm. The setup amplifies any EMG signal activity that can be generated in the body part muscles, where the electrodes are placed. Afterwards, the system encompasses an electrical muscle stimulator along with biofeedback, which assist in teaching muscle contraction again. The muscles are strengthened more, when they contract regularly.

### 7.3.2 Balance control of elderly women to prevent consequences of osteoporosis

Patients with low bone mass have high probability of a fracture caused even by a small wound. The greatest result of osteoporosis is a femur fracture, which is linked with death and morbidity. A way of avoiding falls for women suffering from osteoporosis is systematic physical training. This can also assist in treatment of other diseases and address the declining forces of the elderly. There are multiple efforts that have conveyed enhancement of balance motor skills, endurance, strength, flexibility and enhancing the density of the bones in older females dealing with physical trainings. The approach mentioned in [71] has used the EMG signal to control the behavior of balance function and the risk of falls in elderly women with osteoporosis by educating a vibration balancer shaft.

# Chapter 8: Hardware Description

In this thesis, two systems are used for recording measurements. A Delsys' EMG system and an implemented autonomous EMG are used in the first and second part, respectively. The following sections will analyze the parts constituting each system.

## 8.1 Delsys' EMG system

The handheld Bagnoli™ electromyography system of Delsys [72], as observed in Figure 66, contains the main amplifier unit, a card for converting analog signal to digital and two surface electrodes. It is made for the sole purpose of research and not for any usage in the diagnosis and treatment of humans.

Figure 66: Delsys' EMG system.

### 8.1.1 Main Amplifier Unit

The main amplifier unit, as depicted in Figure 67, operates with a battery of 9V, has two channels of EMG and seems useful in terms of portability and space. Active sensors are specifically designed for optimal recording of EMG signal on the skin surface, while by rejecting the common noise signals, they yield an excellent signal-over noise. By an appropriate selection switch, they select a gain between 100, 1,000 or 10,000, to obtain signals of different widths. Each output channel is isolated up to the 3.75 kV (RMS) and satisfies the medical requirements for CE marking and 510K pre-approval. There is an LED that indicates if the system is turned on. The specifications of the main amplifier unit are summarized in Figure 68.

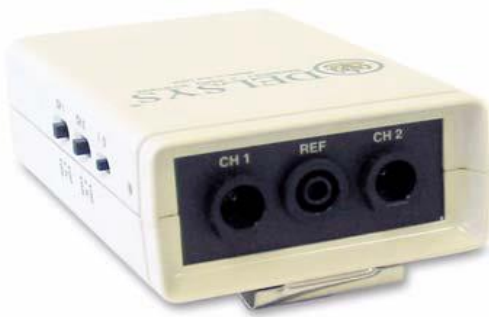

| | |
|---|---|
| Number of Channels | 2 Analog EMG |
| Overall Amplification per Channel | 100, 1000, 10000 |
| Max. Output Voltage Range | ± 5 Volts |
| Channel Frequency Response | 20±5 Hz to 450±50 Hz, 80 dB/decade |
| EMG Sensors | DE-2.1 (single differential) or DE-3.1 (double differential) |
| System Noise (R.T.I.) | <1.2 μV(rms) for the specified bandwidth |
| Power Requirements | 9 VDC, 10mA (quiescent) |
| Channel Output Isolation | 3750 V(rms) @ 60 Hz for 60 sec. |
| Operating Temperature | 15°C to 40°C |
| Case Dimensions | 114 mm x 71 mm x 33 mm |
| Weight | 0.2 kg |

Figure 67: Main Amplification Unit.    Figure 68: Specifications of the main amplification unit.

### 8.1.2 NI USB-6009 DAQ using it as A/D converter

The system NI USB-6009 includes a USB microcontroller, an 8-channel ADC of 14bits and two DACs, which can be explained from Figure 69.

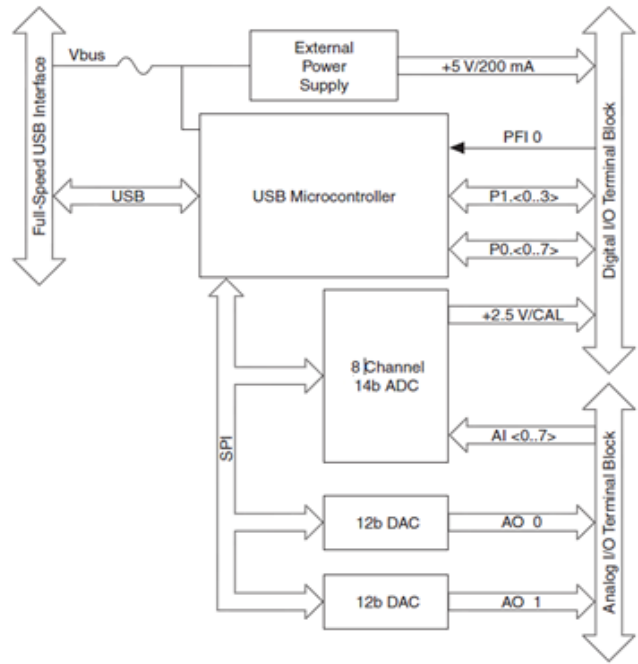

Figure 69: Block diagram of NI USB-6009 DAQ device.

The NI USB-6009 system level diagram is shown in Figure 70. The eight analog channels can be used either as four differential or eight single analog inputs.

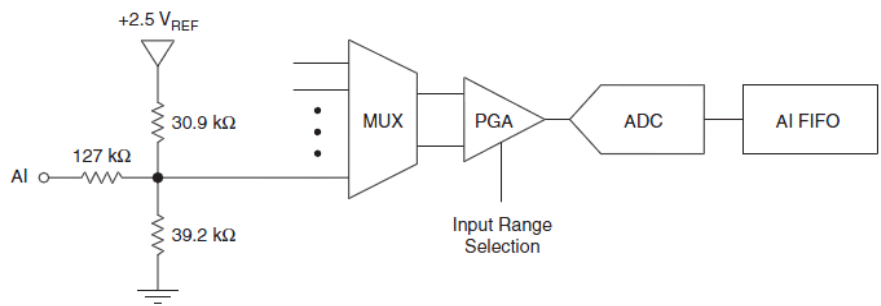

Figure 70: NI USB-6009 analog input circuit diagram.

The main units encompassed in the NI USB-6009 analog part are the following:

- *MUX*: The multiplexer allows one of the eight analog input channels each time since the unit is equipped with only one ADC.
- *PGA*: Stands for the programmable gain amplifier. If it is set in differential mode, it can provide multiple values of gain ranging from 1 to 20, whereas in single mode it can only provide a gain of 1. The gain adapts automatically to the voltage range of the application.
- *ADC*: It transforms the analog input into a digital value.
- *AI FIFO*: The buffer (first-in-first-out or FIFO) assists in capturing all the data acquired by single or multiple conversions.

The unit is equipped with two autonomous analog output channels, as illustrated in Figure 71. The output ranges from 0 to 5 V. There is a timed software control for the analog output updates. The GND stands for the ground signal for the analog outputs.

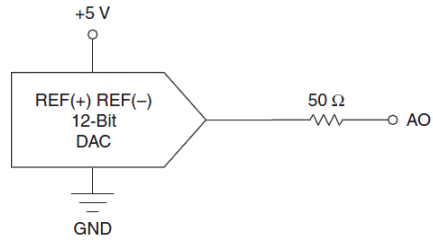

Figure 71: NI USB-6009 analog output circuit diagram.

A DAC is the main unit responsible for converting the digital value into an analog voltage in the NI USB-6009, as shown in Figure 72,. Each analog output line is equipped with a DAC. Figure 73 and Figure 74 show the interface cable and device specifications respectively.

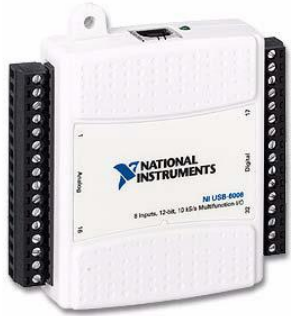

Figure 72: NI USB-6009 DAQ unit.

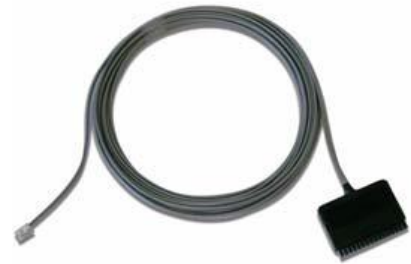

Figure 73: Output cable Bagnoli-2 USB-6009.

| Feature | NI USB-6009 |
|---|---|
| AI resolution | 14 bits differential, 13 bits single-ended |
| Maximum AI sample rate, single channel* | 48 kS/s |
| Maximum AI sample rate, multiple channels (aggregate)* | 48 kS/s |
| DIO configuration | Each channel individually programmable as open collector or active drive |

* System-dependent.

Figure 74: NI USB-6009 DAQ unit specifications.

### 8.1.3 DE-2.1 EMG sensor

The EMG Sensor (DE-2.1), as depicted in Figure 75, conducts subtraction of electromyographic potentials acquired at two different sites (skin surface of the muscle under examination). The potentials generated by the muscle activity are recorded along with a ground reference electrode, as shown in Figure 76, which is placed away from the bioelectric source.

The parallel-bar geometry of the contact sensor guarantees signal stability, repeatability among different acquisitions and optimal representation of the frequency content. This multipurpose sensor can be used in multiple EMG applications and multiple muscle sizes.

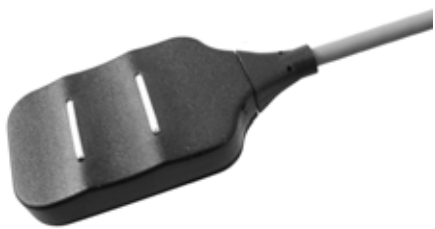

Figure 75: A Differential Surface EMG Sensor (DE-2.1).

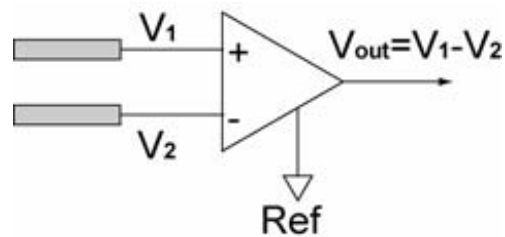

Figure 76: Amplifier Block.

The sensor is completely sealed by a sturdy polycarbonate. There is an internal shield for ambient noise rejection. The contacts bars use 99.9% pure silver and their dimensions are 1 cm (length) x 1 mm (diameter) and their distance is 1 cm for detecting the signal in an optimal way, as shown in Figure 77. There is also attention on promoting the skin contact and in parallel decreasing any adverse effect that sweat can cause by using a curved geometry between the bars. The usefulness of each sensor part and the electrical specifications are summed up in Figure 78 and Figure 79, respectively.

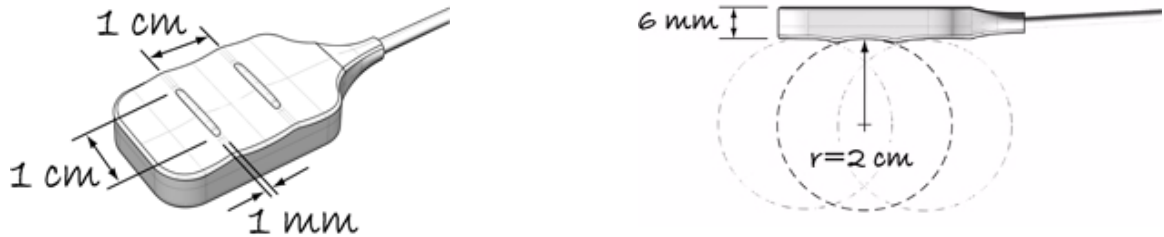

Figure 77: DE-2.1 EMG Sensor Geometry.

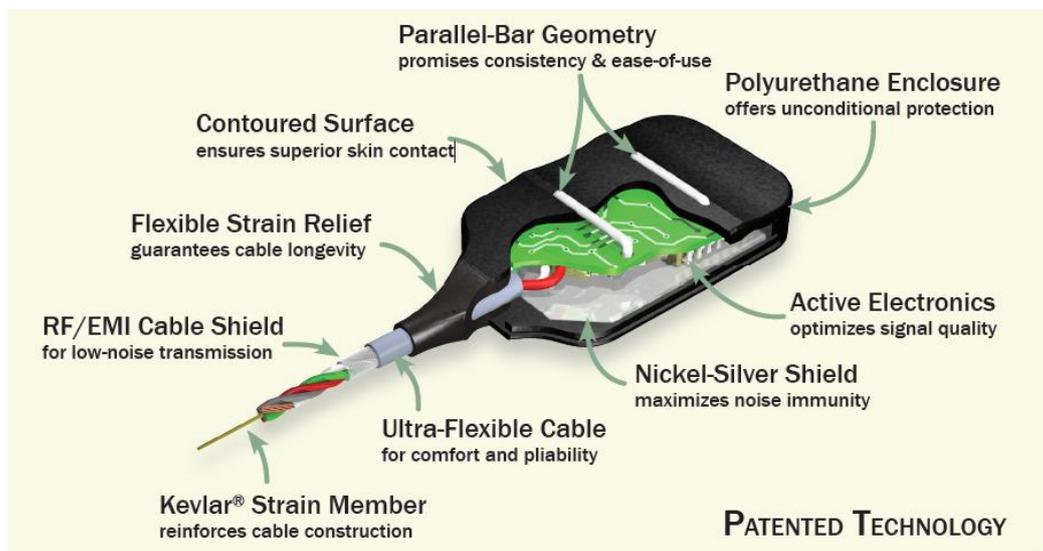

Figure 78: Usefulness of each part of the sensor.

| Preamplifier Gain | 10 V/V ±1% |
|---|---|
| Bandwidth | open |
| Noise | 1.2 μV (RMS, R.T.I.) |
| CMRR (0-500 Hz) | -92 dB (typical) |
| Power Consumption | 20 mW (typical) |
| Input Impedance | >$10^{15}$Ω // 0.2pF |

Figure 79: Electrical specifications of DE-2.1 EMG sensor.

## 8.2 Autonomous EMG system

The autonomous system electromyography needed the following parts in order to be constructed:

3 EMG sensors ([EMG Sensors v2](#))    Microcontroller [Arduino Uno](#)

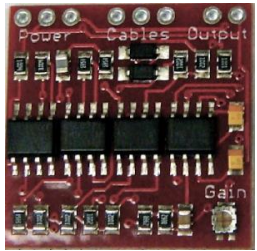    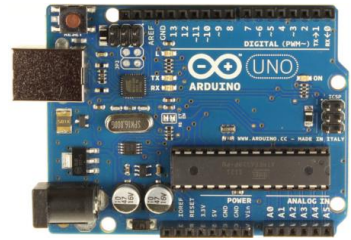

Cables of EMG sensors ([EMG Sensor Cables](#))    System Link Cables
and EMG electrodes ([EMG Electrodes](#))

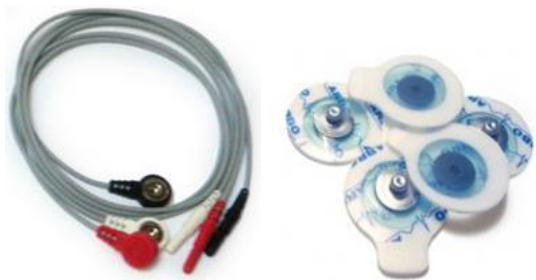    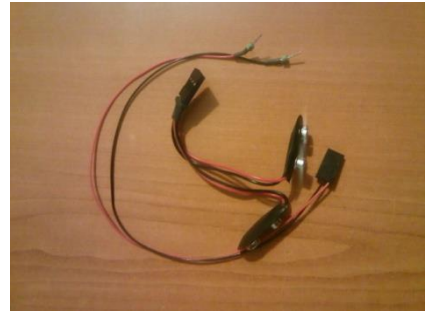

6 9Volt batteries, 2 for each sensor    Orthopedic glove LP
because ±$V_{cc}$ supply is needed.

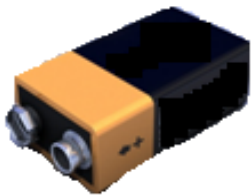    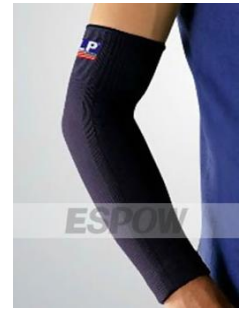

The total cost of construction was $ 135 for the sensors ($ 45 / sensor), $ 30 for Arduino Uno, $ 45 for cables of EMG sensor ($ 15 / triad cables), $ 15 for package of 50 electrodes and € 9 for the orthopedic glove. Therefore, the total cost is about $ 250 if we consider the cost for the 9V batteries, the 9V battery clips and the cables.

The output of each sensor SIG is connected to one of the analog pins A0-A5 and the GND pin into one of the ground pins of Arduino. Moreover, the sensor wires are connected to the central pin of the MID, END, REF. Despite their professional coating, the wires are needed to be cut at the edges, because they cannot be placed next to each other on the pins of the sensor. The power sensor is supplied by the batteries at pins Power (-Vs, GND, +Vs) through the clip of 9V battery, in which the positive part of one battery and the negative of the other are joined and connected to the GND and the negative of the first in -Vs and the other's positive in +Vs. For supplying energy to Arduino two ways can be employed, one via the USB cable via the computer and the other via 9V battery connected with special clip to the power connector. The second way helps to have autonomy of energy but first is necessary in the beginning to upload the code. The connectivity of the system is depicted in Figure 80 and the full implementation in Figure 81.

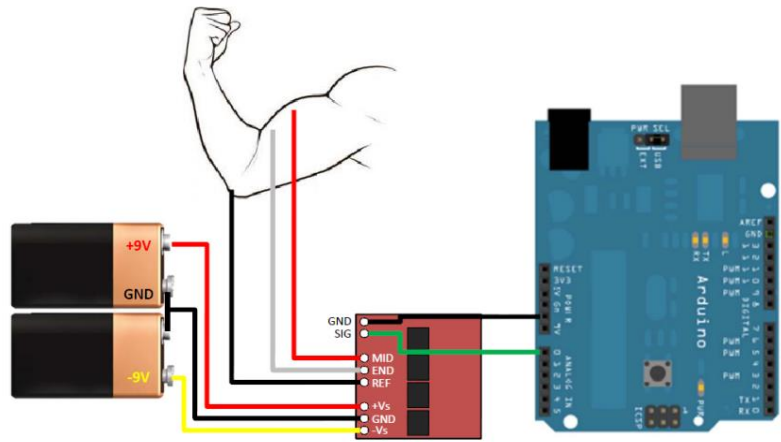

Figure 80: Schematic system interface.

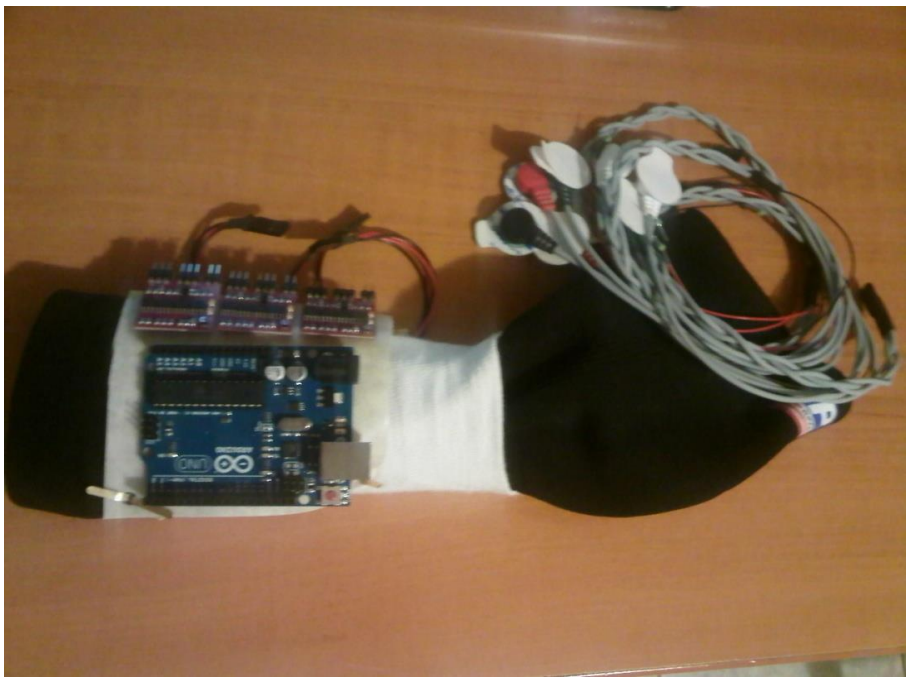

Figure 81: Autonomous EMG system.

### 8.2.1 EMG sensors

The EMG sensor [74], with the block diagram to be depicted in Figure 82, is a circuit in SMD board that contains the following elements:

Circuit Chips
- 3x General-Purpose Operational Amplifiers (TL072 IC)
- 1x Differential Amplifier of gain 10 (INA106 IC)

Capacitors
- 1 x 1.0 uF Ceramic Disk
- 1 x 0.01 uF Ceramic Disk
- 2 x 1.0 uF Tantalum

Resistors
- 1 x 1 kΩ 1%
- 1 x 100 kΩ Trimmer
- 6 x 10 kΩ 1%
- 2 x 80.6 kΩ 1%
- 2 x 1 MΩ 1%
- 3 x 150 kΩ 1%

Diode
- 2 x 1N4148

| Parameter | Min | TYP | Max |
|---|---|---|---|
| Power Supply Voltage (Vs) | ±5V | ±5V | ±18V |
| Gain Setting, Gain = 1650*(X /1 kΩ) | 0.01 Ω (0.0165x) | 20 kΩ (33000x) | 100 kΩ (165000x) |
| Output Signal Voltage (Rectified & Smoothed) | 0V | -- | +Vs |
| Differential Input Voltage | 0 mV | 2-5mV | 10 mV |

Figure 82: Specifications of EMG sensor.

The Figure 83 shows the schematic diagram of the circuit, where they can be distinguished all the elements mentioned above:

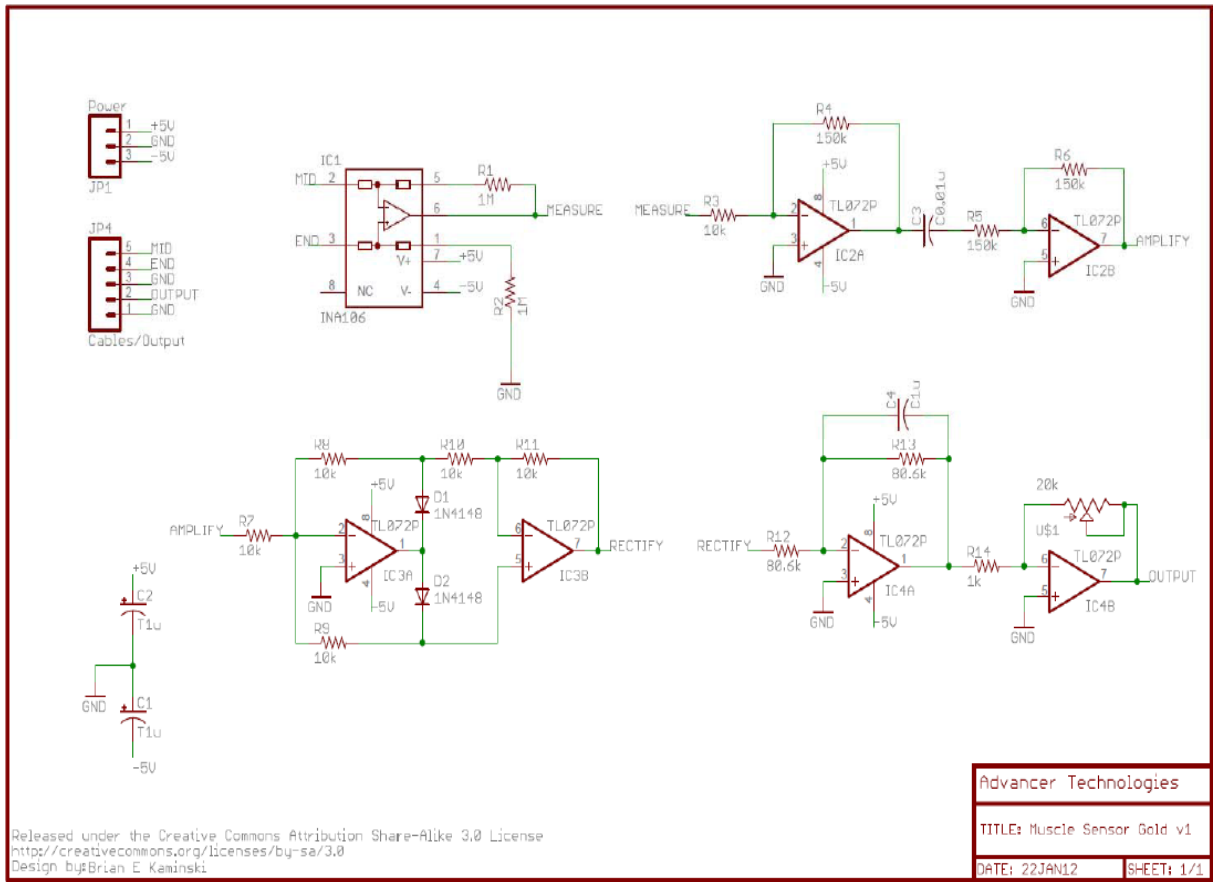

Figure 83: Schematic diagram of the EMG sensor circuit.

### 8.2.2 Arduino Uno Microcontroller

#### 8.2.2.1 Overview

Arduino is a microcontroller intended to be used in conjunction with electronic devices in interdisciplinary work making it more handy for the public. It uses an 8-bit microcontroller Atmel AVR ATmega328, with all the hardware being built around it. Figure 84 is shows the pin mapping with the Arduino. There are 6 analog inputs, 14 digital I/O pins, where 6 of them can also provide a PWM output and a 16 MHz ceramic resonator. It also includes an ICSP header, a power jack, a reset button and a USB connection. Its specifications are summarized in Figure 86.

The software part uses a standard compiler of «Wiring» programming language that is designed specifically for the Arduino, and a boot loader running on the microprocessor. Furthermore, there is "Processing" language, which operates on the host computer in conjunction with the used Arduino to add more features via computer, such as creating GUI [76].

### 8.2.2.2 Inputs and Outputs

There are 14 digital pins on the microcontroller operating at 5V, which can be used either as input or as output by the Wiring functions: digitalWrite, digitalRead, and pinMode. The maximum current that can be provided or received is 40 mA. Each pin is equipped with an internal pull-up resistor of 20-50 kΩ, which in not connected by default. Except their generic operation, some of them have a more specific operation as described below [76]:

- *External Interrupts*: The pins 2 and 3 can be used to generate an interrupt by a low value, a clock event, or a value change.
- *Serial*: The pins 0 (RX) and 1 (TX) are used for receiving (RX) and transmitting (TX) TTL serial data. They are wired with the corresponding pins of the ATmega8U2 USB-to-TTL Serial chip.
- *SPI*: The pins 10 (SS), 11 (MOSI), 12 (MISO) and 13 (SCK) are set up for SPI communication using the SPI library.
- *PWM*: The pins 3, 5, 6, 9, 10, and 11 provide 8-bit PWM output by using the analogWrite function.
- *LED*: The digital pin 13 is connected an onboard LED, which turns ON and OFF when the pin is HIGH and LOW, respectively.

Additionally, there are 6 analog inputs (A0 - A5) with a 10 bit-resolution ($2^{10}$=1024 values). Furthermore, the following pins have a particular functionality:

- *AREF*: It is the reference voltage for the analog inputs and is used with analogReference.
- *TWI*: A4 or SDA pin and A5 or SCL pin. TWI communication needs the use of the Wire library.
- *Reset:* It resets the microcontroller when its LOW.

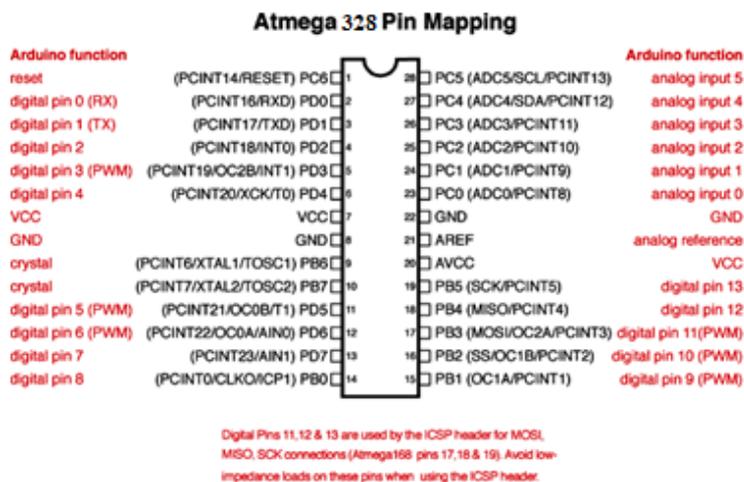
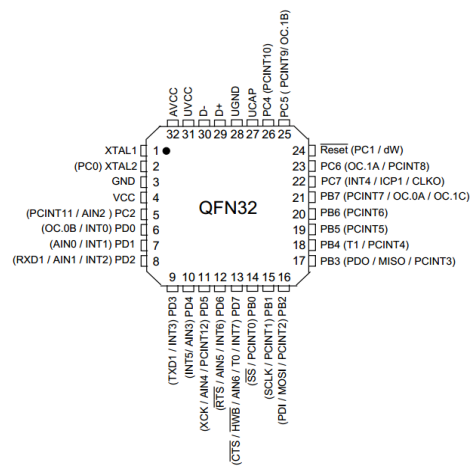

Figure 84: Connection of Arduino and ATmega328 Pins.    Figure 85: ATmega16U2 Pins.

### 8.2.2.3 Memory

The ATmega328 includes a Flash memory of 32 KB (0.5 KB for the bootloader), a SRAM of 2 KB and an EEPROM of 1 KB [76].

### 8.2.2.4 Power

The board can be powered either via an external power supply or the USB connection, which is automatically selected. The power supply can be either a battery or an AC-to-DC adapter. Arduino is equipped with a power jack receiving a 2.1mm center-positive plug as input. The battery leads can also use the two board pins, VIN and GND from the POWER connector.

The external supply can vary from 6-20 V. The appropriate operation region is 7 to 12 V [76]. This is because of two reasons. In the case where $V_{dd}$ < 7V, the 5V pin will supply less voltage leading to an unstable board. In the case where $V_{dd}$ > 12V, there is a chance that the voltage regulator will overheat leading to a possible board damage. The power pins are as follows:

- *VIN*: This pin operates as a power input voltage for the board from an external source. It can be used either to supply power or to read the voltage provided through the power jack.
- 5V: It provides a 5V output, which is provided by a regulator. The power supplied by the DC power jack (DC power) and the VIN pin can vary from 7 to 12V and 5V got the USB connector. It is not advisable to supply voltage via the 3.3V and 5V pins since the regulator is bypassed, which can lead to board damage.
- 3V3: This pin provides a regulated 3.3V with a 50 mA maximum draw of current.
- GND: Ground pin.
- IOREF: This pin provides a voltage reference based on which the microcontroller's operation is relied on. This pin is read by a shield, which selects the suitable power level or enables voltage translators for a power operation of 3.3V or 5V.

### 8.2.2.5 Communication

There are multiple ways of communication between the Arduino UNO with another Arduino or in general a microcontroller, a computer. Serial communication UART TTL (5V) is one of the ways through the 0 (RX) and 1 (TX) on digital pins. The USB serial communication is performed via the onboard ATmega16U2, which shows up to the computer as a virtual communication port. Standard USB COM drivers, without the need of any external, are used for the '16U2 firmware, as depicted in Figure 85. However, a .inf file is essential on Windows. A serial monitor is available for sending and receiving textual data via the Arduino. There are two LEDs, RX and TX, assisting the visual inspection of the data transmission through USB connection to the computer and USB-to-serial chip [76].

### 8.2.2.6 Physical Characteristics

The dimensions for the Arduino Uno PCB board are 2.7 in (length) x 2.1 in (width). The power jack and the USB connector extend these dimensions. There are four holes, one per corner, that assist in mounting the board. The distance among the digital pins 7 and 8 is 160 mil, which is not a 100 mil multiple as it happens for the other pins [76].

| Microcontroller | ATmega328 |
|---|---|
| Operating Voltage | 5V |
| Input Voltage (recommended) | 7-12V |
| Input Voltage (limits) | 6-20V |
| Digital I/O Pins | 14 (of which 6 provide PWM output) |
| Analog Input Pins | 6 |
| DC Current per I/O Pin | 40 mA |
| DC Current for 3.3V Pin | 50 mA |
| Flash Memory | 32 KB (ATmega328) of which 0.5 KB used by bootloader |
| SRAM | 2 KB (ATmega328) |
| EEPROM | 1 KB (ATmega328) |
| Clock Speed | 16 MHz |

Figure 86: Arduino Uno Specifications.

## Chapter 9: Experimental Procedure

Both systems analyzed in the previous chapter were used to perform experiments. The measurements are divided into two main parts. The Delsys' EMG system is used in the first part and the implemented one in the second part. The following sections will analyze the experiments and classification results of movements.

Initially, every part of the experimental process uses two basic movements, opening and closing the hand. The success rate is 100% with little information, using only one electrode and exporting two features, variation and WAMP, as shown on the vertical and horizontal axis of Figure 87 respectively. The vectors for the open and close hand posture are in red and blue region respectively. Vectors of each movement have similar signal characteristics that differ from the vectors of the other category and, thus, they are easily separable.

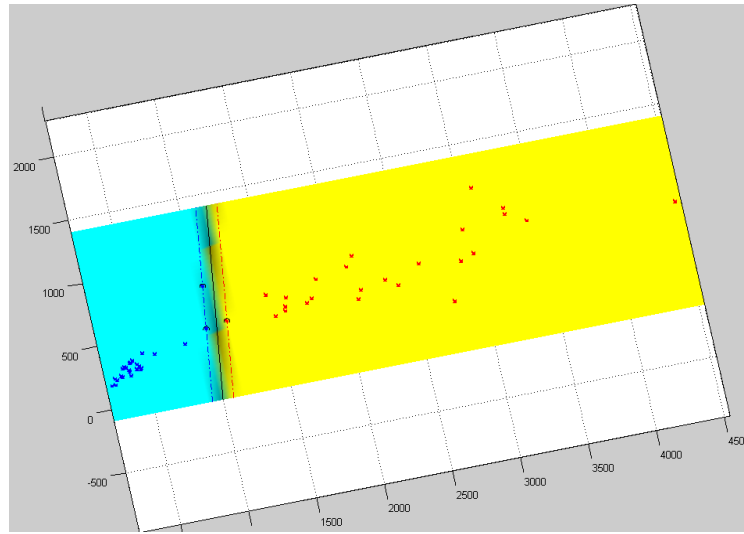

**Figure 87: Classification of movements, open and close hand posture, with the training algorithm of SVM, SMO.**

Afterwards, the previous two movements and the six-main grasping, as shown in Figure 88, are used in the experiments. There are movements in which the involved signal features cause issues in classification, as will be seen in the following experiments.

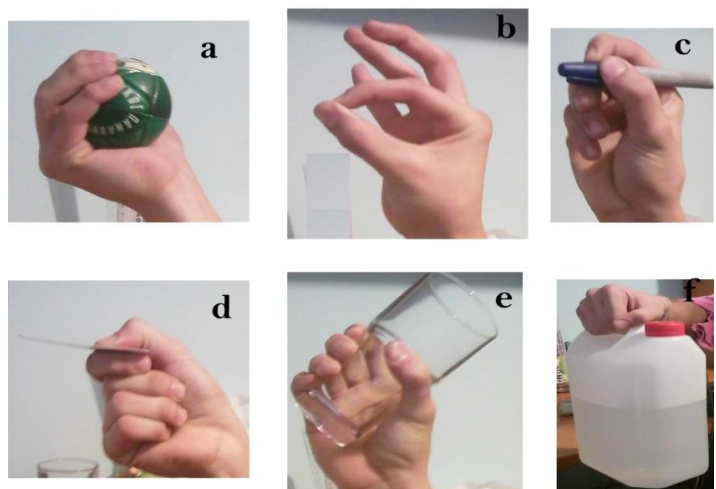

**Figure 88: The 6 basic hand movements.**

The basic hand movements are:
  a) Spherical (S): holding an elastic ball.
  b) Tip (T): the touch of the thumb to the index.
  c) Palmar (P): holding a pen.
  d) Lateral (L): holding a credit card.

e) Cylindrical (C): holding a glass of water.
f) Hook (H): holding the handle of 4L water bottle.

The electrodes after several experimentations are placed on: *Extensor carpi radialis* muscles *and Flexor carpi ulnaris*, as shown in Figure 89. Of course, this is not completely accurate because during the placement of the electrodes, it is not easy to be placed in an exact position.

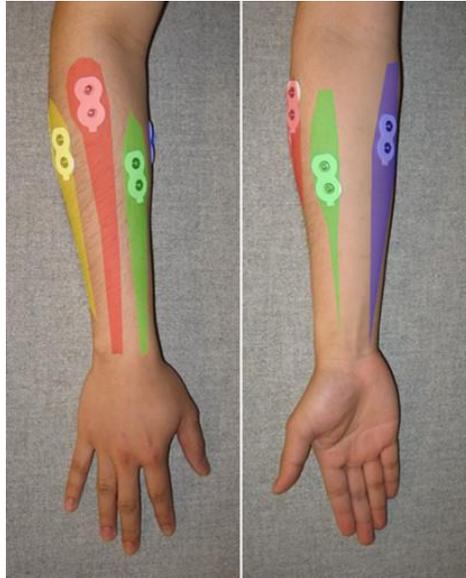

Figure 89: Flexor carpi ulnaris (red) and Extensor carpi radialis (blue) muscles.

Furthermore, the electrodes must be placed in the forearm with an elastic material and pressed in order not to lose the contact with forearm and the original position that have been set. The elbow should not be in exact contact with the desk due to the possible subject's fatigue. Thus, a spongy material should be placed between the elbow of the subject and desk.

## 9.1 Part A of experimental procedure: Use of Delsys' EMG system

In this part of the experimental procedure, the Delsys' EMG system is used for acquiring the signal through an interface created in LabView software, with the code to be summarized in Appendix A, by which the signal is recorded and stored in a text file. The interface includes the display of a single signal, the RMS value for each electrode and allows modification of the gain and the number of samples per second. Moreover, two filters are used: a band-pass Butterworth filter with a low cutoff frequency at 15Hz and a high at 500Hz. Also, a notch filter of 50Hz is used to eliminate the power supply interference, the values of which can also be modified. The text files are used as input in Matlab for further processing, with the code to be summarized in Appendix B.

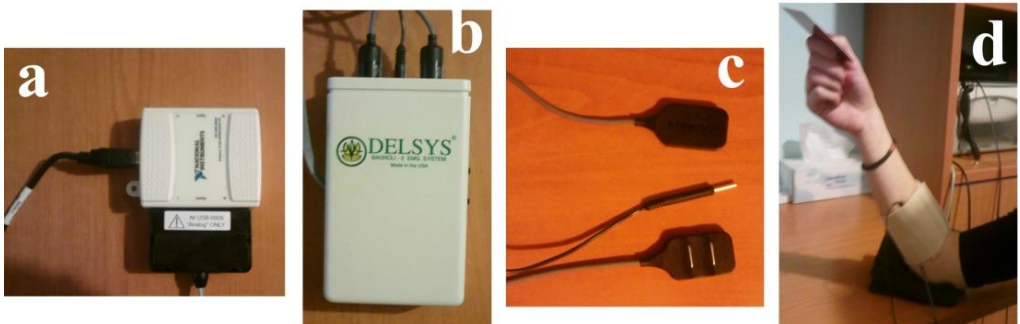

Figure 90: The experimental setup.

Figure 90 shows: a) NI analog to digital converter unit, b) a Delsys' EMG system of two channels, c) the two differential and one reference EMG electrodes and d) the setup used in a subject.

### 9.1.1 1st stage of measurements

In this stage, a 22-year-old subject performed 25 measurements for each movement using two sensors for receiving data. The total time for each measurement is 5 secs. SVM is used and more specifically Sequential Minimal Optimization, SMO, for training SVM. The code that is used can be found in [8]. Table 1 summarizes the classification rates using data from one electrode with two extracted characteristics of the signal: *variance and WAMP*.

Table 1: Classification rates for 2 classes.

| *Movements* | Hook | Cylindrical | Spherical | Tip | Palmar | Lateral |
|---|---|---|---|---|---|---|
| Hook | * | * | * | * | * | * |
| Cylindrical | 70% | * | * | * | * | * |
| Spherical | 98% | 98% | * | * | * | * |
| Tip | 96% | 86% | 100% | * | * | * |
| Palmar | 86% | 70% | 100% | 90% | * | * |
| Lateral | 70% | 86% | 98% | 98% | 98% | * |

As it can be observed, apart from the combinations of hook vs lateral, hook vs cylindrical and Palmar vs Cylindrical, which have low classification rates, all the other combinations approach the absolute success rate in some cases, as in Figure 91 and Figure 92.

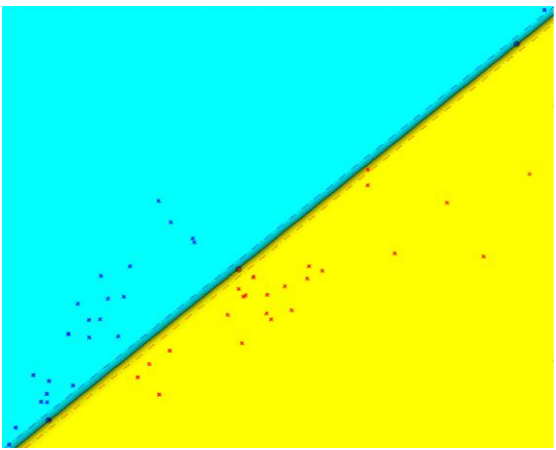

Figure 91: Spherical (Blue) - Hook (Red).

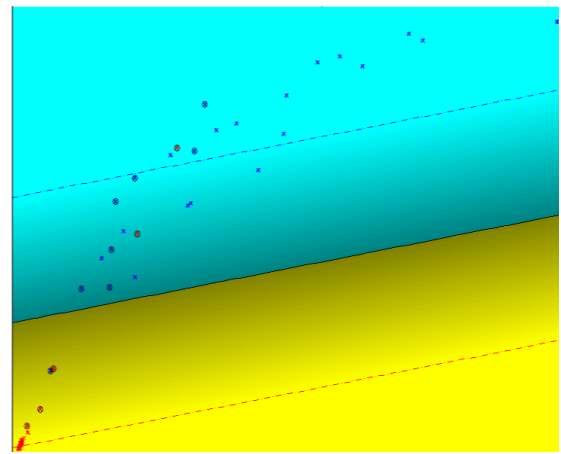

Figure 92: Tip (Blue) – Lateral (Red).

SVM algorithm is developed for separating two categories. For the SVM implementation, functions of Bioinformatics Toolbox in Matlab (svmtrain, svmclassify) are used. For our multiclass problem, the strategy of one-against-all is adopted [77]. The 20 samples are used for training and 5 for classification. Table 2 summarizes the results for various training methods in the svmtrain function.

Table 2: Classification rates (%) with various training methods with the use of raw signal.

| | Kernel Functions | | | | |
|---|---|---|---|---|---|
| Method Categories | Linear | Quadratic | Polynomial | RBF | MLP |
| S-T-P-C-H | 72 | 76 | 68 | 72 | 68 |
| S-T-L-C-H | 80 | 68 | 72 | 72 | 60 |
| S-P-L-C-H | 80 | 72 | 68 | 72 | 44 |
| S-T-P-L-C | 84 | 88 | 68 | 88 | 52 |
| S-P-L-C-H | 76 | 76 | 68 | 84 | 48 |
| T-P-L-C-H | 72 | 68 | 68 | 72 | 32 |
| S-T-P-L-C-H | 76.7 | 76.7 | 66.7 | 73.3 | 30 |

The available Kernel Functions are:
- Linear kernel.
- Quadratic kernel.
- Radial basis function (RBF).
- Polynomial kernel.
- Multilayer Perceptron kernel (MLP).

### 9.1.2 2nd stage of measurements

In this stage, a 22-year-old person performed 100 samples for each of the six movements by using two sensors for acquiring data. The total time for each measurement is 5 secs extracting five features: *Absolute mean, variance, Wavelength, Skewness, and Kurtosis.* This experiment's objective is to evaluate the robustness of the measurements in three consecutive days. The concept for this experiment is an application of prosthetic robotic hand control system, where the electrodes are not always placed exactly in the same position but in a similar. Three different methods are used for finding the separating hyperplane through SVM: radial basis function (RBF), SMO, least squares (LS). Table 3 and Table 4 summarize the classification results for using the raw signal and its RMS value, respectively.

Table 3: Classification rates (%) with various training methods for 3 days of measurements using raw signal.

| Method | RBF | | | SMO | | | LS | | |
|---|---|---|---|---|---|---|---|---|---|
| **Day** Categories | 1st | 2nd | 3rd | 1st | 2nd | 3rd | 1st | 2nd | 3rd |
| S-T-P-C-H | 88 | 81 | 90 | 89 | 82 | 83 | 82 | 86 | 82 |
| S-T-L-C-H | 86 | 93 | 89 | 88 | 97 | 81 | 80 | 95 | 80 |
| S-P-L-C-H | 92 | 85 | 91 | 95 | 89 | 89 | 94 | 92 | 94 |
| S-T-P-L-C | 76 | 77 | 87 | 86 | 82 | 83 | 90 | 84 | 90 |
| S-P-L-C-H | 72 | 76 | 85 | 84 | 79 | 74 | 81 | 81 | 81 |
| T-P-L-C-H | 59 | 68 | 74 | 55 | 62 | 62 | 56 | 57 | 56 |
| S-T-P-L-C-H | 75 | 79.2 | 87.5 | 86.6 | 83.3 | 73.3 | 79.2 | 84.2 | 79.2 |

Table 4: Classification rates (%) with various training methods for 3 days of measurements using the RMS value of raw signal.

| Method | RBF | | | SMO | | | LS | | |
|---|---|---|---|---|---|---|---|---|---|
| **Day** Categories | 1st | 2nd | 3rd | 1st | 2nd | 3rd | 1st | 2nd | 3rd |
| S-T-P-C-H | 66 | 78 | 88 | 80 | 90 | 94 | 80 | 91 | 93 |
| S-T-L-C-H | 74 | 83 | 86 | 87 | 97 | 92 | 91 | 98 | 95 |
| S-P-L-C-H | 76 | 83 | 86 | 89 | 92 | 90 | 92 | 94 | 87 |
| S-T-P-L-C | 68 | 70 | 79 | 79 | 87 | 85 | 88 | 87 | 83 |
| S-P-L-C-H | 67 | 69 | 74 | 74 | 85 | 84 | 85 | 86 | 82 |
| T-P-L-C-H | 64 | 69 | 65 | 68 | 64 | 67 | 63 | 66 | 58 |
| S-T-P-L-C-H | 73.3 | 80.8 | 89.2 | 75 | 81.7 | 80 | 79.2 | 81.7 | 84.2 |

As it can be easily observed, the results of the raw signal and its RMS value, using successive windows for its value calculation, are similar with small differences. Moreover, the results for the three consecutive days differ by a small percentage and exceed 80% in most methods. Also, it can be observed that the rates in the categories T-P-L-C-H are quite low and this is based on the absence of spherical movement which has may prevented some classes to be mixed. Robustness is achieved since the results are similar with high accuracy rates.

### 9.1.3 3rd stage of measurements

In the *first* part of data acquisition, five healthy subjects (three females and two males) around the same age (20 to 22-year-old) performed the six basic movements using two sensors for acquiring the data. The total time for each measurement is 6 sec. There is a stage of pre-processing to select the beginning of the signal when the muscle is contracted. So, the data in which the muscles are inactive are removed. The average IEMG value is calculated with a 40-msec sliding window and a threshold is set. When a sample of the signal is above this threshold, this one and the remaining is the useful information (Figure 93).

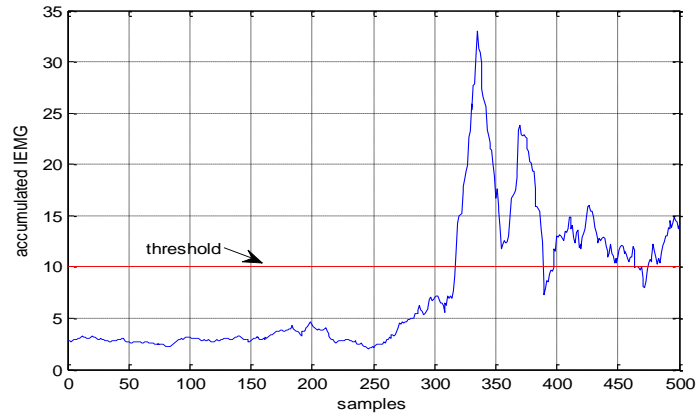

**Figure 93: Set threshold in the IEMG values of the signal and start of measurements in the 320 sample due to contracted muscle.**

There are two main techniques for windowing data: overlapping and non-overlapping (adjacent) windowing [78]. The overlapping windowing is selected with 300-msec window of (150 data points) and a 30-msec time leap (or 270-msec overlap) (Figure 94). EMD is applied on each segment for IMF extraction, as shown in Figure 95 and Figure 96, by using the EMD Toolbox of Matlab [79]. Therefore, we have a total of 40 characteristics for each signal, 8 for the raw signal, 24 (3x8) for each IMF and 8 for the residue. These features are: *mean, variance, Zero Crossing, Slope Sign Changes, Waveform Length, Willison Amplitude, Kurtosis and Skewness*.

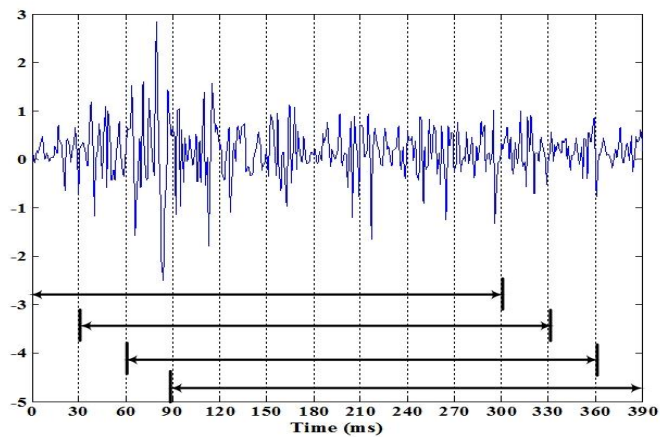

**Figure 94: Overlapping windows of 300 msec duration.**

The main objective is to evaluate if the usage of the extracted features based on EMD can increase the classification accuracy. The choice of the most optimum classifier in not the main goal. Hence, a simple Bayes linear classifier, which forms decision boundaries among sections of linear (hyper)planes in the feature space. It follows a notion that can be applied in most of real life data: "a simple linear surface can do surprisingly well as an estimate of the true decision surface" [80]. Alternatively, each testing sample (each feature vector $x$) will be allocated to a category based on the corresponding discriminant function's maximum value:

$$i = \arg \max_{i} \{2 ln P(\omega_i) - (x - \mu_i)^T C^{-1}(x - \mu_i)\}$$

where $P(\omega_i)$ is the prior probability and $\boldsymbol{\mu}_i$ is the mean of class $i$, while $\boldsymbol{C}$ is the estimated covariance matrix, which assumed common for all classes. Even though this hypothesis is not completely accurate for multiple practical applications, the classifier performs efficiently. Another advantage of selecting this specific classifier is the fact that there is no need for parameter tuning, which is a major challenge in machine learning [81].

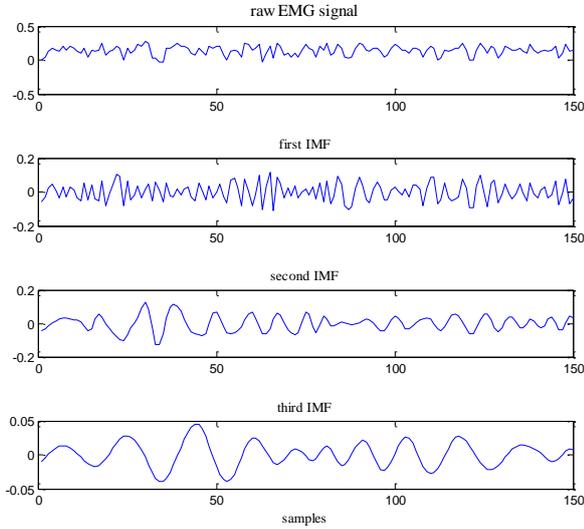

**Figure 95: Subplots of the EMG signal and the first three IMFs for the Lateral movement.**

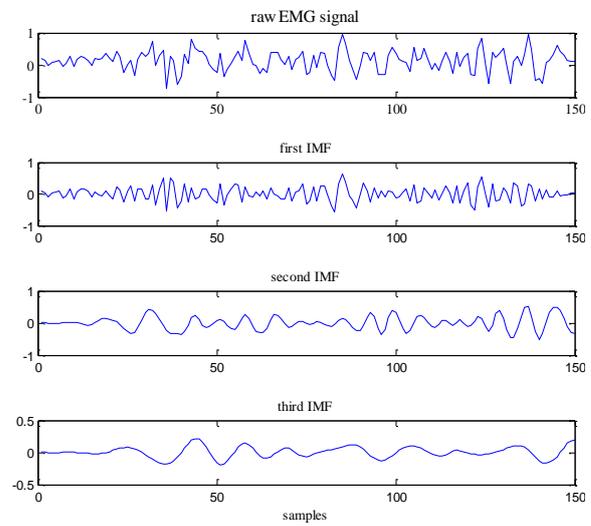

**Figure 96: Subplots of the EMG signal and the first three IMFs for the Cylindrical movement.**

Experimental Studies of 1st Part: Each subject's data is used separately without mixing it with the others. The performance is estimated based on the 5 x 2 cross-validation (CV) method [82]. It means that 15 of the recordings are selected randomly to train the algorithm and the other 15 to test each of the movements. Afterwards, these two sets are swapped (the testing set has become the training set and vice versa) with the whole procedure to be reoccur five times. Then, the average classification results for each subject for all the six movements are depicted in the Table 5. The algorithm performance is tested with the use of features extracted from; i) raw EMG signals, ii) the first IMF, and iii) both. The results for the remaining IMFs are not depicted because the classification accuracy declines.

The first column of Table 5 shows the average results from the use of raw signal for feature extraction while the third column displays the average results after the export of IMFs and the use of 40 features. The results are markedly improved by 1 to 5 percentage points. It worth to observe and the second column which contains only the first IMF. The results are reasonable to be lower as it includes a part of the information of the raw signal.

**Table 5: Average Classification accuracy (%) for 5 subjects.**

| Subject | Classification accuracy (average) | | |
|---|---|---|---|
| | *Raw EMG extracted features* | *First IMF extracted features* | *All extracted features* |
| Male 1 | 86.92 | 78.03 | 90.42 |
| Male 2 | 92.38 | 84.97 | 94.80 |
| Female 1 | 85.24 | 83.32 | 87.25 |
| Female 2 | 83.88 | 78.94 | 88.05 |
| Female 3 | 84.82 | 77.68 | 85.53 |

Experimental Studies of 2nd Part: In the second phase of this experiment, an additional person (Female 4) at the same age with the other subjects is added in measurements following the exact experiment. The different stage is to reduce the size of the signal features for each person prior to the classification of data, which used the same classifier. This is achieved by two different methods which are reported in Chapter 4. Therefore, after using the EMD improved results, we will now try to combine it with the above methods and the use of additional features in the frequency domain.

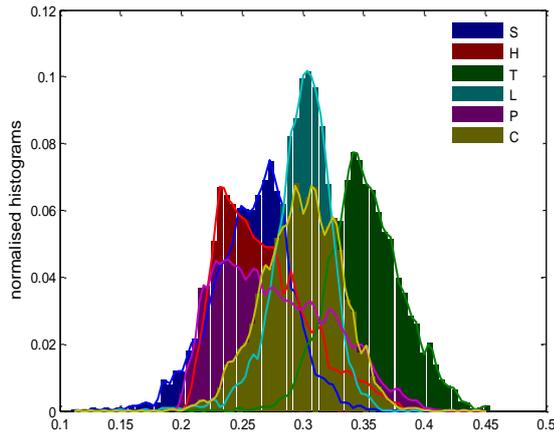

**Figure 97: Normalized Histograms ("empirical pdfs") of the IF median of the first IMF.**

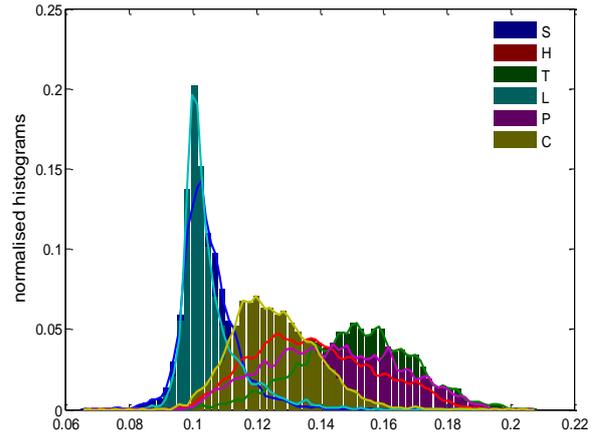

**Figure 98: Normalized Histograms ("empirical pdfs") of the IF median of the second IMF.**

It is worth noting that more features, from frequency domain, are added: the median, the kurtosis and the standard deviation of the instantaneous frequency. As seen in the above figures, Figure 97 and Figure 98, some movements have different standard and in the frequency domain. For example, in Man 1, the frequency information usage of the first IMF can separate the cylindrical movement from the spherical while the second IMF can separate the hook from the spherical.

Two loops (an "inner" and an "outer") of validation are applied to disengage the selection of the a) the retained Principal Components (PCs) and/or b) the included features from the performance estimation. The inner loop is employed for the procedure tuning (number of preserved features and PCs) and the outer for evaluating its performance. Thus, half of the testing samples are used for training in the inner loop and the rest for testing in the outer one for each of the 6 movements, which sets are swapped afterwards. The next step is to divide the data set into 70% for training and 30% for testing. The best configuration of PCs is extracted after repeating 10 times the inner loop in terms of the average classification performance. There is no option for threshold selection. All possible numbers (1-96) of the remaining features are tested.

Therefore, the parameters selection is not associated with the performance assessment [83] and, thus, we avoid excessively optimistic results on the potential of our method. Finally, the outer loop process is repeated five times, practically implemented a 5x2 CV [83]. Classification results are concluded in Table 6 and the combined confusion matrices corresponding to each of the two algorithms of dimension reduction, PCA and RELIEF, are presented in Table 7.

The structure of the aggregated matrix of RELIEF is the same as for PCA displaying analogous classification behavior, as observed in Figure 99 and Figure 100. There is no important statistical difference (significance level a=0.05) between the two methods of dimensionality reduction after the use of the Wilcoxon's Signed-Rank test [82].

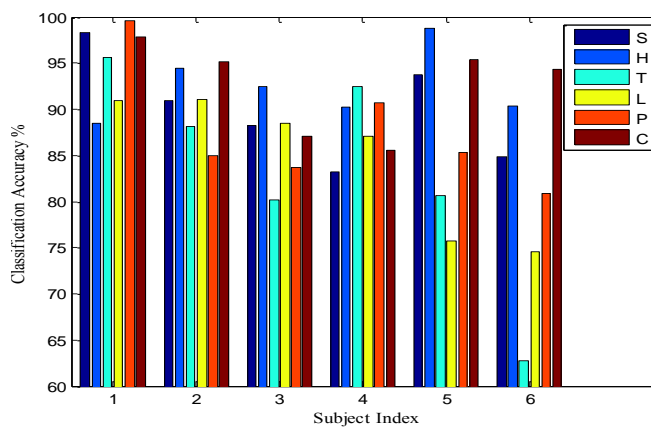

**Figure 99: Hand movement classification performance by PCA.**

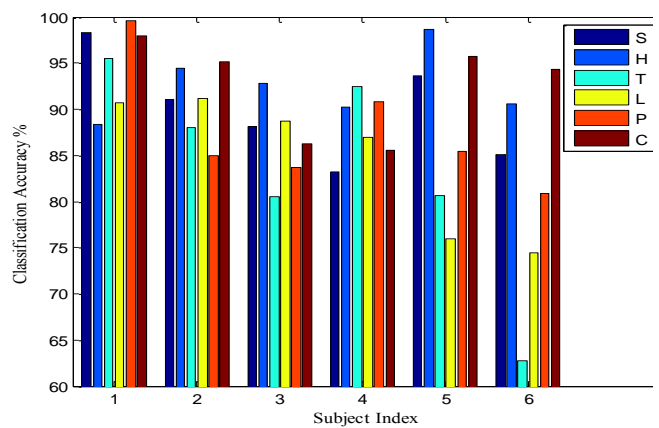

**Figure 100: Hand movement classification performance by RELIEF.**

More than 90 PCs are preserved in most of the cases except from Female 1, which retains around 85 PCs. There is no distinct advantage of one method relative to the other. Thus, the features convey complementary information. Furthermore, adding the frequency related features in the general feature set leads to an increase of the overall performance for 0.5-1% apart from Female 1, for which there is no change in the results.

As it can be observed from the confusion matrices, there is a high mixing among T vs L, T vs P and S vs C for almost all subjects (except for Male 1 for both cases and Female 3 for assigning C movements case).

Table 6: Classification results (%) for every subject using PCA and RELIEF respectively.

| Subject | Classification Results | | | | | | | Subject | Classification Results | | | | | | |
|---|---|---|---|---|---|---|---|---|---|---|---|---|---|---|---|
| | S | H | T | L | P | C | Overall | | S | H | T | L | P | C | Overall |
| Male 1 | 98.3 | 88.5 | 95.6 | 90.9 | 99.6 | 97.8 | 95.2 | Male 1 | 98.3 | 88.4 | 95.5 | 90.7 | 99.6 | 97.9 | 95.1 |
| Male 2 | 90,9 | 94.5 | 88.1 | 91.1 | 85.0 | 95.2 | 90.9 | Male 2 | 91.1 | 94.5 | 88.0 | 91.2 | 85.0 | 95.1 | 90.9 |
| Female 1 | 88.2 | 92.5 | 80.2 | 88.5 | 83.7 | 87.1 | 86.7 | Female 1 | 88.1 | 92.8 | 80.5 | 88.7 | 83.7 | 86.3 | 86.7 |
| Female 2 | 83.2 | 90.2 | 92.6 | 87.1 | 90.7 | 85.6 | 88.0 | Female 2 | 83.2 | 90.2 | 92.5 | 87.0 | 90.8 | 85.5 | 88.1 |
| Female 3 | 93.8 | 98.8 | 80.6 | 75.7 | 85.3 | 95.4 | 88.8 | Female 3 | 93.6 | 98.6 | 80.7 | 76.0 | 85.4 | 95.7 | 88.9 |
| Female 4 | 84.9 | 90.3 | 62.8 | 74.6 | 80.9 | 94.3 | 81.7 | Female 4 | 84.9 | 90.6 | 62.7 | 74.5 | 80.9 | 94.3 | 81.7 |

Table 7: Aggregated confusion matrices for all subjects using PCA and RELIEF.

MALE 1 (PCA)

| | | Predicted Class | | | | | |
|---|---|---|---|---|---|---|---|
| | | S | H | T | L | P | C |
| True Class | S | 21289 | 0 | 0 | 1 | 276 | 89 |
| | H | 51 | 19431 | 1 | 1427 | 0 | 1055 |
| | T | 50 | 149 | 18182 | 224 | 132 | 283 |
| | L | 0 | 1355 | 198 | 1715 | 86 | 74 |
| | P | 73 | 0 | 1 | 0 | 21579 | 17 |
| | C | 124 | 237 | 76 | 12 | 3 | 20398 |

MALE 1 (RELIEF)

| | | Predicted Class | | | | | |
|---|---|---|---|---|---|---|---|
| | | S | H | T | L | P | C |
| True Class | S | 21288 | 1 | 0 | 1 | 276 | 89 |
| | H | 51 | 19420 | 0 | 1451 | 0 | 1043 |
| | T | 50 | 150 | 18169 | 229 | 136 | 286 |
| | L | 0 | 1377 | 208 | 17120 | 86 | 74 |
| | P | 73 | 0 | 1 | 0 | 21579 | 17 |
| | C | 115 | 234 | 72 | 11 | 3 | 20415 |

MALE 2 (PCA)

| | | Predicted Class | | | | | |
|---|---|---|---|---|---|---|---|
| | | S | H | T | L | P | C |
| True Class | S | 17826 | 130 | 371 | 51 | 54 | 1163 |
| | H | 539 | 21449 | 46 | 602 | 1 | 63 |
| | T | 3 | 0 | 16637 | 1337 | 825 | 93 |
| | L | 0 | 44 | 1017 | 16985 | 588 | 11 |
| | P | 0 | 0 | 1898 | 896 | 15866 | 0 |
| | C | 597 | 0 | 268 | 31 | 15 | 18049 |

MALE 2 (RELIEF)

| | | Predicted Class | | | | | |
|---|---|---|---|---|---|---|---|
| | | S | H | T | L | P | C |
| True Class | S | 17856 | 133 | 365 | 47 | 50 | 1144 |
| | H | 539 | 21444 | 47 | 604 | 2 | 64 |
| | T | 3 | 0 | 16629 | 1343 | 824 | 96 |
| | L | 0 | 44 | 1006 | 16997 | 587 | 11 |
| | P | 0 | 0 | 1884 | 906 | 15870 | 0 |
| | C | 604 | 0 | 275 | 38 | 13 | 18030 |

FEMALE 1 (PCA)

| | | Predicted Class | | | | | |
|---|---|---|---|---|---|---|---|
| | | S | H | T | L | P | C |
| True Class | S | 18371 | 673 | 68 | 14 | 13 | 1701 |
| | H | 506 | 19615 | 160 | 59 | 279 | 596 |
| | T | 0 | 76 | 15999 | 1751 | 1759 | 345 |
| | L | 0 | 0 | 1546 | 17114 | 685 | 0 |
| | P | 28 | 245 | 1630 | 1113 | 16800 | 254 |
| | C | 850 | 1124 | 432 | 54 | 199 | 17896 |

FEMALE 1 (RELIEF)

| | | Predicted Class | | | | | |
|---|---|---|---|---|---|---|---|
| | | S | H | T | L | P | C |
| True Class | S | 18372 | 688 | 35 | 13 | 19 | 1713 |
| | H | 481 | 19706 | 153 | 56 | 294 | 525 |
| | T | 0 | 60 | 16055 | 1724 | 1771 | 320 |
| | L | 0 | 1 | 1572 | 17162 | 610 | 0 |
| | P | 29 | 267 | 1657 | 1075 | 16796 | 246 |
| | C | 882 | 1191 | 483 | 47 | 193 | 17759 |

FEMALE 2 (PCA)

| | | Predicted Class | | | | | |
|---|---|---|---|---|---|---|---|
| | | S | H | T | L | P | C |
| True Class | S | 18295 | 1052 | 2 | 0 | 23 | 2628 |
| | H | 1252 | 19666 | 61 | 0 | 4 | 817 |
| | T | 0 | 1 | 17472 | 276 | 1116 | 0 |
| | L | 0 | 0 | 264 | 16194 | 2142 | 0 |
| | P | 0 | 0 | 861 | 882 | 17052 | 0 |
| | C | 3031 | 192 | 0 | 0 | 19 | 19233 |

FEMALE 2 (relief)

| | | Predicted Class | | | | | |
|---|---|---|---|---|---|---|---|
| | | S | H | T | L | P | C |
| True Class | S | 18320 | 1033 | 2 | 0 | 23 | 2622 |
| | H | 1241 | 19678 | 61 | 0 | 4 | 816 |
| | T | 0 | 1 | 17465 | 273 | 1126 | 0 |
| | L | 0 | 0 | 265 | 16189 | 2146 | 0 |
| | P | 0 | 0 | 857 | 880 | 17058 | 0 |
| | C | 3032 | 195 | 0 | 0 | 19 | 19229 |

| FEMALE 3 (PCA) | | | | | | |
|---|---|---|---|---|---|---|
| | Predicted Class | | | | | |
| | | S | H | T | L | P | C |
| True Class | S | 20149 | 224 | 0 | 0 | 0 | 1107 |
| | H | 200 | 21818 | 0 | 0 | 43 | 24 |
| | T | 0 | 0 | 14990 | 2785 | 825 | 0 |
| | L | 0 | 0 | 2141 | 14092 | 2367 | 0 |
| | P | 0 | 4 | 774 | 2013 | 16643 | 66 |
| | C | 413 | 56 | 3 | 0 | 508 | 20465 |

| FEMALE 3 (RELIEF) | | | | | | |
|---|---|---|---|---|---|---|
| | Predicted Class | | | | | |
| | | S | H | T | L | P | C |
| True Class | S | 20106 | 250 | 0 | 0 | 0 | 1124 |
| | H | 220 | 21796 | 0 | 0 | 44 | 25 |
| | T | 0 | 0 | 15016 | 2771 | 813 | 0 |
| | L | 0 | 0 | 2106 | 14145 | 2349 | 0 |
| | P | 0 | 4 | 770 | 2003 | 16652 | 71 |
| | C | 432 | 55 | 4 | 0 | 502 | 20452 |

| FEMALE 4 (PCA) | | | | | | |
|---|---|---|---|---|---|---|
| | Predicted Class | | | | | |
| | | S | H | T | L | P | C |
| True Class | S | 18912 | 415 | 71 | 630 | 100 | 2137 |
| | H | 238 | 19451 | 0 | 3 | 45 | 1793 |
| | T | 169 | 0 | 12305 | 5002 | 2038 | 66 |
| | L | 112 | 53 | 3453 | 14771 | 1411 | 0 |
| | P | 7 | 0 | 1407 | 875 | 15788 | 1433 |
| | C | 372 | 299 | 11 | 16 | 506 | 19896 |

| FEMALE 4 (RELIEF) | | | | | | |
|---|---|---|---|---|---|---|
| | Predicted Class | | | | | |
| | | S | H | T | L | P | C |
| True Class | S | 18941 | 409 | 72 | 640 | 105 | 2098 |
| | H | 232 | 19509 | 0 | 3 | 48 | 1738 |
| | T | 171 | 0 | 12288 | 5042 | 2015 | 64 |
| | L | 123 | 48 | 3453 | 14761 | 1415 | 0 |
| | P | 7 | 0 | 1412 | 879 | 15774 | 1438 |
| | C | 381 | 306 | 12 | 18 | 492 | 19891 |

## 9.2 Part B of experimental procedure: Use of autonomous EMG system

In this part of the experimental process, we use the autonomic EMG system which is implemented as part of this thesis. In order to receive the signal, code in the "Processing" programming language is developed, working together with the operation of Arduino, which receives the signal from the analog inputs. The sampling rate relies on the input of the function delay, which receives as input the ms that the programmer wants to delay loop function.

The Simplot application [84] is used to display the signal during acquisition from Arduino, as depicted in Figure 101. Simplot can show graphically the incoming signal from the corresponding COM port of the computer. The text files are imported to Matlab for further processing.

The k-NN algorithm is developed in Matlab code and finds the virtual center in a vector for each category. Then, these vectors are inserted in a table in "Wiring" code that implements the k-means algorithm. The output of this code is the predicted movement that the user of the device has conducted.

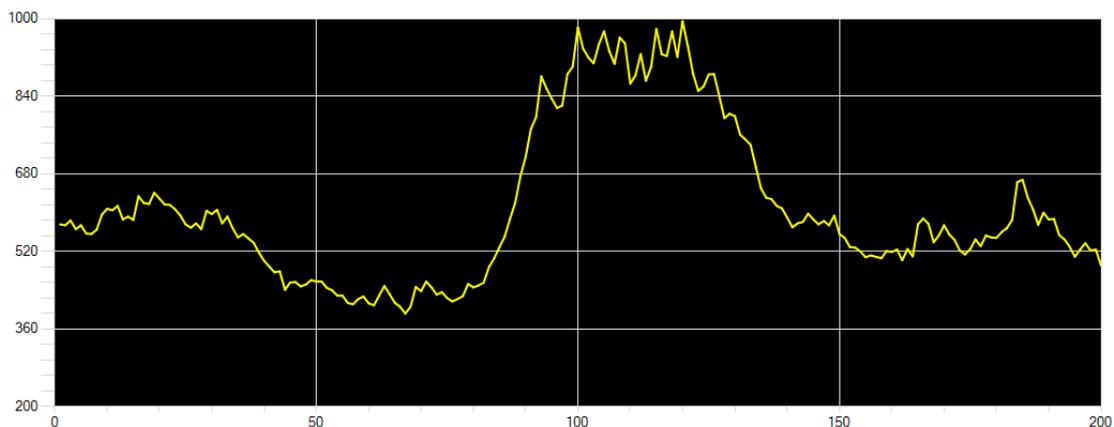

Figure 101: Signal from one sensor in the Simplot.

### 9.2.1 1st stage of measurements

In this stage, a 22-year-old person performed 15 measurements for each of eight movements, six basic movement, opening and closing the hand, using one sensor for receiving data. Ten of the examples are used for training the system and find the virtual center for each category and the other 5 examples for testing.

The total duration of each measurement is 4 secs with 100 samples per sec, i.e. delay(10) (100Hz), having six characteristics of the raw signal: *maximum value, absolute average, waveform length, number of slope changes, Willison Amplitude and standard deviation.*

The success rates are 62.5% and 63.3% for 8 and 6 movements respectively. For fewer categories the success rate increases.

### 9.2.2 2nd stage of measurements

In this stage, a 22-year-old person performed 15 measurements for each of eight movements, six basic movement, opening and closing the hand, using three sensors for receiving data. Ten of the examples are used for training the system and find the virtual center for each category and the other five examples for testing.

The total duration of each measurement is 2 secs with 100 samples per sec, i.e. delay(10) (100Hz), with some signals to be shown in Figure 102 and Figure 103. Six characteristics of the raw signal are extracted: *maximum value, absolute average, waveform length, number of slope changes, Willison Amplitude and standard deviation.*

In the second measurement stage, three sensors are used because the system is fully constructed. The data from one sensor are discarded, because there is a problem with the values provided as most of them touched the peak ($2^{10}-1 = 1023$). This problem arose because of the position of the electrodes or due to the gain of the sensor. Therefore, after using the signals from the two sensors, the success rate is 65% and 76.6% for eight and six movements respectively. For a smaller number of categories, the success rate increased significantly. For example, the success rate for 5 moves ranged from 80% to 88%.

The key finding is that for shorter time than 2 secs, the rates are similar. Moreover, after the change of variables' values a, b, the time window changes. Since the goal is a practical application, the time window decreased 15 to 20 msec and the results are slightly worse, about -3% to -5%. Therefore, with the addition of an extra sensor in measurements, the results will rise from 73.3% to higher rates.

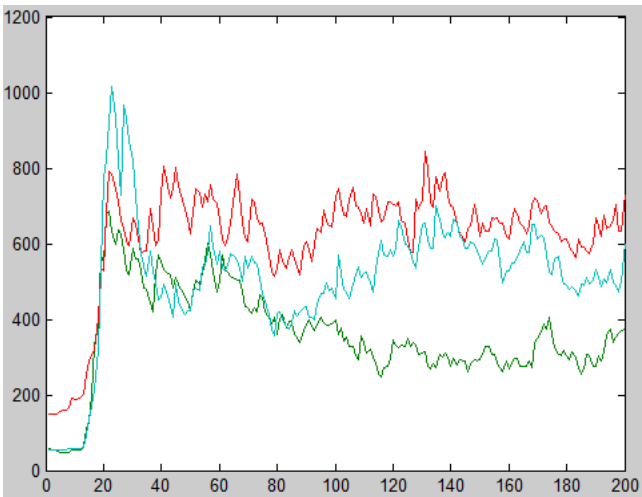
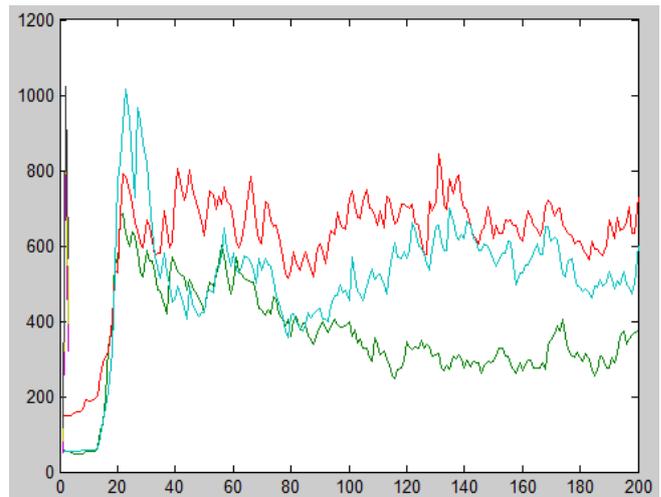

Figure 102: Signal of hook movement from 3 sensors.   Figure 103: Signal of cylindrical movement from 3 sensors.

# Chapter 10: Epilogue

## 10.1 Conclusions

This thesis is a product of interdisciplinary research that extends in the fields of medicine and engineering. This fact is a challenge for a student facing this problem from the electrical and computer engineering perspective because there are multiple difficulties to be faced due to the knowledge gap for a new subject. Projects in a multidisciplinary level are, for sure, the future of research and particularly when it has to deal with the biomedical field, which is related to the immediate improvement of people's life.

The main purpose of undertaking the diploma thesis was the recognition of basic hand movements and the construction of an application. Along with future work, it would be possible to improve the daily life of a hand amputee in conjunction, of course, with a hand exoskeleton.

In the initial experimental stage, the Delsys' EMG system has given the opportunity to experiment with the placement of sensors at various points on the forearm. Except from the study, which was necessary for learning the positions of the muscles and the way to use each one, the conclusion is that when it comes to experimental procedure, only the basic knowledge about this part of science and preoccupation with experiment is the base for the desired results. Moreover, it is very difficult or impossible to place the electrodes on a particular single muscle of the forearm, since multiple muscles are involved. Therefore, the sensor acquires the signal resulting from the contraction of several muscles creating noise, which is detrimental to the results, but unavoidable, considering that multiple muscles exist in such a constrained space.

To mitigate the impact of noise on classification results, the extraction of features from the raw signal, which has not undergone any processing, was initially preferred. SVM exceled in classification accuracy in comparison with the other methods of pattern recognition. The proper selection of features has contributed significantly to obtain the highest possible classification accuracy. The optimum choice could be characterized more as an art [52] and several times is based on empirical observations rather than a particular way. Moreover, apart from an experiment in an individual for a single day, an experiment of three consecutive days was applied to test the robustness of the measurements by placing each time the electrodes in similar positions. In real application, the exact position would be quite challenging. The results are similar, with slight variations by using similar extracted features along with SVM, which succeeded over k-NN and PNN in the previous stage. After 100 measurements for each movement, it is shown on a larger scale that although each signal has a special "signature", it can be mixed several times with other classes causing errors of misclassification.

In the next stage, specific objectives are set: a) conduction of tests to other subjects and b) reduction of the time window from 5 sec to 300 msec. These objectives are designed to test whether it is possible to have both electrodes at specific place on the forearm of several people at a time and with a short-time window to make it feasible for a real application. In this stage, the signal was first analyzed by the EMD method in various IMFs increasing the vector of features because except from the raw signal, the same features are extracted from the IMFs and the residue. Therefore, the dimensionality reduction of the features is important, which is achieved by PCA and Relief algorithm. The results are markedly improved with the use of the above methods compared to using only the primary signal.

In the final stage of thesis, an autonomous EMG system was implemented to conduct experiments in a real application using the Arduino Uno microcontroller. The implementation was challenging as there are several limitations, such as in the memory, which caused changes in the plans for the code that would be used. The goal was to train the system to a computer and the extracted parameters would be used as input in the classification method in the code, which will be executed in the microcontroller. After several failed attempts to expand the memory via the Wifi-Shield containing microSD card and use of PROGMEM variable modifier to be stored the values of variables in the flash memory and not in the SRAM, it was decided not to extend the memory but to consider the restrictions. For this reason, a simple classification algorithm k-means using Euclidean distance was selected. This was due to the data formation in different clusters after inspecting them graphically. The results are encouraging since 75% of accuracy was achieved for the 6 basic movements, using two electrodes, and a time window of 150ms once the muscle started to contract.

## 10.2 Future work

As in any research activity, also for this thesis, the time constraint specifies the end of an effort. Therefore, it is possible to enumerate several extensions and ideas for conducting further experiments and improving the current application.

- The limitation of microcontroller's memory can lead to the use of another that will allow the use of more complex algorithms for increasing the classification accuracy.
- The readout circuit should be changed as it has a high DC component, which can cause problems in measurements. Moreover, the $\pm V_s$ creates the need for an extra battery, which increases the system's weight making it uncomfortable for the user. Finally, by the current development of electronics, the power supply could perhaps be reduced to lower Volt levels.
- Active electrodes should be used in the second system since the passive contain sticker which makes repositioning an issue.
- A user interface for recording the movements should be implemented. The subjects will be asked to perform specific grasping based on the image or message that appears on the screen. This will lead to transform the recording time into a more pleasant procedure.
- Another classification algorithm should be used in Arduino with fewer or equal requirements in terms of memory.
- Experiment with more healthy subjects should be conducted using the autonomous EMG system in order to observe the success rates and allow each person to control the robotic hand.
- Experiment with hand amputees should be performed to evaluate the rates of successful classification. This data should be processed appropriately and along with a proper feature extraction, it should be able to provide a solution for a real application of controlling the robotic hand.

## Chapter 7

## Chapter 8

## Chapter 9

## Appendix

# Sources of reprinted images

## Chapter 2

**Figure 1:** http://www.medicalook.com/human_anatomy/organs/Antebrachium.html
**Figure 2:** K. Pittara, "Association of dietary factors with the syndrome of delayed muscle soreness (DOMS)", Bachelor Thesis, Harokopio University of Athens, 2009.
**Figure 3:** http://www.studyblue.com/notes/note/n/muscle-1/deck/4309721
**Figure 4:** K. Pittara, "Association of dietary factors with the syndrome of delayed muscle soreness (DOMS)", Bachelor Thesis, Harokopio University of Athens, 2009.
**Figure 5:** https://commons.wikimedia.org/wiki/File:Gray417_color.PNG
**Figure 6:** http://www.slideshare.net/redfoxer09/the-antebrachium
**Figure 7:** http://schools.wikia.com/wiki/Pulleys
**Figure 8:** Naoki Fukaya, Shigeki Toyama, Tamim Asfour, Rüdiger Dillmann, "Design of the TUAT/Karlsruhe Humanoid Hand", 2000.

## Chapter 3

**Figure 9:** http://pictureofgoodelectroniccircuit.blogspot.fi/2010/04/phase-and-function-of-analog-signal-or.html
**Figure 10:** http://www.swharden.com/blog/2009-01-15-circuits-vs-software/
**Figure 11:** http://prixray.com/
**Figure 12:** http://www.elements4health.com/pet-scans-reveal-plaques-and-tangles-in-alzheimers.html
**Figure 13:** http://nucleus.iaea.org/HHW/Technologists/NuclearMedicineTech/Educationalresources
**Figure 14:** H. Kolářová, Biosignals, Lecture, Department of Medical Biophysics, Palacký University, Olomouc.
**Figure 15:** H. Kolářová, Biosignals, Lecture, Department of Medical Biophysics, Palacký University, Olomouc.
**Figure 16:** http://www.ptb.de/cms/en/publikationen/online-magazin/in-focus/biomagnetism/linkesmenue/biomagnetic-signals-of-the-human-body.html
**Figure 17:** http://www.google.com/patents/EP1552718B1?cl=en
**Figure 18:** H. Kolářová, Biosignals, Lecture, Department of Medical Biophysics, Palacký University, Olomouc.
**Figure 19:** H. Kolářová, Biosignals, Lecture, Department of Medical Biophysics, Palacký University, Olomouc.
**Figure 20:** H. Kolářová, Biosignals, Lecture, Department of Medical Biophysics, Palacký University, Olomouc.
**Figure 21:** A. Skodras, B. Anastasopoulos, "Digital Signal and Image Processing ", Hellenic Open University, 2003.
**Figure 22:** G. Moustakides, "Basic Techniques of Digital Signal Processing", Tziolas Publications, 2003.
**Figure 23:** A. Skodras, B. Anastasopoulos, "Digital Signal and Image Processing ", Hellenic Open University, 2003.
**Figure 24:** A. Skodras, B. Anastasopoulos, "Digital Signal and Image Processing ", Hellenic Open University, 2003.

## Chapter 4

**Figure 25:** P. Flandrin, P. Goncalves, "Empirical Mode Decompositions as Data-Driven Wavelet-Like Expansions," Int. J. of Wavelets, Multires. and Info. Proc., Vol. 2, No. 4, pp. 477-496., 2004.
**Figure 26:** http://perso.ens-lyon.fr/patrick.flandrin/emd.html
**Figure 27:** http://whatilearned.wikia.com/wiki/File:Kurtosis.jpg
**Figure 28:** http://en.wikipedia.org/wiki/Skewness
**Figure 29:** http://www.nlpca.org/pca_principal_component_analysis.html

## Chapter 5

**Figure 30:** http://users.rcn.com/jkimball.ma.ultranet/BiologyPages/E/ExcitableCells.html
**Figure 31:** J. Clark, J. Webster, "Medical Instrumentation Application and Design", 4th Edition, Wiley, 2009.

**Figure 74:** BagnoliTM 2-Channel Handheld EMG System, User's Guide, Delsys
**Figure 75:** BagnoliTM 2-Channel Handheld EMG System, User's Guide, Delsys
**Figure 76:** BagnoliTM 2-Channel Handheld EMG System, User's Guide, Delsys
**Figure 77:** BagnoliTM 2-Channel Handheld EMG System, User's Guide, Delsys
**Figure 78:** EMG Sensors Product Sheet, Delsys
**Figure 79:** EMG Sensors Product Sheet, Delsys
**Figure 80:** Muscle Sensor v2, Manual, Advancer Technologies
**Figure 82:** Muscle Sensor v2, Manual, Advancer Technologies
**Figure 83:** Muscle Sensor v2, Manual, Advancer Technologies
**Figure 84:** ATmega328, Datasheet
**Figure 85:** ATmega16U2, Datasheet
**Figure 86:** http://arduino.cc/en/Main/arduinoBoardUno

# Appendix A

## LabView Code

The LabView code, used for recording the signal and storing it in a text file, is analyzed in the schematic diagram of Figure 104 and the interface in Figure 109 [85].

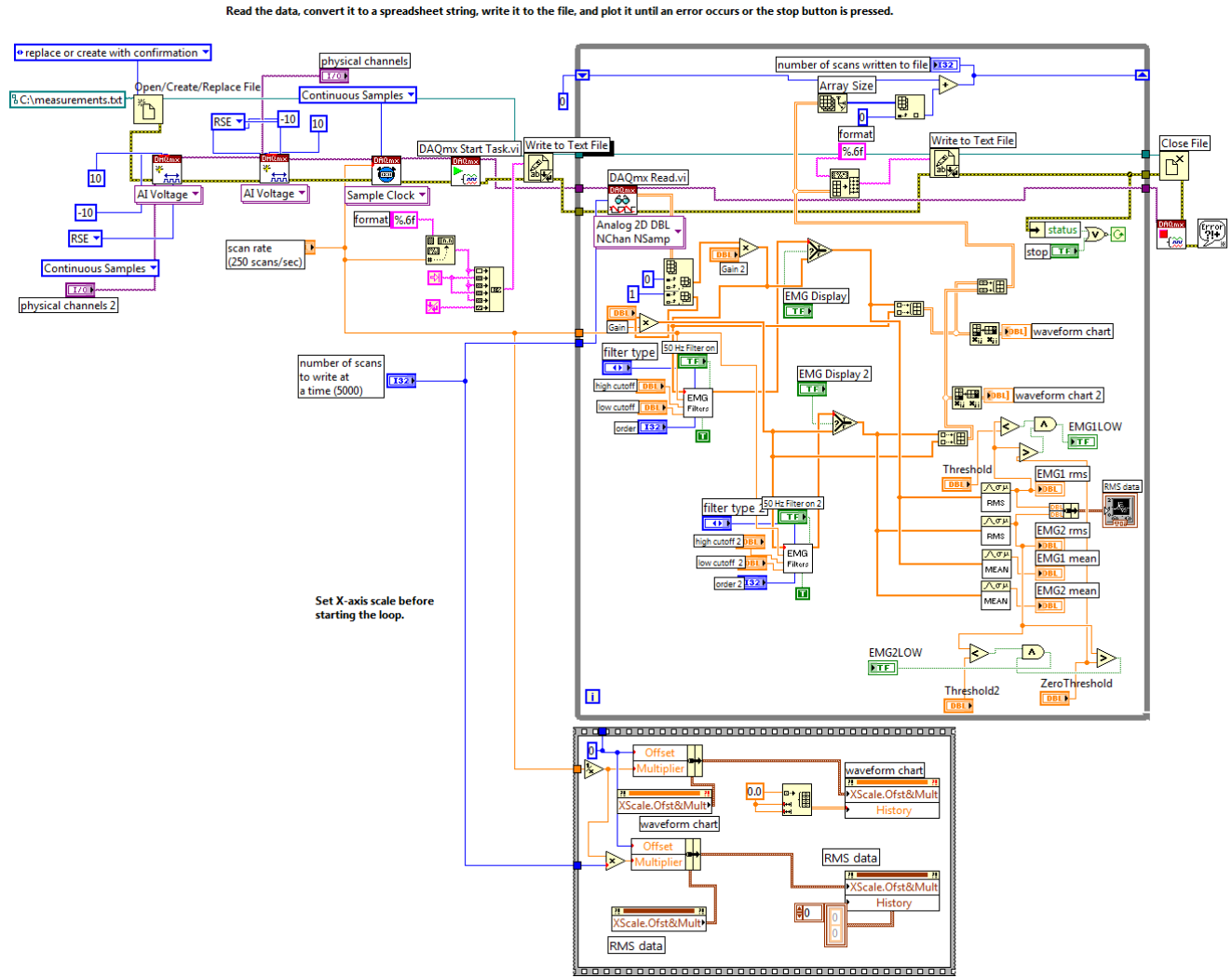

Figure 104: Recording and storage of EMG signal.

In Figure 105, Figure 106, Figure 107 and Figure 108 are shown the parts of the code with the different function that performs each one of them.

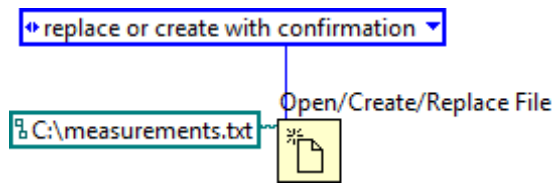

Figure 105: Create new txt file.

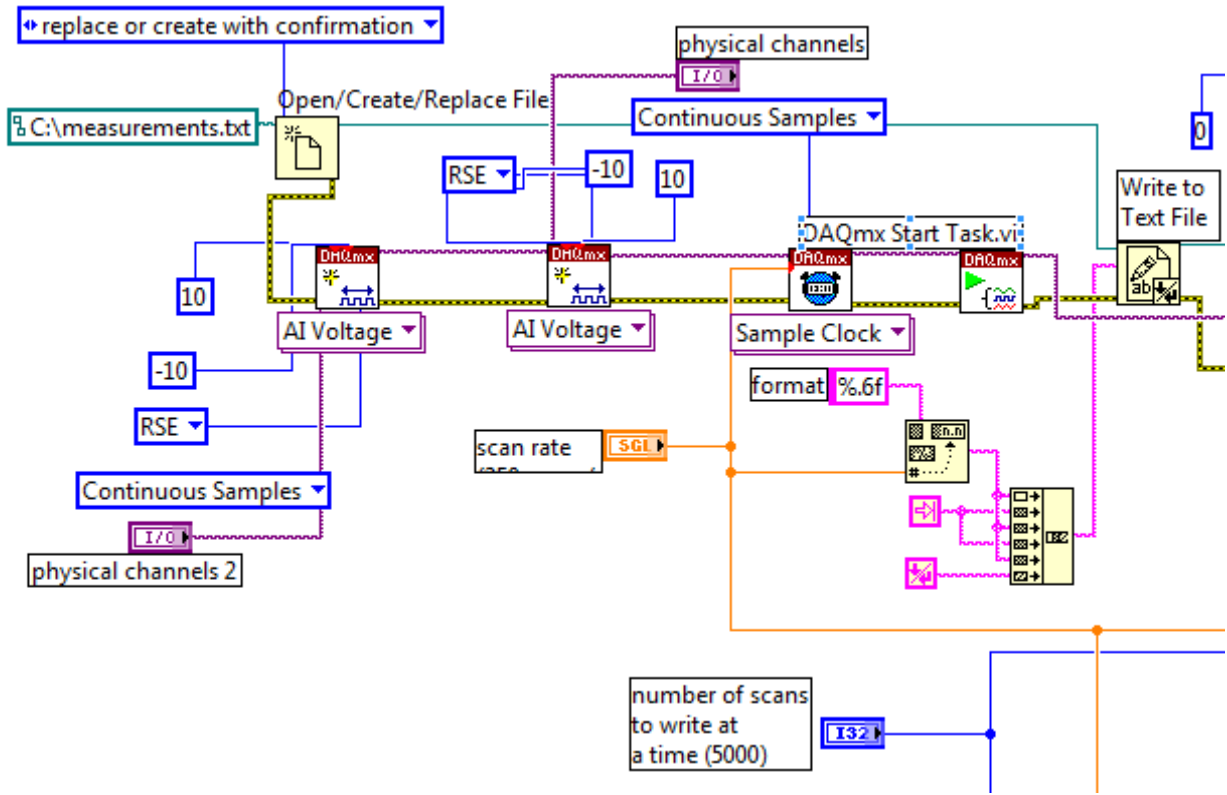

Figure 106: Configuration and start recording.

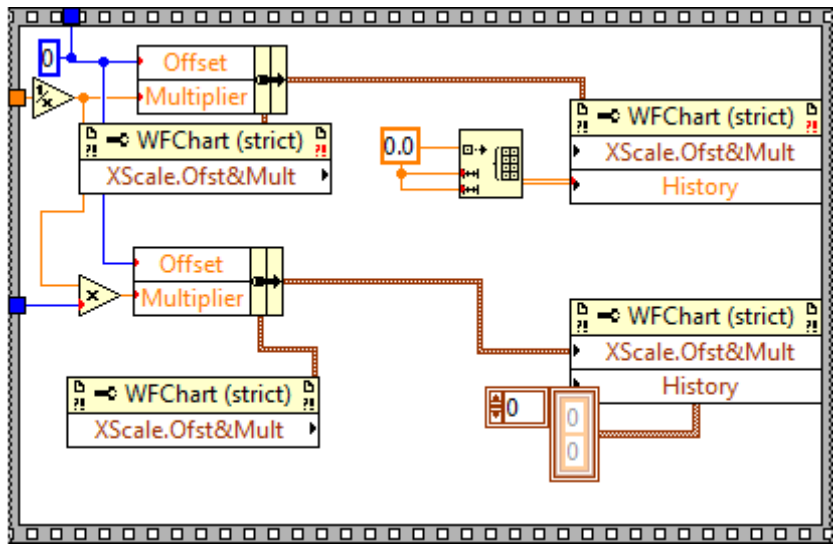

Figure 107: Set X-axis scale before starting the loop.

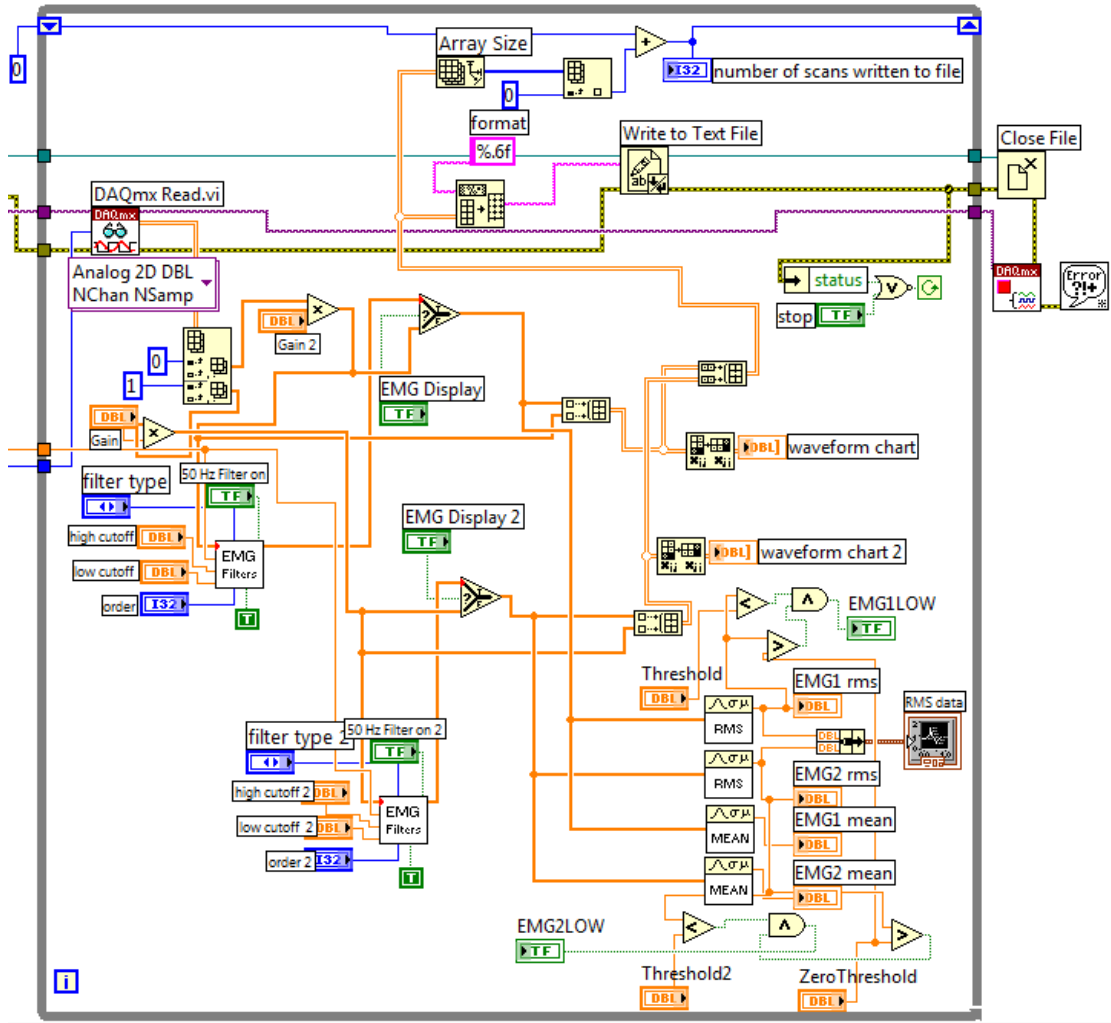

Figure 108: Reading data, converting to string and storing it in a file after the display charts until you stop the process by pressing Stop.

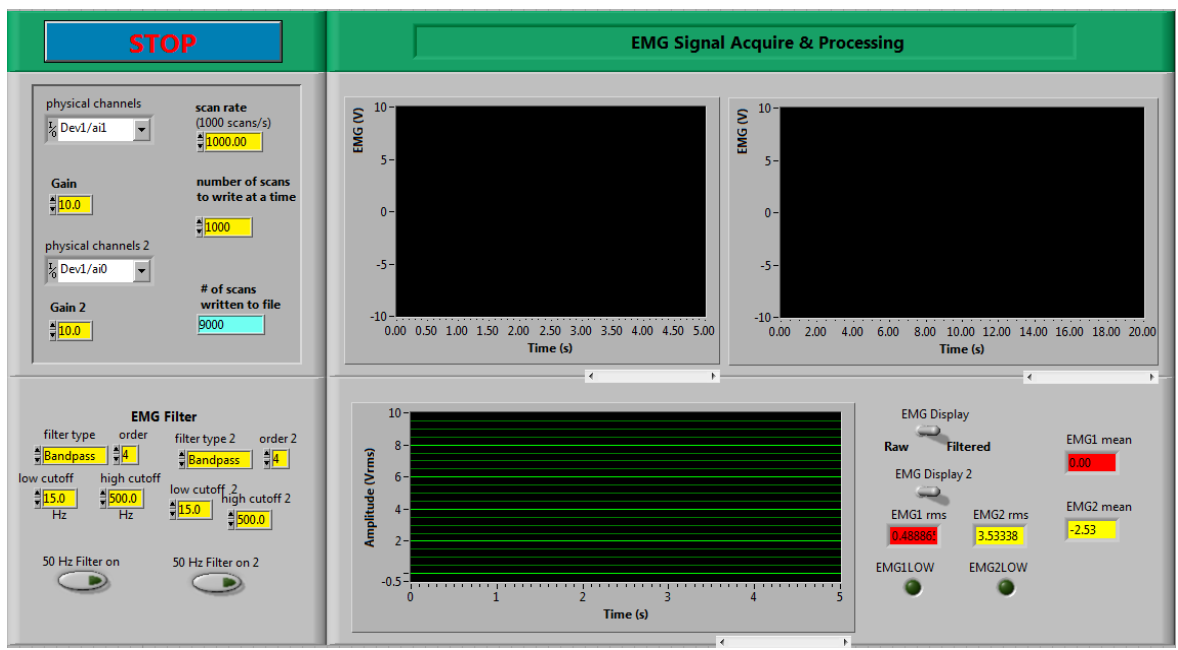

Figure 109: Interface of LabView Code.

# Appendix B

## Code in Matlab

## A' part of experiments (3rd stage)

## Use of EMD and consequently the IMF: first_attempt.m

```matlab
load('subject1.mat');
cumIEMS1=zeros(1,400);
cumIEMS2=zeros(1,400);
Window1=20;
windoW=150;
data_delete=500;
TS=500;
%% pl
    TEMP1=pl_1;
    TEMP2=pl_2;
for k=1:30

for i=1:TS

    cumIEMS1(k,i)=sum(abs(TEMP1(k,i:Window1+i-1)));
    cumIEMS2(k,i)=sum(abs(TEMP2(k,i:Window1+i-1)));

end
    I1=find(cumIEMS1(k,:)>10);
    I2=find(cumIEMS2(k,:)>10);
    if isempty(I1)
       I1=TS;
    end
     if isempty(I2)
       I2=TS;
    end
    I=min(I1(1),I2(1));

count=0;
temp=TEMP1(k,I(1):end-data_delete);
FEATURE1=zeros(length(1:15:length(temp)-windoW),49);
for seg=1:15:length(temp)-windoW
    count=count+1;
%    imf=emdc([],temp(seg:seg+windoW-1),[0.05 0.05],4);
imf=emd(temp(seg:seg+windoW-1),'MAXMODES',3,'FIX',10);
    [A,ff,tt] = hhspectrum(imf(1:end-1,:));
%    feat_pl{k,count}=[median(ff'),std(imf(1:end-1,:)')];
FEATURE1(count,:)=[sEMG(temp(seg:seg+windoW-
1)),median(ff'),std(ff'),kurtosis(ff'),sEMG(imf(1,:)),sEMG(imf(2,:)),sEMG(imf(3,:)),sEMG(imf(4,:))];
end %seg

count=0;
temp=TEMP2(k,I(1):end-data_delete);
```

```
FEATURE2=zeros(length(1:15:length(temp)-windoW),49);
for seg=1:15:length(temp)-windoW
    count=count+1;
%     imf=emdc([],temp(seg:seg+windoW-1),[0.05 0.05],4);
imf=emd(temp(seg:seg+windoW-1),'MAXMODES',3,'FIX',10);
% imf=emd(temp(seg:seg+windoW-1));
    [A,ff,tt] = hhspectrum(imf(1:end-1,:));
%     feat_pl{k,count}=[median(ff'),std(imf(1:end-1,:)')];
FEATURE2(count,:)=[sEMG(temp(seg:seg+windoW-
1)),median(ff'),std(ff'),kurtosis(ff'),sEMG(imf(1,:)),sEMG(imf(2,:)),sEMG(imf(3,:)),sEMG(imf(4,:))];
end %seg
feat_pl{k}=[FEATURE1,FEATURE2];
end
% Similarly for the other movements
save subject1featn3 feat_sf feat_ag feat_ak feat_pal feat_pl feat_kul
```

## 5 x 2 CV (cross-validation) : general5x2cv.m

```
clear all
load subject1featn3 % Similarly for the other subjects

CC1=zeros(6);
CC2=zeros(6);
CC3=zeros(6);
CC4=zeros(6);
for outer=1:5
    rand('seed',outer)
    I=randperm(30);

    SF1=[];AG1=[];AK1=[];PAL1=[];PL1=[];KUL1=[];
    SF2=[];AG2=[];AK2=[];PAL2=[];PL2=[];KUL2=[];
    for i=1:15
      SF1=[SF1;feat_sf{I(i)}];
      AG1=[AG1;feat_ag{I(i)}];
      AK1=[AK1;feat_ak{I(i)}];
      PAL1=[PAL1;feat_pal{I(i)}];
      PL1=[PL1;feat_pl{I(i)}];
      KUL1=[KUL1;feat_kul{I(i)}];
      SF2=[SF2;feat_sf{I(i+15)}];
      AG2=[AG2;feat_ag{I(i+15)}];
      AK2=[AK2;feat_ak{I(i+15)}];
      PAL2=[PAL2;feat_pal{I(i+15)}];
      PL2=[PL2;feat_pl{I(i+15)}];
      KUL2=[KUL2;feat_kul{I(i+15)}];
    end
%% fist fold
A=[SF1;AG1;AK1;PL1;PAL1;KUL1];
```

```
    B=[SF2;AG2;AK2;PL2;PAL2;KUL2];
    test=A;
    train=B;
    [train,ps]=mapstd(train(:,:)');
    train=train';
    test=mapstd('apply',test(:,:)',ps)';

    test(:,[47,96])=[];
    train(:,[47,96])=[];
    labels_test=[ones(size(SF1,1),1);2*ones(size(AG1,1),1);3*ones(size(AK1,1),1);4*ones(size(PL1,1),1);5*ones(size(PAL1,1),1);6*ones(size(KUL1,1),1)];
    labels_train=
    [ones(size(SF2,1),1);2*ones(size(AG2,1),1);3*ones(size(AK2,1),1);4*ones(size(PL2,1),1);5*ones(size(PAL2,1),1);6*ones(size(KUL2,1),1)]

    %% classifier section

    %% second fold
    A=[SF1;AG1;AK1;PL1;PAL1;KUL1];
    B=[SF2;AG2;AK2;PL2;PAL2;KUL2];
    test=B;
    train=A;
    test(:,[47,96])=[];
    train(:,[47,96])=[];
    [train,ps]=mapstd(train(:,:)');
    train=train';
    test=mapstd('apply',test(:,:)',ps)';
    labels_train=[ones(size(SF1,1),1);2*ones(size(AG1,1),1);3*ones(size(AK1,1),1);4*ones(size(PL1,1),1);5*ones(size(PAL1,1),1);6*ones(size(KUL1,1),1)];
    labels_test=
    [ones(size(SF2,1),1);2*ones(size(AG2,1),1);3*ones(size(AK2,1),1);4*ones(size(PL2,1),1);5*ones(size(PAL2,1),1);6*ones(size(KUL2,1),1)]

    %% classifier section

    end
```

### Linear Bayes Normal Classifier (LDC) for a' part of 3$^{rd}$ stage: testing_ldc.m

```
    clear all
    load subject1featn3 % Similarly for the other subjects
    CC1=zeros(6);
    CC2=zeros(6);
    CC3=zeros(6);
    for outer=1:5
        I=randperm(30);
```

```
    SF1=[];AG1=[];AK1=[];PAL1=[];PL1=[];KUL1=[];
    SF2=[];AG2=[];AK2=[];PAL2=[];PL2=[];KUL2=[];
    for i=1:15
      SF1=[SF1;feat_sf{I(i)}];
      AG1=[AG1;feat_ag{I(i)}];
      AK1=[AK1;feat_ak{I(i)}];
      PAL1=[PAL1;feat_pal{I(i)}];
      PL1=[PL1;feat_pl{I(i)}];
      KUL1=[KUL1;feat_kul{I(i)}];

      SF2=[SF2;feat_sf{I(i+15)}];
      AG2=[AG2;feat_ag{I(i+15)}];
      AK2=[AK2;feat_ak{I(i+15)}];
      PAL2=[PAL2;feat_pal{I(i+15)}];
      PL2=[PL2;feat_pl{I(i+15)}];
      KUL2=[KUL2;feat_kul{I(i+15)}];

    end
%%
A=[SF1;AG1;AK1;PL1;PAL1;KUL1];
B=[SF2;AG2;AK2;PL2;PAL2;KUL2];
test=A;
train=B;

% ind1=[1:8,[1:8]+51];
% ind2=[21:28,[21:28]+51];
% ind3=[29:36,[29:36]+51];
% ind4=[37:44,[37:44]+51];
% ind5=[9:20+[9:20]+51];

ind1=[1:8,[1:8]+49];
ind2=[18:25,[18:25]+49];
ind3=[26:33,[26:33]+49];
ind4=[34:41,[34:41]+49];
ind5=[42:49,[42:49]+49];

SW1=size(SF2,1)/size(train,1)*cov(SF2(:,ind1))+size(AG2,1)/size(train,1)*cov(AG2(:,ind1))+size(AK2,1)
/size(train,1)*cov(AK2(:,ind1))+size(PL2,1)/size(train,1)*cov(PL2(:,ind1))+size(PAL2,1)/size(train,1)*co
v(PAL2(:,ind1))+size(KUL2,1)/size(train,1)*cov(KUL2(:,ind1));
SW2=size(SF2,1)/size(train,1)*cov(SF2(:,ind2))+size(AG2,1)/size(train,1)*cov(AG2(:,ind2))+size(AK2,1)
/size(train,1)*cov(AK2(:,ind2))+size(PL2,1)/size(train,1)*cov(PL2(:,ind2))+size(PAL2,1)/size(train,1)*co
v(PAL2(:,ind2))+size(KUL2,1)/size(train,1)*cov(KUL2(:,ind2));
SW3=size(SF2,1)/size(train,1)*cov(SF2(:,ind3))+size(AG2,1)/size(train,1)*cov(AG2(:,ind3))+size(AK2,1)
/size(train,1)*cov(AK2(:,ind3))+size(PL2,1)/size(train,1)*cov(PL2(:,ind3))+size(PAL2,1)/size(train,1)*co
v(PAL2(:,ind3))+size(KUL2,1)/size(train,1)*cov(KUL2(:,ind3));
SW4=size(SF2,1)/size(train,1)*cov(SF2(:,ind4))+size(AG2,1)/size(train,1)*cov(AG2(:,ind4))+size(AK2,1)
/size(train,1)*cov(AK2(:,ind4))+size(PL2,1)/size(train,1)*cov(PL2(:,ind4))+size(PAL2,1)/size(train,1)*co
v(PAL2(:,ind4))+size(KUL2,1)/size(train,1)*cov(KUL2(:,ind4));
```

```matlab
SM1=cov(train(:,ind1));
SM2=cov(train(:,ind2));
SM3=cov(train(:,ind3));
SM4=cov(train(:,ind4));

[trace(inv(SW1)*SM1) trace(inv(SW2)*SM2) trace(inv(SW3)*SM3) trace(inv(SW4)*SM4)]
[det(inv(SW1)*SM1) det(inv(SW2)*SM2) det(inv(SW3)*SM3) det(inv(SW4)*SM4)]

% train=mapstd(train(:,[ind1,ind2,ind3])')';
% test=mapstd(test(:,[ind1,ind2,ind3])')';
[train,ps]=mapstd(train(:,:)');
train=train';
test=mapstd('apply',test(:,:)',ps)';

train1=train(:,ind1);
test1=test(:,ind1);
train2=train(:,ind2);
test2=test(:,ind2);
train3=train(:,[ind1,ind2,ind3,ind4,ind5]);
test3=test(:,[ind1,ind2,ind3,ind4,ind5]);
TEST_prtools1=dataset(test1,
[ones(size(SF1,1),1);2*ones(size(AG1,1),1);3*ones(size(AK1,1),1);4*ones(size(PL1,1),1);5*ones(size(PAL1,1),1);6*ones(size(KUL1,1),1)])

TRAIN_prtools1=dataset(train1,
[ones(size(SF2,1),1);2*ones(size(AG2,1),1);3*ones(size(AK2,1),1);4*ones(size(PL2,1),1);5*ones(size(PAL2,1),1);6*ones(size(KUL2,1),1)])
W1=ldc(TRAIN_prtools1);
[C1,NE1,LABLIST1] = confmat(getlabels(TEST_prtools1),TEST_prtools1*W1*labeld)
%%
TEST_prtools2=dataset(test2,
[ones(size(SF1,1),1);2*ones(size(AG1,1),1);3*ones(size(AK1,1),1);4*ones(size(PL1,1),1);5*ones(size(PAL1,1),1);6*ones(size(KUL1,1),1)])

TRAIN_prtools2=dataset(train2,
[ones(size(SF2,1),1);2*ones(size(AG2,1),1);3*ones(size(AK2,1),1);4*ones(size(PL2,1),1);5*ones(size(PAL2,1),1);6*ones(size(KUL2,1),1)])
W2=ldc(TRAIN_prtools2);
[C2,NE2,LABLIST2] = confmat(getlabels(TEST_prtools2),TEST_prtools2*W2*labeld)

TEST_prtools3=dataset(test3,
[ones(size(SF1,1),1);2*ones(size(AG1,1),1);3*ones(size(AK1,1),1);4*ones(size(PL1,1),1);5*ones(size(PAL1,1),1);6*ones(size(KUL1,1),1)])

TRAIN_prtools3=dataset(train3,
[ones(size(SF2,1),1);2*ones(size(AG2,1),1);3*ones(size(AK2,1),1);4*ones(size(PL2,1),1);5*ones(size(PAL2,1),1);6*ones(size(KUL2,1),1)])
W3=ldc(TRAIN_prtools3);
[C3,NE3,LABLIST3] = confmat(getlabels(TEST_prtools3),TEST_prtools3*W3*labeld)
```

```
CC1=CC1+C1;
CC2=CC2+C2;
CC3=CC3+C3;
% Wpca=pca(TRAIN_prtools,30)
% W=knnc(TRAIN_prtools,11);
% % W=fisherc(TRAIN_prtools*Wpca)
% %
% % W=qdc(TRAIN_prtools);
% % W=baggingc(TRAIN_prtools,ldc,100,[])
%    [C,NE,LABLIST] = confmat(getlabels(TEST_prtools),TEST_prtools*Wpca*W*labeld)
A=[SF1;AG1;AK1;PL1;PAL1;KUL1];
B=[SF2;AG2;AK2;PL2;PAL2;KUL2];
test=B;
train=A;

% ind1=[1:8,[1:8]+51]; % ind2=[21:28,[21:28]+51];% ind3=[29:36,[29:36]+51];
% ind4=[37:44,[37:44]+51]; % ind5=[9:20+[9:20]+51];

ind1=[1:8,[1:8]+49];
ind2=[18:25,[18:25]+49];
ind3=[26:33,[26:33]+49];
ind4=[34:41,[34:41]+49];
ind5=[42:49,[42:49]+49];

SW1=size(SF2,1)/size(train,1)*cov(SF2(:,ind1))+size(AG2,1)/size(train,1)*cov(AG2(:,ind1))+size(AK2,1)/size(train,1)*cov(AK2(:,ind1))+size(PL2,1)/size(train,1)*cov(PL2(:,ind1))+size(PAL2,1)/size(train,1)*cov(PAL2(:,ind1))+size(KUL2,1)/size(train,1)*cov(KUL2(:,ind1));
SW2=size(SF2,1)/size(train,1)*cov(SF2(:,ind2))+size(AG2,1)/size(train,1)*cov(AG2(:,ind2))+size(AK2,1)/size(train,1)*cov(AK2(:,ind2))+size(PL2,1)/size(train,1)*cov(PL2(:,ind2))+size(PAL2,1)/size(train,1)*cov(PAL2(:,ind2))+size(KUL2,1)/size(train,1)*cov(KUL2(:,ind2));
SW3=size(SF2,1)/size(train,1)*cov(SF2(:,ind3))+size(AG2,1)/size(train,1)*cov(AG2(:,ind3))+size(AK2,1)/size(train,1)*cov(AK2(:,ind3))+size(PL2,1)/size(train,1)*cov(PL2(:,ind3))+size(PAL2,1)/size(train,1)*cov(PAL2(:,ind3))+size(KUL2,1)/size(train,1)*cov(KUL2(:,ind3));
SW4=size(SF2,1)/size(train,1)*cov(SF2(:,ind4))+size(AG2,1)/size(train,1)*cov(AG2(:,ind4))+size(AK2,1)/size(train,1)*cov(AK2(:,ind4))+size(PL2,1)/size(train,1)*cov(PL2(:,ind4))+size(PAL2,1)/size(train,1)*cov(PAL2(:,ind4))+size(KUL2,1)/size(train,1)*cov(KUL2(:,ind4));

SM1=cov(train(:,ind1));
SM2=cov(train(:,ind2));
SM3=cov(train(:,ind3));
SM4=cov(train(:,ind4));

[trace(inv(SW1)*SM1) trace(inv(SW2)*SM2) trace(inv(SW3)*SM3) trace(inv(SW4)*SM4)]
[det(inv(SW1)*SM1) det(inv(SW2)*SM2) det(inv(SW3)*SM3) det(inv(SW4)*SM4)]

% train=mapstd(train(:,[ind1,ind2,ind3])')';
% test=mapstd(test(:,[ind1,ind2,ind3])')';

[train,ps]=mapstd(train(:,:)');
```

```
    train=train';
    test=mapstd('apply',test(:,:)',ps)';
    train1=train(:,ind1);
    test1=test(:,ind1);
    train2=train(:,ind2);
    test2=test(:,ind2);
    train3=train(:,[ind1,ind2,ind3,ind4,ind5]);
    test3=test(:,[ind1,ind2,ind3,ind4,ind5]);

    TRAIN_prtools1=dataset(train1,
    [ones(size(SF1,1),1);2*ones(size(AG1,1),1);3*ones(size(AK1,1),1);4*ones(size(PL1,1),1);5*ones(size(
    PAL1,1),1);6*ones(size(KUL1,1),1)])

    TEST_prtools1=dataset(test1,
    [ones(size(SF2,1),1);2*ones(size(AG2,1),1);3*ones(size(AK2,1),1);4*ones(size(PL2,1),1);5*ones(size(
    PAL2,1),1);6*ones(size(KUL2,1),1)])
    W1=ldc(TRAIN_prtools1);
    [C1,NE1,LABLIST1] = confmat(getlabels(TEST_prtools1),TEST_prtools1*W1*labeld)

    TRAIN_prtools2=dataset(train2,
    [ones(size(SF1,1),1);2*ones(size(AG1,1),1);3*ones(size(AK1,1),1);4*ones(size(PL1,1),1);5*ones(size(
    PAL1,1),1);6*ones(size(KUL1,1),1)])

    TEST_prtools2=dataset(test2,
    [ones(size(SF2,1),1);2*ones(size(AG2,1),1);3*ones(size(AK2,1),1);4*ones(size(PL2,1),1);5*ones(size(
    PAL2,1),1);6*ones(size(KUL2,1),1)])
    W2=ldc(TRAIN_prtools2);
    [C2,NE2,LABLIST2] = confmat(getlabels(TEST_prtools2),TEST_prtools2*W2*labeld)

    TRAIN_prtools3=dataset(train3,
    [ones(size(SF1,1),1);2*ones(size(AG1,1),1);3*ones(size(AK1,1),1);4*ones(size(PL1,1),1);5*ones(size(
    PAL1,1),1);6*ones(size(KUL1,1),1)])

    TEST_prtools3=dataset(test3,
    [ones(size(SF2,1),1);2*ones(size(AG2,1),1);3*ones(size(AK2,1),1);4*ones(size(PL2,1),1);5*ones(size(
    PAL2,1),1);6*ones(size(KUL2,1),1)])
    W3=ldc(TRAIN_prtools3);
    [C3,NE3,LABLIST3] = confmat(getlabels(TEST_prtools3),TEST_prtools3*W3*labeld)

    CC1=CC1+C1;CC2=CC2+C2;CC3=CC3+C3;
end

TTT=[sum(sum(eye(6).*CC1))/(sum(sum(CC1))),sum(sum(eye(6).*CC2))/(sum(sum(CC2))),sum(sum(ey
e(6).*CC3))/(sum(sum(CC3)))]*100
```

## Linear Bayes Normal Classifier (LDC) with use of PCA for minimizing the dimensionality of feature vector for b' part of 3$^{rd}$ stage: testing_pca_ldc.m

```matlab
clear all
load subject1featn3 % Similarly for the other subjects

CC1=zeros(6);
CC2=zeros(6);
CC3=zeros(6);
CC4=zeros(6);
for outer=1:5
   rand('seed',outer);
   I=randperm(30);

   SF1=[];AG1=[];AK1=[];PAL1=[];PL1=[];KUL1=[];
   SF2=[];AG2=[];AK2=[];PAL2=[];PL2=[];KUL2=[];
   for i=1:15
      SF1=[SF1;feat_sf{I(i)}];
      AG1=[AG1;feat_ag{I(i)}];
      AK1=[AK1;feat_ak{I(i)}];
      PAL1=[PAL1;feat_pal{I(i)}];
      PL1=[PL1;feat_pl{I(i)}];
      KUL1=[KUL1;feat_kul{I(i)}];

      SF2=[SF2;feat_sf{I(i+15)}];
      AG2=[AG2;feat_ag{I(i+15)}];
      AK2=[AK2;feat_ak{I(i+15)}];
      PAL2=[PAL2;feat_pal{I(i+15)}];
      PL2=[PL2;feat_pl{I(i+15)}];
      KUL2=[KUL2;feat_kul{I(i+15)}];

   end

A=[SF1;AG1;AK1;PL1;PAL1;KUL1];
B=[SF2;AG2;AK2;PL2;PAL2;KUL2];
test=A;
train=B;

test(:,[47,96])=[];
train(:,[47,96])=[];

[train,ps]=mapstd(train(:,:)');
train=train';
test=mapstd('apply',test(:,:)',ps)';

labels_train=[ones(size(SF2,1),1);2*ones(size(AG2,1),1);3*ones(size(AK2,1),1);4*ones(size(PL2,1),1);
5*ones(size(PAL2,1),1);6*ones(size(KUL2,1),1)];
CPCA=zeros(10,96)
```

```matlab
for inner=1:10
   rand('seed',inner);
   INDin=randperm(length(labels_train));
   train_in=train(INDin(1:round(0.7*length(INDin))),:);
   test_in=train(INDin(1+round(0.7*length(INDin)):end),:);
   LABELS_train=labels_train(INDin(1:round(0.7*length(INDin))));
   LABELS_test=labels_train(INDin(1+round(0.7*length(INDin)):end));

   for PCAcount=1:96

      TEST_prtools1in=dataset(test_in,LABELS_test);

TRAIN_prtools1in=dataset(train_in,LABELS_train);
Wpca=pca(TRAIN_prtools1in,PCAcount);
W1in=ldc(TRAIN_prtools1in*Wpca);
[C1,NE1,LABLIST1] = confmat(getlabels(TEST_prtools1in),TEST_prtools1in*Wpca*W1in*labeld);

CPCA(inner,PCAcount)=NE1;
   end
end

[YYYY, INPCA]=min(sum(CPCA)); clear CPCA;
TEST_prtools1=dataset(test,
[ones(size(SF1,1),1);2*ones(size(AG1,1),1);3*ones(size(AK1,1),1);4*ones(size(PL1,1),1);5*ones(size(PAL1,1),1);6*ones(size(KUL1,1),1)])

TRAIN_prtools1=dataset(train,
[ones(size(SF2,1),1);2*ones(size(AG2,1),1);3*ones(size(AK2,1),1);4*ones(size(PL2,1),1);5*ones(size(PAL2,1),1);6*ones(size(KUL2,1),1)])

Wpca=pca(TRAIN_prtools1,INPCA);
W1=ldc(TRAIN_prtools1*Wpca);
[C1,NE1,LABLIST1] = confmat(getlabels(TEST_prtools1),TEST_prtools1*Wpca*W1*labeld);
CALL{outer,1}=C1;
PCAmemory{outer,1}=INPCA;

W12=ldc(TRAIN_prtools1);
[C2,NE1,LABLIST1] = confmat(getlabels(TEST_prtools1),TEST_prtools1*W12*labeld);
CALL2{outer,2}=C2;

CC1=CC1+C1;
CC2=CC2+C2;

%% second fold
A=[SF1;AG1;AK1;PL1;PAL1;KUL1];
B=[SF2;AG2;AK2;PL2;PAL2;KUL2];
test=B; train=A;
test(:,[47,96])=[]; train(:,[47,96])=[];
```

```matlab
% ind1=[1:8,[1:8]+51];
% ind2=[21:28,[21:28]+51];
% ind3=[29:36,[29:36]+51];
% ind4=[37:44,[37:44]+51];
% ind5=[9:20+[9:20]+51];

[train,ps]=mapstd(train(:,:)');
train=train';
test=mapstd('apply',test(:,:)',ps)';

labels_train=[ones(size(SF1,1),1);2*ones(size(AG1,1),1);3*ones(size(AK1,1),1);4*ones(size(PL1,1),1);
5*ones(size(PAL1,1),1);6*ones(size(KUL1,1),1)];
CPCA=zeros(10,96);
for inner=1:10
   rand('seed',inner);
   INDin=randperm(length(labels_train));
   train_in=train(INDin(1:round(0.7*length(INDin))),:);
   test_in=train(INDin(1+round(0.7*length(INDin)):end),:);
   LABELS_train=labels_train(INDin(1:round(0.7*length(INDin))));
   LABELS_test=labels_train(INDin(1+round(0.7*length(INDin)):end));

   for PCAcount=1:96
     TRAIN_prtools1in=dataset(train_in,LABELS_train);
Wpca=pca(TRAIN_prtools1in,PCAcount);
W1in=ldc(TRAIN_prtools1in*Wpca);

[C1,NE1,LABLIST1] = confmat(getlabels(TEST_prtools1in),TEST_prtools1in*Wpca*W1in*labeld);

CPCA(inner,PCAcount)=NE1;
   end
end

[YYYY, INPCA]=min(sum(CPCA)); clear CPCA;
TRAIN_prtools1=dataset(train,
[ones(size(SF1,1),1);2*ones(size(AG1,1),1);3*ones(size(AK1,1),1);4*ones(size(PL1,1),1);5*ones(size(
PAL1,1),1);6*ones(size(KUL1,1),1)])

TEST_prtools1=dataset(test,
[ones(size(SF2,1),1);2*ones(size(AG2,1),1);3*ones(size(AK2,1),1);4*ones(size(PL2,1),1);5*ones(size(
PAL2,1),1);6*ones(size(KUL2,1),1)])

Wpca=pca(TRAIN_prtools1,INPCA);
W1=ldc(TRAIN_prtools1*Wpca);
[C1,NE1,LABLIST1] = confmat(getlabels(TEST_prtools1),TEST_prtools1*Wpca*W1*labeld);
CALL{outer,2}=C1;

W12=ldc(TRAIN_prtools1);
[C2,NE1,LABLIST1] = confmat(getlabels(TEST_prtools1),TEST_prtools1*W12*labeld);
CALL2{outer,2}=C2;
```

```
    PCAmemory{outer,2}=INPCA;
    CC1=CC1+C1;
    CC2=CC2+C2;
    end

    TTT=[sum(sum(eye(6).*CC1))/(sum(sum(CC1))),sum(sum(eye(6).*CC2))/(sum(sum(CC2))),
       sum(sum(eye(6).*CC3))/(sum(sum(CC3))),sum(sum(eye(6).*CC4))/(sum(sum(CC4)))]*100
```

### Linear Bayes Normal Classifier (LDC) with use of RELIEF for minimizing the dimensionality of feature vector for b' part of 3$^{rd}$ stage: testing_ relief _ldc.m

```
    clear all
    load subject1featn3 % Similarly for the other subjects

    CC1=zeros(6);
    CC2=zeros(6);
    CC3=zeros(6);
    CC4=zeros(6);
    for outer=1:5
       rand('seed',outer);
       I=randperm(30);
       SF1=[];AG1=[];AK1=[];PAL1=[];PL1=[];KUL1=[];
       SF2=[];AG2=[];AK2=[];PAL2=[];PL2=[];KUL2=[];
       for i=1:15
          SF1=[SF1;feat_sf{I(i)}];
          AG1=[AG1;feat_ag{I(i)}];
          AK1=[AK1;feat_ak{I(i)}];
          PAL1=[PAL1;feat_pal{I(i)}];
          PL1=[PL1;feat_pl{I(i)}];
          KUL1=[KUL1;feat_kul{I(i)}];

          SF2=[SF2;feat_sf{I(i+15)}];
          AG2=[AG2;feat_ag{I(i+15)}];
          AK2=[AK2;feat_ak{I(i+15)}];
          PAL2=[PAL2;feat_pal{I(i+15)}];
          PL2=[PL2;feat_pl{I(i+15)}];
          KUL2=[KUL2;feat_kul{I(i+15)}];

       end
    %%
    A=[SF1;AG1;AK1;PL1;PAL1;KUL1];
    B=[SF2;AG2;AK2;PL2;PAL2;KUL2];
    test=A;
    train=B;
    test(:,[47,96])=[];
    train(:,[47,96])=[];
    [train,ps]=mapstd(train(:,:)');
```

```
train=train';
test=mapstd('apply',test(:,:)',ps)';

labels_train=[ones(size(SF2,1),1);2*ones(size(AG2,1),1);3*ones(size(AK2,1),1);4*ones(size(PL2,1),1);
5*ones(size(PAL2,1),1);6*ones(size(KUL2,1),1)];
CPCA=zeros(10,96)
for inner=1:10
   rand('seed',inner);
   INDin=randperm(length(labels_train));
   train_in=train(INDin(1:round(0.7*length(INDin))),:);
   test_in=train(INDin(1+round(0.7*length(INDin)):end),:);
   LABELS_train=labels_train(INDin(1:round(0.7*length(INDin))));
   LABELS_test=labels_train(INDin(1+round(0.7*length(INDin)):end));

   WRel=RELIEF(train_in,LABELS_train,5000);
   [WRel2, INDrel]=sort(WRel,'descend');

   for PCAcount=1:96

      TEST_prtools1in=dataset(test_in(:,INDrel(1:PCAcount)),LABELS_test);

TRAIN_prtools1in=dataset(train_in(:,INDrel(1:PCAcount)),LABELS_train);
% Wpca=pca(TRAIN_prtools1in,PCAcount);
W1in=ldc(TRAIN_prtools1in);
[C1,NE1,LABLIST1] = confmat(getlabels(TEST_prtools1in),TEST_prtools1in*W1in*labeld);

CPCA(inner,PCAcount)=NE1;
   end
end

[YYYY, INPCA]=min(sum(CPCA)); clear CPCA;

WRel=RELIEF(train,[ones(size(SF2,1),1);2*ones(size(AG2,1),1);3*ones(size(AK2,1),1);4*ones(size(PL
2,1),1);5*ones(size(PAL2,1),1);6*ones(size(KUL2,1),1)],5000);
[WRel2, INDrel]=sort(WRel,'descend');

TEST_prtools1=dataset(test(:,INDrel(1:INPCA)),
[ones(size(SF1,1),1);2*ones(size(AG1,1),1);3*ones(size(AK1,1),1);4*ones(size(PL1,1),1);5*ones(size(
PAL1,1),1);6*ones(size(KUL1,1),1)])

TRAIN_prtools1=dataset(train(:,INDrel(1:INPCA)),
[ones(size(SF2,1),1);2*ones(size(AG2,1),1);3*ones(size(AK2,1),1);4*ones(size(PL2,1),1);5*ones(size(
PAL2,1),1);6*ones(size(KUL2,1),1)])

% Wpca=pca(TRAIN_prtools1,INPCA);
W1=ldc(TRAIN_prtools1);
[C1,NE1,LABLIST1] = confmat(getlabels(TEST_prtools1),TEST_prtools1*W1*labeld);
CALL{outer,1}=C1;
PCAmemory{outer,1}=INPCA;
```

```matlab
%%
TEST_prtools12=dataset(test,
[ones(size(SF1,1),1);2*ones(size(AG1,1),1);3*ones(size(AK1,1),1);4*ones(size(PL1,1),1);5*ones(size(
PAL1,1),1);6*ones(size(KUL1,1),1)])

TRAIN_prtools12=dataset(train,
[ones(size(SF2,1),1);2*ones(size(AG2,1),1);3*ones(size(AK2,1),1);4*ones(size(PL2,1),1);5*ones(size(
PAL2,1),1);6*ones(size(KUL2,1),1)])

W12=ldc(TRAIN_prtools12);
[C2,NE1,LABLIST1] = confmat(getlabels(TEST_prtools12),TEST_prtools12*W12*labeld);
CALL2{outer,2}=C2;

CC1=CC1+C1;
CC2=CC2+C2;
%% second fold
A=[SF1;AG1;AK1;PL1;PAL1;KUL1];
B=[SF2;AG2;AK2;PL2;PAL2;KUL2];
test=B;
train=A;

test(:,[47,96])=[];
train(:,[47,96])=[];
% ind1=[1:8,[1:8]+51];% ind2=[21:28,[21:28]+51];% ind3=[29:36,[29:36]+51];
% ind4=[37:44,[37:44]+51];% ind5=[9:20+[9:20]+51];

[train,ps]=mapstd(train(:,:)');
train=train';
test=mapstd('apply',test(:,:)',ps)';

labels_train=[ones(size(SF1,1),1);2*ones(size(AG1,1),1);3*ones(size(AK1,1),1);4*ones(size(PL1,1),1);
5*ones(size(PAL1,1),1);6*ones(size(KUL1,1),1)];
CPCA=zeros(10,96);
for inner=1:10
    rand('seed',inner);
    INDin=randperm(length(labels_train));
    train_in=train(INDin(1:round(0.7*length(INDin))),:);
    test_in=train(INDin(1+round(0.7*length(INDin)):end),:);
    LABELS_train=labels_train(INDin(1:round(0.7*length(INDin))));
    LABELS_test=labels_train(INDin(1+round(0.7*length(INDin)):end));

    WRel=RELIEF(train_in,LABELS_train,5000);
    [WRel2, INDrel]=sort(WRel,'descend');

    for PCAcount=1:96

        TEST_prtools1in=dataset(test_in(:,INDrel(1:PCAcount)),LABELS_test);
        TRAIN_prtools1in=dataset(train_in(:,INDrel(1:PCAcount)),LABELS_train);
```

```matlab
    % Wpca=pca(TRAIN_prtools1in,PCAcount);
    W1in=ldc(TRAIN_prtools1in);
    [C1,NE1,LABLIST1] = confmat(getlabels(TEST_prtools1in),TEST_prtools1in*W1in*labeld);

     CPCA(inner,PCAcount)=NE1;
    end
end
[YYYY, INPCA]=min(sum(CPCA)); clear CPCA;

WRel=RELIEF(train,[ones(size(SF1,1),1);2*ones(size(AG1,1),1);3*ones(size(AK1,1),1);4*ones(size(PL
1,1),1);5*ones(size(PAL1,1),1);6*ones(size(KUL1,1),1)],5000);
[WRel2, INDrel]=sort(WRel,'descend');
TRAIN_prtools1=dataset(train(:,INDrel(1:INPCA)),
[ones(size(SF1,1),1);2*ones(size(AG1,1),1);3*ones(size(AK1,1),1);4*ones(size(PL1,1),1);5*ones(size(
PAL1,1),1);6*ones(size(KUL1,1),1)])

TEST_prtools1=dataset(test(:,INDrel(1:INPCA)),
[ones(size(SF2,1),1);2*ones(size(AG2,1),1);3*ones(size(AK2,1),1);4*ones(size(PL2,1),1);5*ones(size(
PAL2,1),1);6*ones(size(KUL2,1),1)])

% Wpca=pca(TRAIN_prtools1,INPCA);
W1=ldc(TRAIN_prtools1);
[C1,NE1,LABLIST1] = confmat(getlabels(TEST_prtools1),TEST_prtools1*W1*labeld);
CALL{outer,2}=C1;

TRAIN_prtools12=dataset(train,
[ones(size(SF1,1),1);2*ones(size(AG1,1),1);3*ones(size(AK1,1),1);4*ones(size(PL1,1),1);5*ones(size(
PAL1,1),1);6*ones(size(KUL1,1),1)])

TEST_prtools12=dataset(test,
[ones(size(SF2,1),1);2*ones(size(AG2,1),1);3*ones(size(AK2,1),1);4*ones(size(PL2,1),1);5*ones(size(
PAL2,1),1);6*ones(size(KUL2,1),1)])

W12=ldc(TRAIN_prtools12);
[C2,NE1,LABLIST1] = confmat(getlabels(TEST_prtools12),TEST_prtools12*W12*labeld);
CALL2{outer,2}=C2;

PCAmemory{outer,2}=INPCA;
%%

CC1=CC1+C1;
CC2=CC2+C2;

end

TTT=[sum(sum(eye(6).*CC1))/(sum(sum(CC1))),sum(sum(eye(6).*CC2))/(sum(sum(CC2))),...
    sum(sum(eye(6).*CC3))/(sum(sum(CC3))),sum(sum(eye(6).*CC4))/(sum(sum(CC4)))]*100
```

## B' part of Experiments (1st and 2nd stage)

## Main Application: 8mov_15.m

```matlab
clear all
clc

% Select specific time period of the data
load data_mov

%for 8 movements
%kinhseis = [an;kl;ag;ak;pal;pl;kul;sf];
%for 6 movements
kinhseis = [ag;ak;pal;pl;kul;sf];
channels=3;

[rows col]= size(kinhseis);
categories=rows/(channels*15);
ratio=15;

% Plot all the signals of the same category in a same diagram and for each channel a figure

for j= 1:categories
  for i= 1:channels
    figure(i)
    subplot(categories,1,j), plot(kinhseis(i+(j-1)*45:3:j*45,:)')
  end
end

% Feature Extraction

for i= 1:rows
   features(i,:)=sEMG_no_can(kinhseis(i,:));
end

% Select specific features that differ in every movement

for i= 1:rows
   sel_features(i,1)=features(i,1);
   sel_features(i,2)=features(i,2);
   sel_features(i,3)=features(i,3);
   sel_features(i,4)=features(i,5);
   sel_features(i,5)=features(i,6);
   sel_features(i,6)=features(i,9);
end

for i= 1:(rows/channels)
   feat1(i,:) = [sel_features(3*i-2,:) sel_features(3*i-1,:) sel_features(3*i,:)];
end
```

```matlab
    feat=round(feat1);

    for i= 1:categories
       beg = (ratio * i) - 4 ;
       telos = ratio * i;
       for j= beg:telos
          for k= 1:col_s
             sel_examples(j-10*i,k)=feat(j,k);
          end
       end
    end
```

\* The K-means algorithm was used also in Matlab to check the success rate of classification.

```matlab
   %% Knn and Clustering with kmeans

   [rows_m col_m]= size(feat);
    % find mean
   for k = 1:categories
      beg = ratio * (k-1)+1 ;
      telos = ratio * k-5;
      for i = 1:col_m
         sum_ave(i) = 0.0;
         for j = beg:telos
            sum_ave(i)= sum_ave(i)+feat(j,i);
         end
         mean(k,i)=sum_ave(i)/(telos-beg+1);
      end
   end
   mean=round(mean);

   fid_mean = fopen('means.txt','w');
   fprintf(fid_mean,'{');
   fprintf(fid_mean,'\n');
  for i= 1:categories
      fprintf(fid_mean,'{');
      fprintf(fid_mean,'%4.0f,', mean(i,:));
      fprintf(fid_mean,'}');
      fprintf(fid_mean,',');
      %fprintf(fid_mean,'\n');
   end
   fprintf(fid_mean,'\n');
   fprintf(fid_mean,'};');

   character=col_m;
   examples_per_cat=5;
   examples= categories*examples_per_cat;
   flips = examples;

   for i = 1:examples

   for i = 1:examples
     dmin = 100000000.0;
     color = cluster_new(i);
     for j = 1:categories
        dx = 0.0;

        for k = 1:character
           dx =dx + (sel_examples(i,k) - mean(j,k))^2;
        end
        if (dx < dmin)
           color = j;
           dmin = dx;
        end
     end

     if (cluster_new(i) ~= color)
       flips=flips+1;
       cluster_new(i) = color;
     end

   results(o,p)=color;
     o=o+1;
     if o==5*categories+1
        o=1;
        p=2;
     end
     count(cluster_new(i))=count(cluster_new(i))+1;

     for j = 1:character
        sum(cluster_new(i),j)  =  sum(cluster_new(i),j)
   + sel_examples(i,j);
     end
   end

   for i = 1:categories
```

```matlab
        cluster_new(i) = 1;
    end
  o=1;p=1;
  while (flips>0)

    flips = 0;

    for j = 1:categories
      count(j) = 0;
      for i = 1:character
        sum(j,i) = 0.0;
      end
    end
```
```matlab
            for j = 1:character
                mean(i,j) = sum(i,j)/count(i);
            end
        end

    end
    res_pos(results(:,1),categories,5)
    res_pos(results(:,2),categories,5)
```

**Function of feature extraction: sEMG_no_can.m**

```matlab
function results=sEMG(emg)

max_value = max(emg);
% A. Intergrated Electromyogram (IEMG)
emg_abs=abs(emg);
IEMG=mean(emg_abs,2);

% B. Zero Crossing (ZC)
sumZC=0;

for i=1:length(emg)-1

    if emg(i)>0 && emg(i+1)<0 || emg(i)<0 && emg(i+1)>0
        f=1;
    else
        f=0;
    end

    sumZC=sumZC+f;

end

ZC=sumZC;

% C. Variance (VAR)
Variance=var(emg);

% D. Slope Sign Changes (SSC)
sumSSC=0;
for i=2:length(emg)-1
    if emg(i+1)>emg(i) && emg(i)<emg(i-1) || emg(i+1)<emg(i) && emg(i)>emg(i-1)
        f=1;
    else
        f=0;
    end
```

```matlab
      sumSSC=sumSSC+f;
   end

   SSC=sumSSC;

   % E. Waveform Lenght (WL)
   sumWL=0;

   for i=1:length(emg)-1
      sumWL=sumWL+abs(emg(i+1)-emg(i));
   end

   WL=sumWL;

   % F. Willison Amplitude (WAMP)
   sumWAMP=0;
   threshold=10; % If the threshold increases, the WAMP decreases

   for i=1:length(emg)-1
      if abs(emg(i)-emg(i+1))>threshold
         f=1;
         sumWAMP=sumWAMP+f;
      end
   end

   WAMP=sumWAMP;

   % G. Skewness
   Skew = skewness(emg(:));

   % H. Kurtosis
   Kurt = kurtosis(emg(:));
   stand_dev=Variance^(0.5);

   results=[max_value IEMG Variance/100 SSC WL/10 WAMP Skew Kurt stand_dev];
```

## Function of calculating the success rate (%) : res_pos.m

```matlab
   function result=res_pos(res_5, movements, examples)
   cnt1=0;
   for j=1:movements
      for i=(1+examples*(j-1)):(examples+examples*(j-1))
         if res_5(i)==j
            cnt1=cnt1+1;
         end
      end
   end
   pos_5=(cnt1/(movements*examples))*100;
   result=pos_5;
```

# Appendix C

## Code in Arduino

### kmeans_clas_v2.ino

```
#define examples  3
#define character 6
#define categories 6
#define chan 3
#define rows 100

int buffer[5];

void setup() {
  Serial.begin(9600);
}

void loop() {
  float features[6][chan];
  int data[chan][rows];
  // mean includes the centers of every category through the method of knn that has been
  //used in Matlab
 int mean[6][6] = {{520,355,5503,2252,81,74,},{79,46,333,415,1,18,},
                   {188,116,1152,980,16,34,}, {352,223,3450,1522,41,59,},
                   {179,119,738,967,15,27,},{248,147,1917,1098,20,44,}};

  int threshold=200;
  int i,j,l;

  //Every number of the variable category is equivalent to a movement
  char* movement[]={"ag", "ak", "pal","pl","kul","sf"};
  unsigned long sum_av;
  int threshold_wamp=10; // If the threshold is increased,  WAMP declines

  do{
     data[0][0] = analogRead(A0);
  }while (data[0][0]<50);
  // Could have been used (data[0][0]<threshold || data[1][0]<threshold || data[2][0]<threshold)
  // but it works with data[0][0]<50, which detects if there is any movement of the hand.
  // If yes, the loop will brake and the data recording will start.

  /****************************Data recording************************/
  Serial.println(" ");
  Serial.println("Begin of Movement");
  //Starts from zero in the plot
  data[0][0] = 0;  data[1][0] = 0;  data[2][0] = 0;

  plot(data[0][0],data[1][0],data[2][0]);
```

```
    for (i = 0; i < rows; i++) {
      data[0][i] = analogRead(A0);
      data[1][i] = analogRead(A2);
      data[2][i] = analogRead(A4);
      plot(data[0][i],data[1][i],data[2][i]);
      delay(10);
    }
    Serial.println(" ");
    Serial.println("End of Movement");

    Serial.println(" ");

    for (i = 0; i < rows; i++) {
      Serial.print(data[0][i]);
      Serial.print(" ");
    }
    Serial.println(" ");

   for (i = 0; i < rows; i++) {
     Serial.print(data[1][i]);
     Serial.print(" ");
   }

   Serial.println(" ");

   for (i = 0; i < rows; i++) {
     Serial.print(data[2][i]);
     Serial.print(" ");
   }

   Serial.println(" ");

   Serial.print(" ");
   Serial.println(" ");

   // Features Extraction  (Max Value, IEMG, Variance, WL, WAMP, Standard Deviation)

   // A. Max Value

    for (j=0; j < chan; j++){
     sum_av=0;
     for (i=0; i < rows; i++){
        if (data[j][i]>features[0][j]) {
           features[0][j] = data[j][i];
        }
        sum_av += data[j][i];
     }
```

```
   //B. Intergrated Electromyogram (IEMG)
     features[1][j] = sum_av/rows;
    }
    // C. Variance, D. Standard Deviation, E. Waveform Length, F. Willison Amplitude (WAMP)
    for (j=0; j < chan; j++){
     sum_av=0;
     for (i=0; i < rows; i++){
        sum_av += (data[j][i] - features[1][j])*(data[j][i] - features[1][j]);
        if (i!=rows-1) {
         features[4][j]+=abs(data[j][i+1]-data[j][i]);
        }

        if (abs(data[j][i]-data[j][i+1])>threshold_wamp){
          features[5][j]++;
        }
     }
     Serial.println(" ");
     Serial.print("Channel --> ");
     Serial.println(j+1);
     Serial.print("max: ");
     Serial.println(features[0][j]);
     Serial.print("average: ");
     Serial.println(features[1][j]);
     features[2][j] = sum_av/rows;
     Serial.print("Variance: ");
     Serial.println(features[2][j]);
     features[3][j] = pow(features[2][j],0.5);
     Serial.print("Standard Deviation: ");
     Serial.println(features[3][j]);
     Serial.print("WL: ");
     Serial.println(features[4][j]);
     Serial.print("WAMP: ");
     Serial.println(features[5][j]);
    }
    /************************kmeans*******************************/
     int cluster[examples], count[categories], category, flips, k;
    int sum[categories][character];
    double dmin, dx;

    for (i = 0; i < examples; i++)
       cluster[i] = 0;
    flips = examples;
    while (flips>0) {
      flips = 0;
      for (j = 0; j < categories; j++) {
        count[j] = 0;
        for (i = 0; i < character; i++)
    sum[j][i] = 0.0;
      }
```

```
   for (i = 0; i < examples; i++) {
      dmin = 100000000.0;
     category = cluster[i];
     for (j = 0; j < categories; j++) {
     dx = 0.0;
     for (k = 0; k < character; k++)
       dx +=  (features[k][i] - mean[j][k])*(features[k][i] - mean[j][k]);
       if (dx < dmin) {
            category = j;
            dmin = dx;
      }
     }

     Serial.print (movement[category]);
     Serial.print (" ");

     if (cluster[i] != category) {
       flips++;
       cluster[i] = category;
     }

     count[cluster[i]]++;

     for (j = 0; j < character; j++)
     sum[cluster[i]][j] += data[i][j];
    }

    for (i = 0; i < categories; i++) {
     for (j = 0; j < character; j++) {
     mean[i][j] = sum[i][j]/count[i];
     }
    }
   }
   Serial.println(" ");
   delay(100);
  }
// The function that extracts the data in the Simplot Software http://www.negtronics.com/simplot
void plot(int data1, int data2, int data3)
{  int pktSize;
   buffer[0] = 0xCDAB;          //SimPlot packet header. Indicates start of data packet
   //Size of data in bytes. Does not include the header and size fields
  buffer[1] = 4*sizeof(int);
  buffer[2] = data1;
  buffer[3] = data2;
 buffer[4] = data3;
 pktSize = 2 + 2 + (4*sizeof(int)); //Header bytes + size field bytes + data
  //IMPORTANT: Change to serial port that is connected to PC
  Serial.write((uint8_t * )buffer, pktSize);
 }
```

# Wiring Code (Arduino) using on-board micro SD of Wi-Fi Shield

Finally, I did not use this code as a part of the main code since I faced problems with SRAM in order to store all data of the movements and then to use it in order to extract features and classify the data.

## ReadWrite_2.ino

```
#include <SD.h>

File myFile;

void setup()
{
  Serial.begin(9600);  // Open serial communications and wait for port to open:
  Serial.print("Initializing SD card...");
  pinMode(10, OUTPUT);
  if (!SD.begin(4))
  {
    Serial.println("initialization failed!");
    return;
  }
  Serial.println("initialization done.");
  myFile = SD.open("test.txt", FILE_WRITE);
  // if the file opened okay, write to it:
  if (myFile) {
    Serial.print("Writing to test.txt...");
  } else {
    Serial.println("error opening test.txt");    // if the file didn't open, print an error
  }
  myFile.close();
}

void loop()
{
  int rows=60;
  int psifia=3;
  int character[rows][psifia], data[rows];
  myFile = SD.open("test.txt", FILE_WRITE);
  String value = "";
  int sensorValue;
  int l=0;

  do{
      sensorValue = analogRead(A0)-24;
      l++;
    }while (sensorValue<100);
  myFile.print(sensorValue);
  myFile.print(',');
```

```
   for (int i = 0; i < rows-1; i++) {
     sensorValue = analogRead(A0) - 24 ; //In order to have as max value the 999 and not 1023.
     myFile.print(sensorValue);
     myFile.print(',');
     l++;
   }
   myFile.print(' ');
   myFile.close();
   Serial.print("l: ");
   Serial.println(l);
   Serial.println(" ");
   Serial.println("Save data from sensors.");
   Serial.println(" ");

   // re-open the file for reading:
   myFile = SD.open("test.txt");

   if (myFile) {
     Serial.println("Open test.txt:");
     Serial.println(" ");
      for (int k = 0; k < rows; k++) {
       for (int i = 0; i < 4; i++) {
        if ( i != psifia ){
          character[k][i] = myFile.read() - '0'; //transform to integer
        }else{
          myFile.read();
         }
        data[k]=100*character[k][0] + 10*character[k][1] + character[k][2];
        }
       }
       Serial.println("Data: ");
       for (int k = 0; k < rows; k++) {
         Serial.print(data[k]);
         Serial.println(" ");
       }
   // close the file:
   myFile.close();
   } else {
   // if the file didn't open, print an error:
   Serial.println("error opening test.txt");
  }
   while(SD.exists("test.txt")) //Delete .txt file, if exists, in order to have only the new data.
   {
     /* If so then delete it */
     Serial.println(" ");
     Serial.println("test.txt already exists...DELETING");
   SD.remove("test.txt");
    }
  }
```

## Processing Code in order to save data from the sensors in .txt file

**exp_data_v1.pde**

```
import processing.serial.*;

Serial mySerial;
PrintWriter output;

int slice_count=400;
int rows=1;

void setup() {
  //Create a new file in the sketch directory
  mySerial = new Serial( this, "COM2", 9600 );  //Serial.list()[0]
  output = createWriter("positions.txt");
}

void draw() {

  String sensorValue = mySerial.readString();

  if ( sensorValue != null ) {
     output.println( sensorValue );
  }

}

void keyPressed() {
  output.flush();  //Writes the remaining data to the file
  output.close();  //Finishes the file
  exit();  //Stops the program
}
```